\documentclass[a4paper,11pt]{article}

\pdfoutput=1 

\usepackage{jheppub} 

\usepackage[T1]{fontenc} 

\usepackage{epsfig}
\usepackage{amssymb}
\usepackage{amsmath}
\usepackage{doi}
\usepackage{dsfont}
\usepackage{setspace}
 \usepackage{slashed}

\usepackage{morefloats}

\preprint{TTK-16-30}

\title{Off-shell Top Quarks with One Jet at the LHC: A Comprehensive
  Analysis at NLO QCD}

\author[a]{G. Bevilacqua,}
\author[b]{H. B. Hartanto,}
\author[b]{M. Kraus\,}
\author[b]{and M.  Worek\,}

\affiliation[a]{MTA-DE Particle Physics Research Group,
University of Debrecen, H-4010 Debrecen, Hungary}
\affiliation[b]{Institute for Theoretical Particle Physics and
  Cosmology, RWTH Aachen University, D-52056 Aachen, Germany}

\emailAdd{\\ giuseppe.bevilacqua@science.unideb.hu}
\emailAdd{ \\ hartanto@physik.rwth-aachen.de}
\emailAdd{ \\ kraus@physik.rwth-aachen.de} 
\emailAdd{ \\ worek@physik.rwth-aachen.de}

\abstract{
We present a comprehensive study of the production of top quark pairs
in association with one hard jet in the di-lepton decay channel at the
LHC. Our predictions, accurate at NLO in QCD, focus on the LHC Run II
with a center-of-mass energy of $13$ TeV. All resonant and
non-resonant contributions at the perturbative order ${\cal
O}(\alpha_s^4 \alpha^4)$ are taken into account, including irreducible
backgrounds to $t\bar{t}j$ production, interferences and off-shell
effects of the top quark and the $W$ gauge boson. We extensively
investigate the dependence of our results upon variation of
renormalisation and factorisation scales and parton distribution
functions in the quest for an accurate estimate of the theoretical
uncertainties. Additionally, we explore a few possibilities for a
dynamical scale choice with the goal of stabilizing the perturbative
convergence of the differential cross sections far away from the
$t\bar{t}$ threshold. Results presented here are particularly relevant
for searches of new physics as well as for precise measurements of the
top-quark fiducial cross sections and top-quark properties at the
LHC.}


\begin{document} 
\maketitle
\flushbottom

%
\section{Introduction}
%

The large top-quark pair production cross section at the LHC, a $pp$
collider, makes it a unique laboratory for studying the behaviour of
QCD at the highest accessible energy to date.  Besides the
determination of the top-quark mass ($m_t$) and the strong coupling
constant ($\alpha_s$) key measurements include fiducial cross
sections, various infra-red safe differential distributions, spin
correlations, inclusive top-quark charge asymmetry as well as leptonic
charge asymmetry together with top-quark couplings to gauge bosons and
the Standard Model (SM) Higgs boson. In addition, work on constraining
parton distribution functions mainly for the dominant gluon-gluon
production channel with the help of the total cross section and
various differential distributions is ongoing.  The decays of
top quarks to charged leptons, neutrinos and b-quarks make this
process a primary source of background in many searches for new
physics.  Therefore, an accurate modelling of top-quark events forms
an important part of the LHC physics programme.  At LHC energies,
however, a large fraction of top-quark pairs is accompanied by
additional hard jets.  To estimate the size of the $t\bar{t}j$
contribution in the inclusive $t\bar{t}$ sample we show in Table
\ref{tab:1} the cross section for the on-shell $pp \to t\bar{t}j$
production at NLO in QCD with various $p_{T , \,j}$ cuts on
the hard jet.  Also shown is its ratio to the inclusive $pp \to
t\bar{t}$ production at the same level of accuracy.  Results are given
for the LHC Run II energy of 13 TeV, the top-quark mass of $m_t =
173.2$ GeV and for $\mu=\mu_R=\mu_F= m_t$.  For parton
distribution functions (PDFs), the  CT14nlo set has been employed. We can
observe that, for example, by requiring a minimal transverse momentum of
$40$ GeV for the additional jets will result in a contribution from
$t\bar{t}j$ events of more than $40\%$.
%
\begin{table}[t!]
\caption{\label{tab:1} \it The NLO cross section for the on-shell $pp
\to t\bar{t}j+X$ production with various values for $p_{T,\,j}$ cut on the
hardest jet. Also shown is its ratios to the NLO cross section for the
on-shell $pp\to t\bar{t} +X$ production.  Results are obtained for the
LHC Run II energy of $13$ TeV, the top-quark mass of $m_t=173.2$ GeV,
$\mu_R=\mu_F =m_t$ and for the CT14nlo PDF set.  }
\begin{center}
\begin{tabular}{ccc}
\hline\hline
&&\\
 $p_{T, \,j}$ ~~[GeV] &$\sigma^{\rm NLO}_{t\bar{t}j}$ ~~[pb] &
 \textsc{Ratio}  ~$[\%]$\\
&&\\
\hline
\hline
&&\\
40 &296.97 $\pm$ 0.29 & 41\\
60 &207.88 $\pm$ 0.19 & 29 \\
80 &152.89 $\pm$ 0.13& 21 \\
100 &115.60 $\pm$ 0.14& 16\\
120 &~~89.05 $\pm$ 0.10& 12\\
&&\\
\hline\hline
\end{tabular}
 \end{center}
\end{table}
%
From an experimental point of view, jets not originating from the
decay of the top quark and top antiquark, but arising from quark and
gluon radiation produced in association with the $t\bar{t}$ system
need to be understood very precisely since their appearance affects the
reconstruction of the $t\bar{t}$ event.  The additional jet activity
can be used to examine the underlying production and decay mechanisms
even further and to design new methods for a sizeable reduction of QCD
backgrounds
\cite{Rainwater:1996ud,DelDuca:2004wt,Figy:2007kv}. Because of its 
large production rate, the $pp\to t\bar{t}j$ process is a sizeable
background process for SM studies or searches for new physics that
involve a production of $W^+ W^-$ gauge boson pairs in association
with jets \cite{Mangano:2008ha,
Englert:2008tn,Englert:2008wp,Campanario:2010mi}. The most prominent
example is  SM Higgs boson production in vector boson fusion,
where $t\bar{t}j$ production is the dominant background process
\cite{Rainwater:1999sd,Kauer:2000hi}. Another example is  the
production of top-quark flavour violating resonances that can be
singly produced in association with the top quark at the LHC
\cite{Gresham:2011dg}. Searches for new heavy resonances, a color
singlet $W^\prime$ or a colour triplet $\phi^a$, produced in
association with the top quark have been performed by both ATLAS and
CMS collaborations. Limits on the mass and the coupling of $W^\prime$
and $\phi^a$ have been set by analyzing the $t+q$ and the $\bar{t}+q$
invariant mass spectrum in $t\bar{t}j$ candidate events
\cite{Aad:2012em,Chatrchyan:2012su}.  A search for new physics can be
also performed by looking for effects on the top-quark dipole moments,
i.e.  chromo-electric dipole and chromo-magnetic dipole
moments, which can be parametrised by adding an effective term to the
top-quark-gluon gauge coupling
\cite{Buchmuller:1985jz,AguilarSaavedra:2008zc,
Aguilar-Saavedra:2014iga}.  Anomalous $t\bar{t}g$ couplings would lead
to a significant modification of the $t\bar{t}$ spin correlation in
$t\bar{t}$ and $t\bar{t}j$ systems that might be visualised in the
normalised cross sections as a function of the difference in azimuthal
angle between the two charged leptons, $|\Delta\phi(\ell \ell)|$, in
the dilepton decay mode \cite{CMS:2014bea,Khachatryan:2016xws}. 
Additionally, $t\bar{t}j$ production can be employed in the top-quark mass
extraction by studying normalised differential cross sections as a
function of the inverse invariant mass of the $t\bar{t}j$ system
\cite{Alioli:2013mxa}. The method has already been successfully used
by experimental groups at the LHC \cite{Aad:2015waa,CMS:2016khu}.
Both the ATLAS and CMS collaborations are carefully examining $pp\to
t\bar{t}$+jets production. The studies performed at the LHC include
measurements of jet activity in top-quark events, measurements of
$t\bar{t}$ production with a veto on additional central jet activity
and measurements of heavy flavor composition of $t\bar{t}$ events
\cite{ATLAS:2012al,Chatrchyan:2014gma,
Aad:2014iaa,Aad:2015yja,Khachatryan:2015mva,
ATLAS-CONF-2015-065,Aaboud:2016omn}.  For example, the ATLAS experiment
has measured using $4.6\, {\rm fb}^{-1}$ of data at $\sqrt{s}=7$ TeV
the fiducial $t\bar{t}$ cross section as a function of the light jet
multiplicity for up to eight jets with jet $p_T$ thresholds of $25$,
$40$, $60$, and $80$ GeV. A precision of the order of $10\%$ has
been obtained for the $\sigma_{t\bar{t}j}$ contribution, while for the
differential cross section as a function of transverse momentum of the
hardest light jet, a precision between $10\%$ and $16\%$ has been
reached. Similar studies have been repeated at $8$ and $13$ TeV with
$20.3 \,{\rm fb}^{-1}$ and $3.2 \, {\rm fb}^{-1}$ of $pp$ collision
data respectively.  In the former case experimental uncertainties
remain the same, whereas in the latter they are quite large, of the
order of $25\%-40\%$, due to low statistics. However, the situation
will improve very soon once more data is analysed. On the other hand,
a very recent CMS study \cite{CMS-PAS-TOP-15-006} at $\sqrt{s} = 8$
TeV with an integrated luminosity of $19.7 \, {\rm fb}^{-1}$ has shown
that the total cross section for $t\bar{t} \,+ \ge 1$ jet production
can be measured with the total experimental uncertainty of the order
of $7\%$.  Additional jet activity in $t\bar{t}$ events has also
been investigated by analysing the so-called gap-fraction
distributions. The ATLAS and CMS collaborations have vetoed
events that contain an additional jet with transverse momentum above a
given threshold in a central rapidity interval.  The fraction of
events surviving the jet veto, i.e. the gap fraction, has been
presented in these studies as a function of the threshold. Owing to
the rich top-quark physics program at the LHC and to the precision,
which has already been achieved, it is of great importance to reduce
uncertainties for the $t\bar{t}j$ process also on the theory side. In
this respect, the need of precise theoretical predictions for various
physical observables in the $pp\to t\bar{t}j$ production process is
indisputable.

The NLO corrections to $pp\to t\bar{t}j+X$ production have first been 
calculated in \cite{Dittmaier:2007wz,Dittmaier:2008uj} for stable
top quarks. Afterwards, LO top-quark decays in the narrow width
approximation (NWA) have been included
\cite{Melnikov:2010iu}. Subsequently, NLO top-quark decays in the NWA,
including $t\to Wbj$, have been added consistently
\cite{Melnikov:2011qx}. A different approach to this process is
described in \cite{Kardos:2011qa,Alioli:2011as,Czakon:2015cla}, where
on-shell $t\bar{t}j+X$ production at NLO QCD is matched to parton
shower programs following either the POWHEG procedure
\cite{Nason:2004rx,Frixione:2007vw,Alioli:2010xd} or the MC$@$NLO one
\cite{Frixione:2002ik}.  Finally, very recently, a complete
description of top-quark pair production in association with a jet in
the dilepton channel has been provided at NLO in QCD
\cite{Bevilacqua:2015qha}. In this calculation all non-resonant
diagrams, interferences and off-shell effects of the top quark have
been consistently taken into account together with non-resonant and
off-shell effects due to the finite $W$ gauge boson width. The
integrated cross section together with the scale dependence of the
total cross section and a few differential cross sections for the LHC
Run I centre-of-mass energy of $\sqrt{s} =8$ TeV have been
studied there.  In this paper we extend our previous study on
the NLO QCD corrections to $pp\to e^+ \nu_e \mu^- \bar{\nu}_\mu
b\bar{b}j+X$ production at the LHC. We shall supplement the previous
discussion with more results for the current LHC centre-of-mass 
energy of $\sqrt{s} = 13$ TeV. To be more precise, we shall present
integrated and differential cross sections and estimate theoretical
uncertainties as provided by the scale variation. Furthermore, we
shall include dynamical scales in our study. Moreover various PDF
parameterisations will be studied and in each case internal PDF
uncertainties will be evaluated.

Let us note at this point, that full off-shell top-quark effects at NLO
have already been considered in the literature for simpler processes,
i.e. top-quark pair production and top-quark pair production in
association with the SM Higgs boson
\cite{Denner:2010jp,Bevilacqua:2010qb,Denner:2012yc,Frederix:2013gra,
Cascioli:2013wga,Heinrich:2013qaa,Denner:2015yca,Denner:2016jyo}.

The article is organised as follows. In the next section we describe
the details of our calculation. All ingredients, methods and Monte
Carlo programs, that are needed for our NLO QCD calculations, are
listed and described briefly.  In Section \ref{Details of the
Calculations} we additionally list all checks that have been performed
to ensure the correctness of our results. Numerical results for the
integrated and differential cross sections for various
renormalisation, $\mu_R$, and factorisation, $\mu_F$, scale choices
are presented in Section \ref{Results for the LHC Run II Energy of 13
TeV}. The theoretical uncertainty of the total cross section,
associated with neglected higher order terms in the perturbative
expansion, which are  estimated by varying the renormalisation and
factorisation scales independently by a factor 2, are also given there.
Additionally, the theoretical uncertainty stemming from various
parameterisations of PDFs are investigated in Section \ref{Results for
the LHC Run II Energy of 13 TeV} together with their internal PDFs
errors. Finally, in Section \ref{Conclusions} we give our
conclusions.

%
\section{Details of the Calculations}
\label{Details of the Calculations}
%

%
\begin{figure}[t!]
\caption{\it A representative set of Feynman diagrams, involving two
(a), one (b) and no top-quark resonances (c), contributing to the
leading order $pp \to e^+\nu_e \mu^- \bar{\nu}_\mu b\bar{b}j+X$
process at ${\cal O}(\alpha_s^3 \alpha^4) $.  The last diagram
(d) with a single $W$ boson resonance contributes to the
off-shell effects of the $W$ gauge boson.  }
\begin{center}
\includegraphics[width=1.0\textwidth]{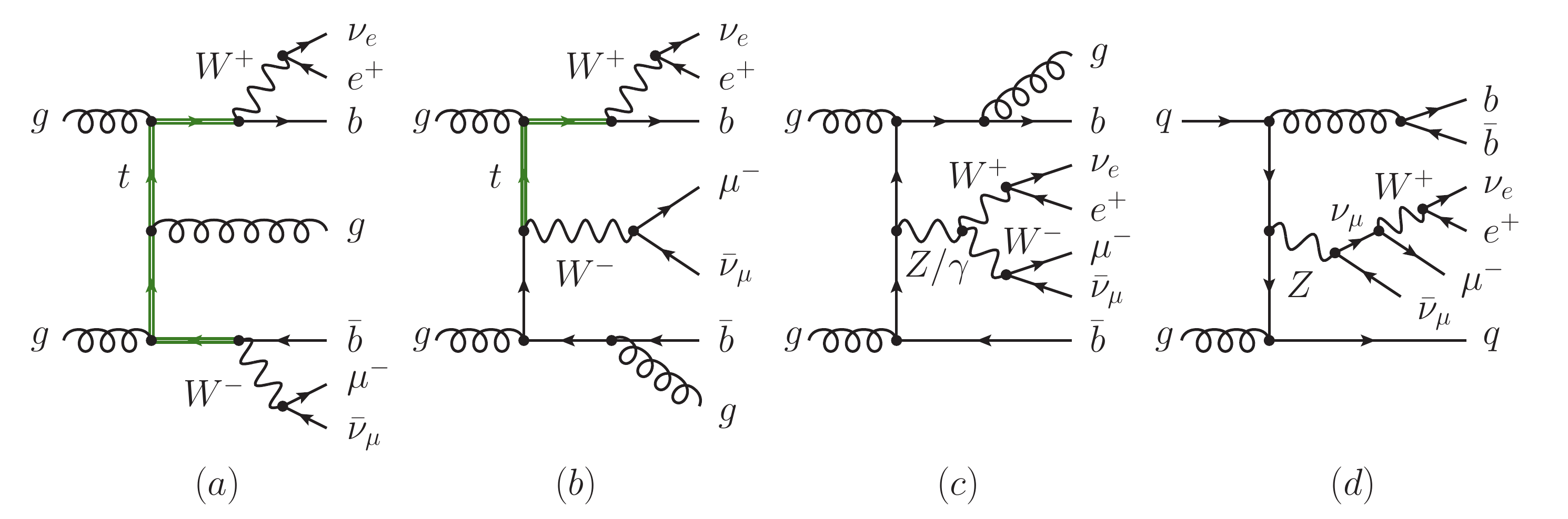}
\end{center}
\label{fig:1}
\end{figure}
%
\begin{figure}[t!]
\caption{\it A representative set of Feynman diagrams, involving
heptagons (a and d), hexagons (e and f), the pentagon diagram (b) and
the box diagram (c) contributing to virtual corrections to the $pp \to
e^+\nu_e \mu^- \bar{\nu}_\mu b\bar{b}j+X$ process at ${\cal
O}(\alpha_s^4 \alpha^4) $.}
\begin{center}
\includegraphics[width=1.0\textwidth]{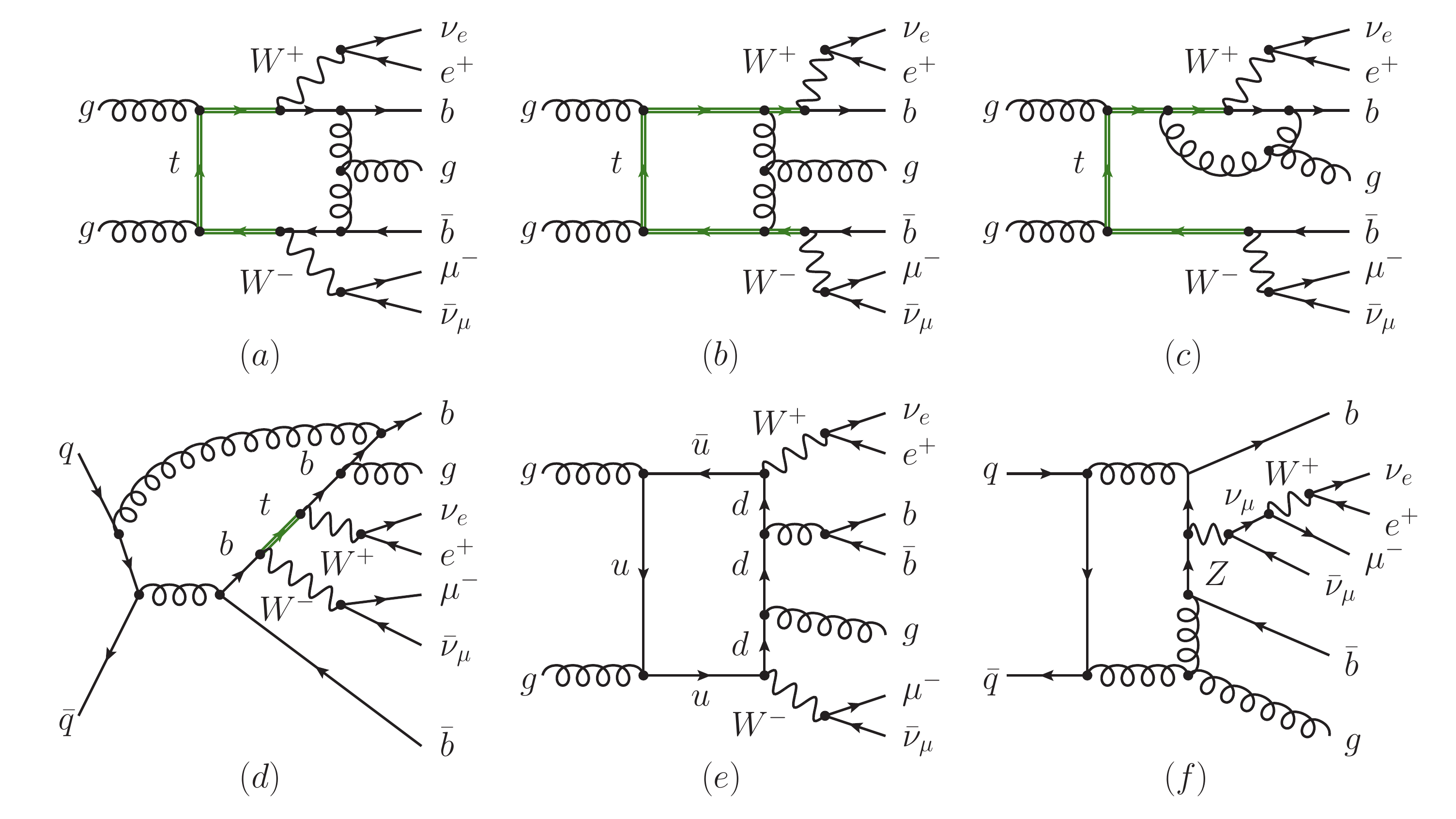}
\end{center}
\label{fig:2}
\end{figure}
%
For the $pp\to e^+\nu_e \mu^-\bar{\nu}_\mu b\bar{b}j+X$ production
process at the leading order (LO) in perturbative expansion and at 
${\cal O}( \alpha_s^3 \alpha^4)$, the contribution from the
following partonic subprocesses need to be taken into account:
\begin{equation}
\begin{split}
gg\to e^+\nu_e
\mu^-\bar{\nu}_\mu b\bar{b}g \,, \\
 gq \to e^+\nu_e
\mu^-\bar{\nu}_\mu b\bar{b}q \,,  \\
g\bar{q}\to e^+\nu_e
\mu^-\bar{\nu}_\mu b\bar{b}\bar{q}\,,  \\
q\bar{q}\to e^+\nu_e
\mu^-\bar{\nu}_\mu b\bar{b}g\,,
\end{split}
\end{equation}
where $q=u,d,c,s$. A representative set of Feynman diagrams
contributing to the process under consideration is depicted in Figure
\ref{fig:1}. In total, the $gg \to e^+\nu_e \mu^-\bar{\nu}_\mu
b\bar{b}g$ subprocess involves 508 tree diagrams, on the other hand 
the $q\bar{q} \to e^+\nu_e \mu^-\bar{\nu}_\mu b\bar{b}g$ subprocess as
well as $gq\to e^+\nu_e \mu^-\bar{\nu}_\mu b\bar{b} q$ and
$g\bar{q}\to e^+\nu_e \mu^-\bar{\nu}_\mu b\bar{b} \bar{q}$ subprocesses, that
are all related by crossing symmetry, comprise 234 tree diagrams
each. Even though we do not actually employ Feynman diagrams in our
calculations, we present them as a measure of the complexity. The
calculation of scattering amplitudes is performed by means of an
automatic off-shell iterative algorithm
\cite{Draggiotis:1998gr,Draggiotis:2002hm,Papadopoulos:2005ky}, which
is implemented within the \textsc{Helac-Dipoles} package
\cite{Czakon:2009ss} and the \textsc{Helac-Phegas} Monte Carlo program
\cite{Kanaki:2000ey,Cafarella:2007pc}. The latter framework has been used to
cross check our LO results.  For the phase-space integration
\textsc{Parni} \cite{vanHameren:2007pt} and \textsc{Kaleu}
\cite{vanHameren:2010gg} have been employed. 

At  NLO, virtual corrections are obtained from the interference of the one-loop
diagrams with the tree level amplitude. They can be classified into
self-energy, vertex-, box-, pentagon-, hexagon- and heptagon-type
corrections. A representative set of one-loop diagrams contributing to
the process is shown in Figure ~\ref{fig:2}. To give an estimate of
the complexity of the calculations we present the number of one-loop
Feynman diagrams as obtained with \textsc{Qgraf}
\cite{Nogueira:1991ex}.  For the dominant gluon-gluon production
channel $39180$ one-loop diagrams have been counted. In more details,
the most complex contributions comprise $120$ heptagons and $1155$
hexagons with tensor integrals up to rank six.  Virtual corrections
are evaluated in $d = 4 - 2\epsilon$ dimensions in the 't
Hooft-Veltman version of the dimensional regularisation
\cite{'tHooft:1972fi} within the Feynman gauge for gauge bosons. The
singularities coming from infrared divergent pieces are canceled by
the corresponding ones arising from the counterterms of the adopted
subtraction scheme integrated over the phase space of the unresolved
parton. The finite contributions of the loop diagrams are evaluated
numerically in $d = 4$ dimensions.  To ensure numerical stability of
our calculations we perform a few tests. Since every partonic
subprocess at ${\cal O}(\alpha_s^4\alpha^4)$ has at least one gluon as
an external particle, we have used the Ward identity test for every
phase space point. Higher precision has been used to recompute events
which fail the gauge-invariance check. As a second test we have
verified cancelation of infrared poles.  We have also cross-checked
our results with the publicly available
\textsc{MadGraph5}${}_{-}$\textsc{aMC@NLO} code \cite{Alwall:2014hca}.
More specifically we have compared results for the virtual NLO
contribution to the squared amplitude, $2\Re \left({\cal M}_{\rm
tree}^*{\cal M}_{\rm one-loop} \right)$, for a few phase-space
points. The calculation of the virtual corrections has been achieved
with the help of the package \textsc{Helac-1Loop}
\cite{vanHameren:2009dr} which incorporates \textsc{CutTools}
\cite{Ossola:2007ax} and \textsc{OneLOop} \cite{vanHameren:2010cp} as
its cornerstones.
%
\begin{figure}[t!]
\caption{\it A representative set of Feynman diagrams involving two
(a), one (b) and no top-quark resonances (c and d) contributing to the
real emission corrections to the $pp \to e^+\nu_e \mu^- \bar{\nu}_\mu
b\bar{b}j+X$ process at ${\cal O}(\alpha_s^4\alpha^4) $. }
\begin{center}
\includegraphics[width=1.0\textwidth]{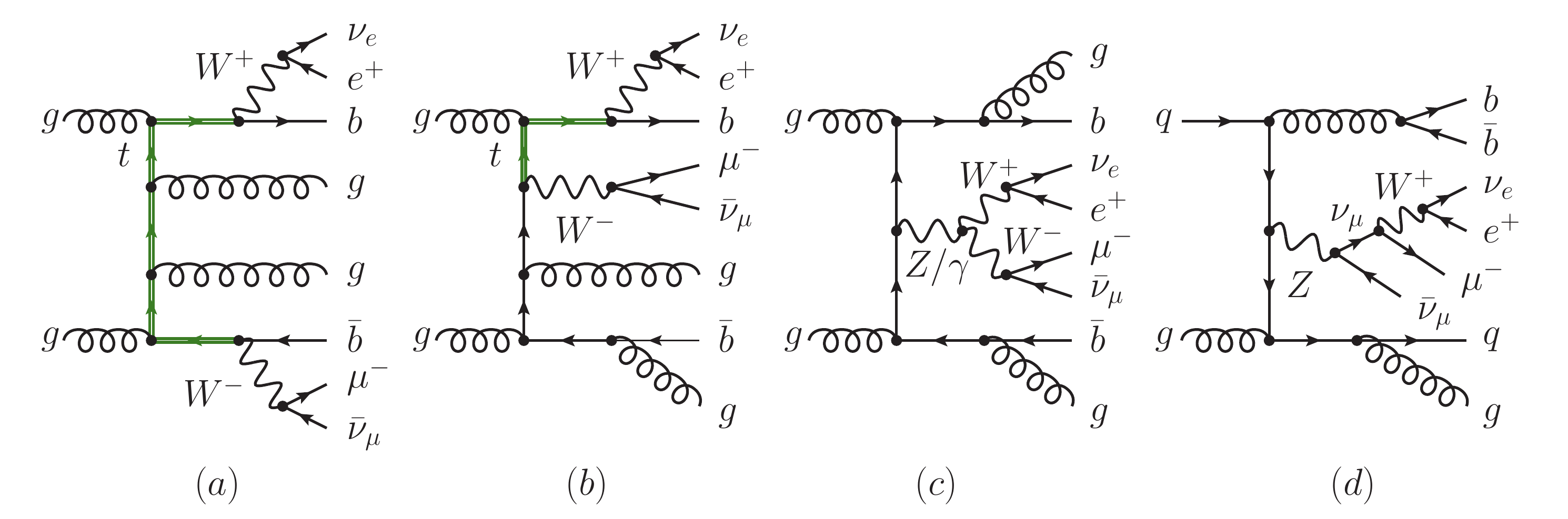}
\end{center}
\label{fig:3}
\end{figure}
%
\begin{table}[t!]
\caption{\label{tab:2}  \it
The list of partonic subprocesses contributing to the subtracted real
emissions for the $pp\to e^+\nu_e \mu^-\bar{\nu}_\mu b\bar{b}j +X$
process. Also shown are the number of Feynman diagrams, as well as the
number of Catani-Seymour and Nagy-Soper subtraction terms.  Notation:
$q$ and $\bar{q}$ stands for $up-$ or $down-$type quark, $Q$ and
$\overline{Q}$ denotes charm or strange quark. }
\begin{center}
\begin{tabular}{cccc}
\hline\hline
&&&\\
 \textsc{Partonic } & \textsc{Number Of } &
\textsc{Number Of  } & \textsc{Number Of  }\\
 \textsc{Subprocess} & \textsc{Feynman Diagrams} &
\textsc{CS Dipoles} & \textsc{NS Subtractions}\\
&&&\\
\hline
\hline
&&&\\
$gg\to e^+\nu_e\mu^-\bar{\nu}_\mu b\bar{b}gg$
&4447&56 &14\\
&&&\\
$gg\to e^+\nu_e\mu^-\bar{\nu}_\mu b\bar{b} q\bar{q}$ &1952& 40&10\\
$gq\to e^+\nu_e\mu^-\bar{\nu}_\mu b\bar{b} gq$ &1952& 40&10\\
$g\bar{q}\to e^+\nu_e\mu^-\bar{\nu}_\mu b\bar{b} g\bar{q}$&1952&
                                                                 40&10\\
$q\bar{q}\to e^+\nu_e\mu^-\bar{\nu}_\mu b\bar{b} gg$&1952& 40&10\\
&&&\\
$qq\to e^+\nu_e\mu^-\bar{\nu}_\mu b\bar{b} qq$&930 &20 &5\\
$q\bar{q}\to e^+\nu_e\mu^-\bar{\nu}_\mu b\bar{b} q\bar{q}$& 930& 16&4\\
$\bar{q}\bar{q}\to e^+\nu_e\mu^-\bar{\nu}_\mu b\bar{b}
  \bar{q}\bar{q}$&930&20&5\\
&&&\\
$qq^\prime \to e^+\nu_e\mu^-\bar{\nu}_\mu b\bar{b} qq^\prime$&501&12
                     &3\\
$q\bar{q}\to e^+\nu_e\mu^-\bar{\nu}_\mu b\bar{b} q^\prime
  \bar{q}^{\,\prime}$&501& 8& 2\\
$q\bar{q}^{\,\prime}\to e^+\nu_e\mu^-\bar{\nu}_\mu b\bar{b}
  q\bar{q}^{\,\prime}$&501&12&3\\
$\bar{q}\bar{q}^{\,\prime}\to e^+\nu_e\mu^-\bar{\nu}_\mu b\bar{b}
  \bar{q}\bar{q}^{\,\prime}$&501&12&3\\
&&&\\
$qQ\to e^+\nu_e\mu^-\bar{\nu}_\mu b\bar{b} qQ$&465&12&3\\
$q\bar{q}\to e^+\nu_e\mu^-\bar{\nu}_\mu b\bar{b}
  Q\overline{Q}$&465&8&2\\
$q\overline{Q}\to e^+\nu_e\mu^-\bar{\nu}_\mu b\bar{b}
  q\overline{Q}$&465&12&3\\
$\bar{q}\overline{Q}\to e^+\nu_e\mu^-\bar{\nu}_\mu b\bar{b}
  \bar{q}\overline{Q}$&465&12&3\\
&&&\\
$qQ\to e^+\nu_e\mu^-\bar{\nu}_\mu b\bar{b} q^\prime Q^\prime$&36&4&1\\
$q\overline{Q}\to e^+\nu_e\mu^-\bar{\nu}_\mu b\bar{b}
  q^\prime\overline{Q}^\prime$&36&4&1\\
$q\bar{q}^{\,\prime}\to e^+\nu_e\mu^-\bar{\nu}_\mu b\bar{b}
  Q\overline{Q}^{\, \prime}$&36&4&1\\
$\bar{q}\overline{Q}\to e^+\nu_e\mu^-\bar{\nu}_\mu b\bar{b} \bar{q}^{\,\prime}
  \overline{Q}^{\,\prime}$&36&4&1\\
&&&\\
$gg\to e^+\nu_e\mu^-\bar{\nu}_\mu b\bar{b} b\bar{b} $&3904&48&12\\
&&&\\
$q\bar{q}\to e^+\nu_e\mu^-\bar{\nu}_\mu b\bar{b} b\bar{b} $&930&16&4\\
&&&\\
\hline\hline
\end{tabular}
 \end{center}
\end{table}
%
The first code contains an implementation of the OPP method for the
reduction of one-loop amplitudes at the integrand level
\cite{Ossola:2006us,Ossola:2008xq,Mastrolia:2008jb,Draggiotis:2009yb}, 
while the second one is dedicated to the evaluation of the one-loop
scalar functions. Renormalisation is done, as usual, by evaluating
tree-level diagrams with counterterms. For our process, we chose to
renormalise the coupling in the $\overline{\rm MS}$ scheme with five
active flavours and the top quark decoupled. The mass renormalisation
is performed in the on-shell scheme. 

The real emission corrections to the LO process arise from tree-level
amplitudes with one additional parton, i.e. an additional gluon, or a
quark anti-quark pair replacing a gluon. For the calculation of the
real emission contributions, the package \textsc{Helac-Dipoles} has
been employed. It implements the massless dipole formalism of Catani
and Seymour \cite{Catani:1996vz}, as well as its massive version as
developed by Catani, Dittmaier, Seymour and Trocsanyi
\cite{Catani:2002hc}, for arbitrary helicity eigenstates and colour
configurations of the external partons \cite{Czakon:2009ss}. Moreover,
a new subtraction formalism, first introduced by Nagy and Soper in the
formulation of an improved parton shower
\cite{Nagy:2007ty,Nagy:2014mqa}, is also included in the
framework. The Nagy-Soper subtraction scheme \cite{Bevilacqua:2013iha}
makes use of random polarisation and colour sampling of the external
partons. A phase space restriction on the contribution of the
subtraction terms is included for both subtraction cases. Also for the
real corrections, we adopt the \textsc{Kaleu} phase-space generator
that is equipped with additional, special channels that proved to be
important for phase-space optimisation.  All possible subprocesses
contributing to the real emission part can be classified into various
categories presented in Table \ref{tab:2}, together with the number of
Feynman diagrams, the Catani-Seymour dipoles and the Nagy-Soper
subtraction terms corresponding to each subprocess. Typical examples
of the real emission graphs are displayed in Figure
\ref{fig:3}. Having two independent subtraction schemes available, we
were able to cross check the correctness of the real corrections by
comparison between the two results.

Finally, since the produced top quarks are unstable particles, the
inclusion of the decays is performed in the complex mass scheme
\cite{Denner:1999gp,Denner:2005fg,Denner:2012yc}, which respects gauge
invariance.  At the amplitude level (at LO and NLO) we simply
incorporate $\Gamma_t$ into the definition of the squared top-quark
mass as follows
\begin{equation}
\mu^2_t=m_t^2-im_t \Gamma_t\,,
\end{equation}
where $\mu^2_t$ is identified with the position of the pole of the top-quark
propagator. All matrix elements are, thus, evaluated using complex
masses and the top-quark mass counter-term $\delta\mu_t$ is related to
the top-quark self-energy at $p^2_t=\mu^2_t$. Another non trivial aspect
of this  substitution consists of the evaluation of one-loop scalar
integrals in the presence of complex masses.  In our case this part is
done by \textsc{OneLOop}, which  supports complex masses.  

To summarise, our computational system is based on
\textsc{Helac-1Loop} and \textsc{Helac-Dipoles}, which are both parts
of the \textsc{Helac-NLO} MC program \cite{Bevilacqua:2011xh}. The
framework relies upon a number of optimizations designed to speed-up
the evaluation of the virtual and real corrections while keeping an 
acceptable numerical precision. Prominent examples are the use of
Monte Carlo sampling over colour configurations and
polarisations/helicities and reweighting techniques for the
calculation of the virtual part, or the adoption of a phase space
restriction for the calculation of the real-emission part. All these
techniques have been extensively used and proved their efficiency in
our previous calculations \cite{Bevilacqua:2009zn,
Bevilacqua:2010ve,Bevilacqua:2010qb,Bevilacqua:2011aa,
Bevilacqua:2012em,Bevilacqua:2013taa,Bevilacqua:2014qfa}. At the same
time, given the complexity of the current project, it has been
necessary to extend our computational framework with new
functionalities and improvements which proved essential for the
feasibility of the calculation.   Without putting too much weight on
technical details, one relevant improvement concerns the optimisation
of the algorithms for the generation of the skeleton files, which
store all the necessary information for the calculation of the
amplitudes in the \textsc{Helac-NLO} system. To be more precise,
skeletons contain the full set of instructions for the recursive
evaluation of amplitudes according to the Dyson-Schwinger algorithm,
together with relevant accessory information such as the number of
external particles, flavour assignments and colour-connection
configurations. This information is evaluated in the form of integers
and stored once for all in skeleton files during the so-called
initialisation phase. In the subsequent phase, skeletons are read to
provide the instructions to return the actual value of the
amplitude. For more details we refer the interested reader to our
previous publications \cite{Cafarella:2007pc,Bevilacqua:2011xh}. It
should be clear that achieving an efficient generation of skeletons is
the fundamental prerequisite for the whole calculation.  
Typically, the combinatorics of diagram topologies become quickly
very complex when the number of external particles increases.  In the
\textsc{Helac-NLO} software, a top-down approach is used to obtain all
currents needed in the Dyson-Schwinger recursive representation of the
amplitude. All possible vertices are first scanned in order to select
all non-zero sub-amplitudes. Afterwards the program checks whether the
selected sub-amplitudes are indeed contributing to the final amplitude
under consideration. The number of loop topologies rapidly increases
with the number of external particles and puts serious challenges
starting from $2 \to 5$ processes, where the efficient selection and
bookkeeping of topologies becomes a critical issue for the feasibility
of the calculation. To this end a few optimizations have been
introduced for the generation of the skeleton files in the
\textsc{Helac-1Loop} program. The most relevant comprises  
an introduction of  the additional filter which performs a pre-selection
of topologies based on the information of particles that are attached
to the loop.  In this way, assuming a specific model, which is the
Standard Model in our case, a large fraction of configurations allowed
by combinatorics can be just discarded a priori  without need to go
through subsequent and more time-consuming steps that scan individual
vertices. A similar approach has also been applied to the treatment of
tree-level processes by looking at the external particle content in
each Dyson-Schwinger current. This increases dramatically the
efficiency of  the generation of skeleton files also in \textsc{Helac-Dipoles}.
Finally, we have exploited the fact that the procedure of computing
skeleton files consists of several independent modules for each loop
topology (i.e. heptagons, hexagons etc.) and colour-connection
configuration. Thus, it is possible to perform parallel runs that are
dedicated to the individual pieces of the skeleton file.  All parts
are put together in the end, which reduces considerably the computing
time. Using these optimisations we have achieved a reduction of one
order of magnitude in the generation of skeleton files for the process
under consideration. Another improvement in \textsc{Helac-NLO}
is the implementation of a new option for selecting automatically the
desired perturbative order in $\alpha$ and $\alpha_s$, preserving at
the same time the structure and the advantages of the Dyson-Schwinger
recursion algorithm for the construction of the amplitudes. This
modification is particularly useful for our project given that we are
interested in mixed contributions, i.e. ${\cal O}(\alpha_s^3\alpha^4)$
at LO and ${\cal O}(\alpha_s^4\alpha^4)$ at NLO. The modifications
summarised above make the calculation feasible. Due to high demands in
terms of CPU time it is, however, very expensive to repeat the
calculation for different choices of scales and PDFs, as is required
for a proper assessment of the theoretical uncertainties. To be able
to study scale and PDF uncertainties in a timely manner, we have made
use of unweighting techniques to produce event samples for the central
scale and PDF set, which are then reweighted to get results for
different sets of scales and PDFs. To be more precise, building on
\cite{Bern:2013zja} we have implemented in \textsc{Helac-NLO} the
apparatus for the generation of Ntuples of events. The Ntuple format
shows a clear advantage for changing kinematical cuts or observables,
which can be obtained without need of any additional rerunning of the
code. Furthermore, any change in scales or PDFs can be accommodated by
simple reweighting, provided that the necessary matrix-element
information is stored in the Ntuples.
%
\begin{table}[t!]
\caption{\label{tab:ntuples} \it Number of events, number of files and
  the averaged number of events per file as well as the total size per
contribution for the different Ntuple samples for the $pp\to e^+\nu_e
\mu^-\bar{\nu}_\mu b\bar{b}j +X$ process.}
  \begin{center}
  \begin{tabular}{lcccc}
    \hline\hline 
   &&&&\\
    \textsc{Contribution} & \textsc{Nr. of Events} & \textsc{Nr. of
                                                     files} 
& \textsc{(avg) events/file} & \textsc{Size} \\
&&&&\\
    \hline\hline
&&&&\\
    Born & ~$21 \times 10^6$ & $60$ & $350 \times 10^3$ & ~~$38$ GB \\
    Born + Virtual & ~$33 \times 10^6$ & $380$ & ~$87 \times 10^3$ & ~~$72$ GB \\
    Integrated dipoles & ~$80\times  10^6$ &  $450$ & $178 \times 10^3$ & ~$160$ GB \\
    Real $+$ Sub. Real & $626 \times 10^6$ & $18000$ & ~$35 \times 10^3$ & $1250$ GB \\
&&&&\\
    \hline\hline
&&&&\\
    Total: & $760 \times 10^6$ & $18890$ & $40 \times 10^3$ & $1520$ GB \\
&&&&\\
 \hline\hline
  \end{tabular}
  \end{center}
  \end{table}
%
Table \ref{tab:ntuples} summarises the total number of Ntuple files
and their sizes, which have been generated for the present
analysis. Except for the virtual part, which is obtained by
reweighting of the (unweighted) Born events, the Ntuples contain
unweighted events. In this way we have minimised the number of events
stored in these files. With the goal of optimising the performance of
the unweighting, we have implemented the so-called partial unweighting
in \textsc{Helac-NLO}. Instead of looking for the maximal weight to
perform the unweighting according to its value we have decided to
choose some approximate $w_{\rm max}$. All events with a weight $w$
lower than a given threshold $w_{\rm max}$ have been unweighted up to
this threshold, while for events with $w > w_{\rm max}$ the event
weights, i.e.  $w/w_{\rm max}$, have been kept. In the end both types
of events, i.e. events with $w=1$ and $w \ne1$, have to be evaluated
together to give a final cross section and its error.  This procedure
has proved particularly helpful for the process at hand, where when
using the standard unweighting procedure some shortcomings are
encountered, as elaborated in the following. Before the unweighting
procedure is performed, a preunweighting phase is done to find the
maximum weight. In order to find the correct maximal weight a huge
number of events need to be evaluated, which for such complicated
final state is time consuming. If the correct maximum weight is found,
which typically is a very large number comparing to the average
weight, the unweighting procedure becomes extremely
inefficient. Moreover, if during the preunweighting phase the maximum
weight found is not the correct one, the unweighting procedure is
spoilt. As a consequence various differential distributions close to
their peaks are not properly described. The partial unweighting helps
to bypass these problems while giving the exact answer at the same
time. In practice, to find the approximate maximal weight we use
$200000$ accepted events in the preunweighting phase. This typically
results in about $1\%-10\%$ of the total number of events carrying a
non-unit weight.

%
\section{Results for the LHC Run II Energy of 13 TeV}
\label{Results for the LHC Run II Energy of 13 TeV}
%

\subsection{Numerical Setup}

We consider the process $pp \to e^+ \nu_e \mu^- \bar{\nu}_\mu
b\bar{b}j+X$ for the LHC Run II energy of $\sqrt{s}=13$ TeV.  We only
simulate decays of the weak bosons to different lepton generations to
avoid virtual photon singularities stemming from quasi-collinear
$\gamma^* \to \ell^\pm \ell^\mp $ decays. These interference effects
are at the per-mille level for inclusive cuts, as checked by an
explicit leading order calculation. The complete cross section with
$\ell= e, \mu$ can be obtained by multiplying the result with a
lepton-flavor factor of 4.  The Cabibbo-Kobayashi-Maskawa (CKM) mixing
of the quark generations is neglected, i.e.  the CKM matrix has a
diagonal form. The SM parameters are given in the $G_\mu$
scheme
\begin{equation}
\begin{array}{ll}
 G_{\mu}=1.16637 \cdot 10^{-5} ~{\rm GeV}^{-2}\,,  
&   \quad \quad \quad \quad 
m_{t}=173.2 ~{\rm GeV} ~ \text{\cite{Agashe:2014kda}}\,,
\vspace{0.2cm}\\
 m_{W}=80.399 ~{\rm GeV}\,, &\quad \quad
\quad \quad
\Gamma_{W} = 2.09875 ~{\rm GeV}\,, 
\vspace{0.2cm}\\
  m_{Z}=91.1876  ~{\rm GeV}\,, &\quad \quad
\quad \quad
\Gamma_{Z} = 2.50848 ~{\rm GeV}\,.
\end{array}
\end{equation}
The electromagnetic coupling is derived from
the Fermi constant $G_\mu$ according to
\begin{equation}
\alpha = \frac{\sqrt{2}}{\pi}  \, G_\mu \, m_W^2 \sin^2\theta_W\,, 
\end{equation}
where $\sin^2\theta_W$ is  the weak mixing angle defined as 
\begin{equation}
\sin^2\theta_W= 1-\frac{m_W^2}{m_Z^2}\,.
\end{equation}
Since we are interested in NLO QCD corrections, electroweak gauge
bosons are treated within the fixed width scheme, thus, we use the
real $W$ and $Z$ boson masses and also $\sin^2\theta_W$ is kept real.
Masses of all other particles (leptons and quarks), including the
bottom quark, are set to zero. We have checked using the integrated
cross section at LO that finite bottom-quark mass  effects lead to a
reduction of the cross section by less than $1\%$. The width of the
top quark for an unstable $W$ boson and the massless bottom quark
according to \cite{Jezabek:1988iv} is given by
\begin{equation}
 \Gamma_{t}^{\rm LO} = 1.47834 ~{\rm GeV}\,, \quad \quad
\quad \quad \quad \quad \quad \quad
 \Gamma_{t}^{\rm NLO} = 1.35146  ~{\rm GeV} \,.
\end{equation}
Since we treat bottom quarks as massless partons there are no diagrams
with Higgs boson exchange  at tree level.  We also neglect closed
fermion loops involving top quarks coupled to the Higgs boson.
Following recommendations of PDF4LHC for the usage of parton
distribution functions (PDFs) suitable for applications at the LHC Run
II \cite{Butterworth:2015oua} we employ CT14 \cite{Dulat:2015mca},
MMHT14 \cite{Harland-Lang:2014zoa} and NNPDF3.0 \cite{Ball:2014uwa} sets.
In particular, we take CT14nlo, NNPDF3.0-nlo-as-0118 and
MMHT14nlo68clas118 at NLO as well as CT14llo, NNPDF3.0-lo-as-0130 and
MMHT14lo68cl at LO.  The running of the strong coupling constant
$\alpha_s$ with two-loop (one-loop) accuracy at NLO (LO) is provided
by the LHAPDF interface \cite{Buckley:2014ana}.  The number of active
flavours is $N_F = 5$.  Contributions induced by the bottom-quark parton
density are neglected.  We have determined that for the integrated LO
cross section neglecting the bottom-quark contribution to PDFs amounts
to less than $0.1\%$. We use the corresponding prescription from each
group to provide the $68\%$ confidence level (C.L.) PDF uncertainties.  Both
CT14 PDFs and MMHT14 PDFs include a central set and error sets in the
Hessian representation. In that case we use the asymmetric expression for
PDF uncertainties \cite{Buckley:2014ana}.  For an observable ${\cal O}$,
given a central PDF member $S_0$ and $2N$ eigenvector PDF members
$S^\pm_i$ $(i=1,\dots,N)$, uncertainties are given by
\begin{equation}
\label{pdfpm}
\begin{split}
\delta {\cal O}_{\rm PDF +} &= \sqrt{\sum_{i=1}^{N}  \left[ \max
\left( {\cal O}(S^+_i)-{\cal O}(S_0), {\cal O}(S^-_i)-{\cal O}(S_0),0\right)
\right]^2}\,,\\
\delta {\cal O}_{\rm PDF -} &= \sqrt{\sum_{i=1}^{N}  \left[ \max
\left( {\cal O}(S_0)-{\cal O}(S^+_i), {\cal O}(S_0)-{\cal O}(S_i^-),0\right)
\right]^2}\,.
\end{split}
\end{equation}
Let us note that for CT14 and MMHT14 we have $2N=56$ and $2N=50$
respectively. Additionally, the CT14 errors are rescaled by a factor
$1/1.645$ since they are provided at $90\%$ C.L. On the other hand
NNPDF3.0 PDFs uses the Monte Carlo sampling method in conjunction with
neural networks. In that case PDF uncertainties are obtained using the
replicas method defined by
\begin{equation}
\label{pdfmc}
\delta {\cal O}_{\rm PDF +} = \delta {\cal O}_{\rm PDF -} =
\delta {\cal O}_{\rm PDF} = \sqrt{\frac{1}{N-1} \sum^N_{i=1} \left[
{\cal O}(S_i) -{\cal O}(S_0)
\right]^2}\,,
\end{equation}
where a set of $N=100$ Monte Carlo PDF members $S_i$ $(i=1,\dots,N)$
has been used.  We also have 
\begin{equation}
{\cal O}(S_0)=  \langle  {\cal O} \rangle = \frac{1}{N} \sum^N_{i=1}
{\cal O}(S_i) \,,
\end{equation}
such that  $\delta {\cal O}_{\rm PDF}$ can be rewritten as 
\begin{equation}
\delta {\cal O}_{\rm PDF} = \sqrt{\frac{N}{N-1} \left[ \langle {\cal
      O}^2\rangle - \langle {\cal O}\rangle^2\right]}\,. 
\end{equation}
Our calculation, like any fixed-order one, contains  a residual
dependence on the renormalisation scale, $\mu_R$, and the
factorisation scale, $\mu_F$, arising from the truncation of the
perturbative expansion. As a consequence, observables depend on the
values of $\mu_R$ and $\mu_F$ that are provided as input parameters.
We assume that the default scale $\mu_R=\mu_F=\mu_0$ is the same for
both the renormalisation and factorisation scales. The scale
systematics, however, is evaluated by varying $\mu_R$ and $\mu_F$
independently in the range
\begin{equation}
\frac{1}{2} \, \mu_0  \le \mu_R\,,\mu_F \le  2 \,  \mu_0\,, \quad \quad 
\quad \quad \quad \quad \quad \quad \frac{1}{2}  \le
\frac{\mu_R}{\mu_F} \le  2 \,,
\end{equation}
which in practise amounts to consider  the following pairs 
\begin{equation}
\label{scan}
\left(\frac{\mu_R}{\mu_0}\,,\frac{\mu_F}{\mu_0}\right) = \Big\{
\left(2,1\right),\left(0.5,1  
\right),\left(1,2\right), (1,1), (1,0.5), (2,2),(0.5,0.5)
\Big\} \,.
\end{equation}
We search for the minimum and maximum of the resulting cross
section. Let us mention here that while calculating the scale
dependence for the NLO cross section we keep $\Gamma^{\rm NLO}_t$
fixed independently of the scale choice.  The error introduced by this
treatment is however of higher orders. We have checked that for the
$pp \to e^+\nu_e\mu^- \bar{\nu}_\mu b\bar{b}+X$ production process,
which is a simpler case, and for two scales $\mu = 0.5 \mu_0$ and
$\mu= 2\mu_0$ with $\mu_0=m_t$ it amounts to $\pm1.5\%$ deviation,
respectively \cite{Bevilacqua:2010qb}. As a natural scale for the
process we choose the mass of the heaviest  particle appearing in the
process, that is the top-quark mass and set $\mu_0=m_t$.  Total cross
sections are mostly influenced by final-state production relatively
close to the threshold as defined by particle masses, which justifies
our choice. However, differential cross sections extend  up
to energy scales that are much larger than the threshold, and show
larger shape distortions in such high-energy regions
\cite{Bevilacqua:2015qha}. Therefore, we examine two additional
choices, namely $\mu_0=E_T/2$ and $\mu_0=H_T/2$, where $E_T$ 
and $H_T$ are defined as
\begin{equation}
\label{scaledef}
\begin{array}{c}
E_T = m_{T,\,t}  + m_{T,\,\bar{t}} =
\sqrt{m_t^2 + p^2_{T,\,t}} + \sqrt{m_t^2 + p^2_{T,\,\bar{t}}}
  \vspace{0.2cm}\\
H_T = p_{T,\,e^+} + p_{T,\,\mu^-} + p_{T,\,j_{b_1}} + p_{T,\,j_{b_2}}
  + \slashed{p}_{T} +
  p_{T,\,j_{1}}\,.
\end{array}
\end{equation}
Here $t$ and $\bar{t}$ are reconstructed from their decay products,
albeit we use bottom-jets denoted as $j_{b_1}$ and $j_{b_2}$ not
bottom quarks in the reconstruction. Additionally, $j_{1}$ is the
first hardest light-jet (jets are ordered in $p_T$) and
$\slashed{p}_T=|{\bf p}_{T,\, \nu_e} +{\bf p}_{T,\,\bar{\nu}_\mu}|$ is
the total missing transverse momentum from escaping neutrinos.  Let us
note here, that for small values of $p_{T,\,t}$ and $p_{T,\,\bar{t}}$,
i.e. close to the $t\bar{t}$ threshold, $E_T/2 \approx m_t$. All final
state partons with pseudorapidity $|\eta|<5$, where $\eta=-\ln
\left(\tan \theta/2\right)$, are recombined into jets via the IR-safe
anti$-k_T$ jet algorithm \cite{Cacciari:2008gp} with the separation
parameter in the rapidity-azimuthal-angle plane set to $R=0.5$. We
require exactly two bottom-jets, at least one light-jet, two charged
leptons and non-zero missing transverse momentum
$\slashed{p}_T$. These final states have to fullfil the following
criteria, which we consider to be very inclusive selection  cuts 
\begin{equation}
\begin{array}{lcl}
 p_{T,\,\ell}>30 ~{\rm GeV}\,,    & & p_{T,\,j}>40 ~{\rm GeV}\,, 
\vspace{0.2cm}\\
\slashed{p}_{T} >40 ~{\rm GeV} \,,   & 
\quad \quad \quad \quad \quad \quad 
& \Delta R_{jj}>0.5\,,
\vspace{0.2cm}\\
\Delta R_{\ell\ell}>0.4 \,,  && 
 \Delta R_{\ell j}>0.4 \,,
\vspace{0.2cm}\\
 |y_\ell|<2.5\,,&&
|y_j|<2.5 \,, 
\end{array}
\end{equation}
where $\ell$ stands for $\mu^-,e^+$ and $j$ corresponds to light- and
bottom-jets. Additionally, the transverse momentum, $p_{T,\, i}$,
rapidity, $y_i$, as well as the separation in the
rapidity-azimuthal-angle-plane, $\Delta R_{ik}$, where $i,k=\ell,j$ are
defined as
\begin{equation}
p_{T,\,i} =\sqrt{p^2_{x,\,i} + p^2_{y,\,i}}\,,  
\end{equation}
\begin{equation}
y_i  =\frac{1}{2} \ln \left( \frac{E_i+p_{z,\,i}}{E_i-p_{z,\,i}}
\right)\,,  
\end{equation}
\begin{equation}
\Delta R_{ik} =  \sqrt{\Delta \phi_{ik}^2 + \Delta y_{ik}^2}\,. 
\end{equation}

\vspace{0.2cm}
%
\subsection{Integrated Cross Sections with Theoretical Uncertainties}
%

We begin the presentation of our results with a discussion of the
integrated cross section using the scale choice
$\mu_F=\mu_R=\mu_0=m_t$. We define the upper and the lower limit of
the scale variation according to Eq.~\eqref{scan} and the PDF
uncertainties are considered to be at the $\pm1\sigma$ level ($68\%$
C.L.). Our results for the integrated cross section with the CT14 PDF
sets and $\mu_0=m_t$ are as follows
\begin{equation}
\begin{split}
\sigma^{\rm LO}_{e^+\nu_e\mu^-\bar{\nu}_\mu b\bar{b}j}  ({\rm CT14},
\mu_0=m_t) &= 
608.09^{+303.52~(+50\%)}_{-188.85~(-31\%)}  ~[{\rm scales}] ~{\rm
  fb}\,,\\
\sigma^{\rm NLO}_{e^+\nu_e\mu^-\bar{\nu}_\mu b\bar{b}j} ({\rm CT14},
\mu_0=m_t) &= 
537.24^{~\,+10.12~(~+2\%)}_{-190.35~(-35\%)} ~[{\rm scales}] ~
 {}^{+17.32~(+3\%)}_{-18.34~(-3\%)}  ~[{\rm PDF}] ~{\rm
  fb}\,.
\end{split}
\end{equation}
For the MMHT14 PDF sets we  have obtained instead 
\begin{equation}
\begin{split}
\sigma^{\rm LO}_{e^+\nu_e\mu^-\bar{\nu}_\mu b\bar{b}j}  ({\rm MMHT14},
\mu_0=m_t) &= 
665.58^{+357.64~(+54\%)}_{-216.08~(-32\%)}  ~[{\rm scales}] ~{\rm
  fb}\,,\\
\sigma^{\rm NLO}_{e^+\nu_e\mu^-\bar{\nu}_\mu b\bar{b}j} ({\rm MMHT14},
\mu_0=m_t) &= 542.56^{~\,+10.02~(~+2\%)}_{-106.46~(-20\%)} ~[{\rm scales}]
~ {}^{+12.31~(+2\%)}_{-11.33~(-2\%)} ~[{\rm PDF}] ~{\rm
  fb}\,.
\end{split}
\end{equation}
And finally, with the NNPDF3.0 PDF sets our results read
\begin{equation}
\begin{split}
\sigma^{\rm LO}_{e^+\nu_e\mu^-\bar{\nu}_\mu b\bar{b}j}  ({\rm NNPDF3.0},
\mu_0=m_t) &=   582.29^{+302.06~(+52\%)}_{-184.75~(-32\%)} ~[{\rm scales}] ~{\rm
  fb}\,,\\
\sigma^{\rm NLO}_{e^+\nu_e\mu^-\bar{\nu}_\mu b\bar{b}j} ({\rm NNPDF3.0},
\mu_0=m_t) &= 559.66^{~\,+10.64~(~+2\%)}_{-111.05~(-20\%)}  ~[{\rm
  scales}]~ {}^{+8.42~(+2\%)}_{-8.42~(-2\%)} ~[{\rm PDF}] ~{\rm
  fb}\,.
\end{split}
\end{equation}
A few comments are in order. To start, at the central value of
the fixed scale, i.e. for $\mu_0=m_t$, we obtain negative and moderate
NLO corrections, which are of the order of $12\%$ for the CT14 PDF
set, $18\%$ for MMHT14 and $4\%$ for NNPDF3.0. Defining scale
uncertainties in a very conservative way, using the lower and upper
bounds of our results, gives us an estimate of $50\%$ for the LO
prediction, independent of the PDF set. After inclusion of the NLO QCD
corrections, they are reduced down to about $20\%$ for MMHT14 PDF and
NNPDF3.0. In case of CT14 PDF the reduction is smaller and the final
theoretical uncertainties are at the $35\%$ level. However in the case
of truly asymmetric uncertainties it is always more appropriate to
symmetrise the errors. After symmetrisation the scale uncertainty at
LO is assessed to be instead of the order of $40\%$. After inclusion
of the NLO QCD corrections, the scale uncertainty is reduced down to
$11\%$ for NNPDF3.0 and MMHT14 and $18\%$ for CT14.  Another source of
uncertainties comes from the PDF parametrisation.  We calculate these
uncertainties as explained in the previous section according to
Eq.~\eqref{pdfpm} and Eq.~\eqref{pdfmc}. They amount to $\pm \,3\%$
for CT14 and $\pm \,2\%$ for MMHT14 and NNPDF3.0. These numbers refer
to the uncertainties at the $68\%$ C.L. for the individual PDF sets,
but do not take into account additional systematics coming from the
underlying assumptions that enter the parametrisation of different PDF
sets, which cannot be quantified within a given scheme.  We see that
CT14, MMHT14 and NNPDF3.0 NLO results differ by $1\%-4\%$, which is
comparable to the individual estimates of PDF systematics. Overall,
the PDF uncertainties for the process under scrutiny are well below
the theoretical uncertainties due to the scale dependence, which
remain the dominant source of the theoretical systematics. In Table
\ref{tab:3} we report the total cross section at LO and NLO for
different cuts on the transverse momentum of the hardest light-jet,
$p_{T ,\, j_1}$. Theoretical uncertainties coming from scale
variation, denoted as $\delta_{scale}$, and from PDFs, denoted as
$\delta_{\rm PDF}$ together with a ${\cal K}-$factor defined as
$\sigma^{\rm NLO}/\sigma^{\rm LO}$ are additionally presented in the
Table \ref{tab:3}.  Within each PDF set we observe a very stable
behaviour of systematics when varying the $p_{T,\,j_1}$ cut within the
$40-120$ GeV range. NLO corrections are also quite stable, changing
the ${\cal K}$-factor by less than $7\%$, $5\%$ and $4\%$ for CT14,
MMHT14 and NNPDF3.0 respectively.
%
\begin{table}[t!]
\caption{ \label{tab:3}\it 
Integrated cross section for the $pp\to e^+\nu_e \mu^- \bar{\nu}_\mu
b\bar{b} j+X$ production process at the LHC with $\sqrt{s}=13$
TeV. Results are evaluated using $\mu_R=\mu_F=\mu_0=m_t$ for three different
PDF sets and five different $p_{T,\,j_1}$ cuts for the hardest light-jet. Also
given are theoretical uncertainties coming from scale variation,
$\delta_{scale}$, and from  PDFs,  $\delta_{\rm PDF}$. In the last
column a ${\cal K}-$factor is shown.}
\begin{center}
\begin{tabular}{cccccccc}
  \hline \hline
&&&&&&&\\
  PDF &$p_{T,\,j_1}$ & $\sigma^{\rm LO}$  [fb]  & $\delta_{scale}$ &
 $\sigma^{\rm NLO}$ [fb] & $\delta_{scale}$ & $\delta_{\rm PDF}$ & ${\cal K}$ \\
&&&&&&&\\
  \hline
\hline
&&&&&&&\\
  CT & $40$ & $608.09$ & $^{+303.52~(+50\%)}_{-188.85~(-31\%)}$ & $537.24$ & $^{+10.12~(+2\%)}_{-190.35~(-35\%)}$ & $^{+17.32~(+3\%)}_{-18.34~(-3\%)}$ & $0.88$\\
       & $60$ & $433.47$ & $^{+220.20~(+51\%)}_{-136.12~(-31\%)}$ & $384.35$ & $^{+6.35~(+2\%)}_{-127.14~(-33\%)}$ & $^{+13.20~(+3\%)}_{-13.54~(-4\%)}$ & $0.89$\\
       & $80$ & $330.55$ & $^{+170.40~(+52\%)}_{-104.76~(-33\%)}$ & $289.15$ & $^{+4.78~(+2\%)}_{-93.77~(-32\%)}$ & $^{+10.44~(+4\%)}_{-10.41~(-4\%)}$ & $0.87$\\
       & $100$ & $261.65$ & $^{+136.64~(+52\%)}_{-83.60~(-32\%)}$ & $223.70$ & $^{+4.01~(+2\%)}_{-73.36~(-33\%)}$ & $^{+8.41~(+4\%)}_{-8.18~(-4\%)}$ & $0.85$\\
       & $120$ & $212.23$ & $^{+112.14~(+53\%)}_{-68.31~(-32\%)}$ &
                                                                    $176.05$ & $^{+3.57~(+2\%)}_{-59.58~(-34\%)}$ & $^{+6.88~(+4\%)}_{-6.53~(-4\%)}$ & $0.83$\\
&&&&&&&\\
  \hline 
\hline
&&&&&&&\\
  MMHT & $40$ & $665.58$ & $^{+357.64~(+54\%)}_{-216.08~(-32\%)}$ & $542.56$ & $^{+10.02~(+2\%)}_{-106.46~(-20\%)}$ & $^{+12.31~(+2\%)}_{-11.33~(-2\%)}$ & $0.82$\\
       & $60$ & $471.36$ & $^{+257.33~(+55\%)}_{-154.52~(-33\%)}$ & $387.34$ & $^{+6.25~(+2\%)}_{-73.95~(-19\%)}$ & $^{+8.97~(+2\%)}_{-8.15~(-2\%)}$ & $0.82$\\
       & $80$ & $357.55$ & $^{+197.80~(+55\%)}_{-118.17~(-33\%)}$ & $290.91$ & $^{+4.71~(+2\%)}_{-58.23~(-20\%)}$ & $^{+6.83~(+2\%)}_{-6.18~(-2\%)}$ & $0.81$\\
       & $100$ & $281.75$ & $^{+157.69~(+56\%)}_{-93.78~(-33\%)}$ & $224.75$ & $^{+3.95~(+2\%)}_{-49.17~(-22\%)}$ & $^{+5.34~(+2\%)}_{-4.82~(-2\%)}$ & $0.80$\\
       & $120$ & $227.63$ & $^{+128.76~(+57\%)}_{-76.26~(-34\%)}$ &
                                                                    $176.59$ & $^{+3.54~(+2\%)}_{-43.14~(-24\%)}$ & $^{+4.25~(+2\%)}_{-3.84~(-2\%)}$ & $0.78$\\
&&&&&&&\\
  \hline
 \hline
&&&&&&&\\
  NNPDF& $40$ & $582.29$ & $^{+302.06~(+52\%)}_{-184.75~(-32\%)}$ & $559.66$ & $^{+10.64~(+2\%)}_{-111.05~(-20\%)}$ & $^{+8.42~(+2\%)}_{-8.42~(-2\%)}$ & $0.96$\\
       & $60$ & $410.73$ & $^{+216.23~(+53\%)}_{-131.50~(-32\%)}$ & $399.81$ & $^{+6.64~(+2\%)}_{-77.17~(-19\%)}$ & $^{+6.06~(+2\%)}_{-6.06~(-2\%)}$ & $0.97$\\
       & $80$ & $310.50$ & $^{+165.46~(+53\%)}_{-100.15~(-32\%)}$ & $300.39$ & $^{+4.99~(+2\%)}_{-60.79~(-20\%)}$ & $^{+4.64~(+2\%)}_{-4.64~(-2\%)}$ & $0.97$\\
       & $100$ & $243.89$ & $^{+131.35~(+54\%)}_{-79.19~(-32\%)}$ & $232.13$ & $^{+4.19~(+2\%)}_{-51.35~(-22\%)}$ & $^{+3.67~(+2\%)}_{-3.67~(-2\%)}$ & $0.95$\\
       & $120$ & $196.46$ & $^{+106.82~(+54\%)}_{-64.16~(-33\%)}$ &
                                                                    $182.46$
                                                                                     & $^{+3.74~(+2\%)}_{-45.06~(-25\%)}$ & $^{+2.97~(+2\%)}_{-2.97~(-2\%)}$ & $0.93$\\
&&&&&&&\\
\hline\hline
\end{tabular}
\end{center}
\end{table}
\begin{table}[t!]
\caption{ \label{tab:4}\it 
Integrated cross section for the $pp\to e^+\nu_e \mu^- \bar{\nu}_\mu
b\bar{b} j+X$ production process at the LHC with $\sqrt{s}=13$
TeV. Results are evaluated using $\mu_R=\mu_F=\mu_0=E_T/2$ for three different
PDF sets and five different $p_{T,\,j_1}$ cuts for the hardest light-jet. Also
given are theoretical uncertainties coming from scale variation,
$\delta_{scale}$, and from  PDFs,  $\delta_{\rm PDF}$. In the last
column a ${\cal K}-$factor is shown.}
\begin{center}
\begin{tabular}{cccccccc}
  \hline\hline
&&&&&&&\\
  PDF &$p_{T,\,j_1}$ & $\sigma^{\rm LO}$~[fb] & $\delta_{scale}$ &
                                                                $\sigma^{\rm
                                                                NLO}$
                                                                [fb] &
                                                                       $\delta_{scale}$
  & $\delta_{\rm PDF}$ & ${\cal K}$ \\
&&&&&&&\\  
\hline\hline
&&&&&&&\\
  CT& $40$  & $493.54$ & $^{+230.40~(+47\%)}_{-147.02~(-30\%)}$ & $544.64$ & $^{+2.95~(+1\%)}_{-117.47~(-22\%)}$ & $^{+18.10~(+3\%)}_{-18.92~(-3\%)}$ & $1.10$ \\
       & $60$  & $347.04$ & $^{+164.28~(+47\%)}_{-104.31~(-30\%)}$ & $387.25$ & $^{+3.23~(+1\%)}_{-75.76~(-20\%)}$  & $^{+13.67~(+4\%)}_{-13.87~(-4\%)}$ & $1.12$\\
       & $80$  & $261.26$ & $^{+125.07~(+48\%)}_{-79.10~(-30\%)}$  & $290.83$ & $^{+2.80~(+1\%)}_{-54.31~(-19\%)}$  & $^{+10.79~(+4\%)}_{-10.63~(-4\%)}$ & $1.11$\\
       & $100$ & $204.16$ & $^{+98.69~(+48\%)}_{-62.20~(-30\%)}$   & $225.43$ & $^{+2.27~(+1\%)}_{-41.32~(-18\%)}$  & $^{+8.73~(+4\%)}_{-8.39~(-4\%)}$   & $1.10$\\
       & $120$ & $163.48$ & $^{+79.69~(+49\%)}_{-50.08~(-31\%)}$   & $178.04$ & $^{+1.76~(+1\%)}_{-32.72~(-18\%)}$  & $^{+7.17~(+4\%)}_{-6.73~(-4\%)}$   & $1.09$\\
 &&&&&&&\\
 \hline\hline
&&&&&&&\\
  MMHT & $40$ & $536.43$ & $^{+268.93~(+50\%)}_{-166.94~(-31\%)}$ & $549.58$ & $^{+3.11~(+1\%)}_{-49.90~(-9\%)}$ & $^{+12.74~(+2\%)}_{-11.61~(-2\%)}$ & $1.02$\\
       & $60$ & $374.58$ & $^{+190.06~(+51\%)}_{-117.46~(-31\%)}$ & $389.97$ & $^{+5.04~(+1\%)}_{-37.67~(-10\%)}$ & $^{+9.20~(+2\%)}_{-8.33~(-2\%)}$ & $1.04$\\
       & $80$ & $280.38$ & $^{+143.64~(+51\%)}_{-88.46~(-32\%)}$ & $292.39$  & $^{+4.13~(+1\%)}_{-28.79~(-10\%)}$ & $^{+7.01~(+2\%)}_{-6.32~(-2\%)}$ & $1.04$\\
       & $100$ & $218.01$ & $^{+112.61~(+52\%)}_{-69.13~(-32\%)}$ & $226.33$ & $^{+2.74~(+1\%)}_{-22.26~(-10\%)}$ & $^{+5.51~(+2\%)}_{-4.95~(-2\%)}$ & $1.04$\\
       & $120$ & $173.79$ & $^{+90.41~(+52\%)}_{-55.36~(-32\%)}$ & $178.48$  & $^{+1.79~(+1\%)}_{-17.26~(-10\%)}$ & $^{+4.41~(+2\%)}_{-3.97~(-2\%)}$ & $1.03$\\
 &&&&&&&\\ 
\hline \hline
&&&&&&&\\
  NNPDF& $40$ & $473.88$ & $^{+223.00~(+47\%)}_{-144.34~(-30\%)}$ & $567.13$ & $^{+3.15~(+1\%)}_{-51.53~(-9\%)}$ & $^{+8.63~(+2\%)}_{-8.63~(-2\%)}$ & $1.20$\\
       & $60$ & $329.81$ & $^{+161.85~(+49\%)}_{-101.15~(-31\%)}$ & $402.67$ & $^{+5.20~(+1\%)}_{-38.96~(-10\%)}$ & $^{+6.21~(+2\%)}_{-6.21~(-2\%)}$ & $1.22$\\
       & $80$ & $246.17$ & $^{+121.86~(+50\%)}_{-75.91~(-31\%)}$ & $302.03$ & $^{+4.26~(+1\%)}_{-29.80~(-10\%)}$ & $^{+4.76~(+2\%)}_{-4.76~(-2\%)}$ & $1.23$\\
       & $100$ & $190.91$ & $^{+95.18~(+50\%)}_{-59.14~(-31\%)}$ & $233.86$ & $^{+2.82~(+1\%)}_{-23.05~(-10\%)}$ & $^{+3.79~(+2\%)}_{-3.79~(-2\%)}$ & $1.22$\\
       & $120$ & $151.82$ & $^{+76.15~(+50\%)}_{-47.21~(-31\%)}$ & $184.48$ & $^{+1.83~(+1\%)}_{-17.88~(-10\%)}$ & $^{+3.09~(+2\%)}_{-3.09~(-2\%)}$ & $1.22$\\
&&&&&&&\\
\hline\hline
\end{tabular}
\end{center}
\end{table}
\begin{table}[t!]
\caption{ \label{tab:5}\it 
Integrated cross section for the $pp\to e^+\nu_e \mu^- \bar{\nu}_\mu
b\bar{b} j+X$ production process at the LHC with $\sqrt{s}=13$
TeV. Results are evaluated using $\mu_R=\mu_F=\mu_0=H_T/2$ for three different
PDF sets and five different $p_{T,\,j_1}$ cuts for the hardest light-jet. Also
given are theoretical uncertainties coming from scale variation,
$\delta_{scale}$, and from  PDFs,  $\delta_{\rm PDF}$. In the last
column a ${\cal K}-$factor is shown.}
\begin{center}
\begin{tabular}{cccccccc}
  \hline\hline
&&&&&&&\\
  PDF &$p_{T,\,j_1}$ & $\sigma^{\rm LO}$ [fb] & $\delta_{scale}$ &
                                                                 $\sigma^{\rm
                                                                     NLO}$
                                                                 [fb]
  & $\delta_{scale} $ & $\delta_{\rm PDF} $ & ${\cal K}$ \\
&&&&&&&\\
  \hline\hline
&&&&&&&\\
  CT & $40$  & $479.38$ & $^{+221.91~(+46\%)}_{-142.05~(-30\%)}$ & $549.65$ & $^{+10.25~(+2\%)}_{- 53.42~(-10\%)}$ & $^{+18.00~(+3\%)}_{-19.15~(-3\%)}$ & $1.15$\\
       & $60$  & $328.60$ & $^{+153.04~(+47\%)}_{- 97.75~(-30\%)}$ & $384.37$ & $^{+11.93~(+3\%)}_{- 40.33~(-10\%)}$ & $^{+13.43~(+3\%)}_{-13.91~(-4\%)}$ & $1.17$\\
       & $80$  & $241.43$ & $^{+113.00~(+47\%)}_{- 72.05~(-30\%)}$ & $286.68$ & $^{+11.23~(+4\%)}_{-31.57~(-11\%)}$ & $^{+10.50~(+4\%)}_{-10.66~(-4\%)}$ & $1.19$\\
       & $100$ & $184.69$ & $^{+ 86.79~(+47\%)}_{ -55.26~(-30\%)}$ & $221.01$ & $^{+9.61~(+4\%)}_{-24.96~(-11\%)}$ & $^{+8.43~(+4\%)}_{-8.37~(-4\%)}$ & $1.20$\\
       & $120$ & $145.11$ & $^{+ 68.43~(+47\%)}_{ -43.52~(-30\%)}$ & $173.90$ & $^{+7.90~(+5\%)}_{-19.90~(-11\%)}$ & $^{+6.88~(+4\%)}_{-6.71~(-4\%)}$ & $1.20$\\
&&&&&&&\\ 
 \hline\hline
&&&&&&&\\
  MMHT & $40$  & $521.08$ & $^{+259.12~(+50\%)}_{-161.36~(-31\%)}$ & $554.61$ & $^{+10.85~(+2\%)}_{- 54.51~(-10\%)}$ & $^{+12.06~(+2\%)}_{-12.22~(-2\%)}$ & $1.06$\\
       & $60$  & $354.08$ & $^{+176.68~(+50\%)}_{-109.89~(-31\%)}$ & $386.98$ & $^{+12.30~(+3\%)}_{-40.98~(-11\%)}$ & $^{+8.58~(+2\%)}_{-8.84~(-2\%)}$ & $1.09$\\
       & $80$  & $258.31$ & $^{+129.23~(+50\%)}_{- 80.30~(-31\%)}$ & $288.13$ & $^{+11.50~(+4\%)}_{-31.99~(-11\%)}$ & $^{+6.43~(+2\%)}_{-6.81~(-2\%)}$ & $1.12$\\
       & $100$ & $196.39$ & $^{+ 98.44~(+50\%)}_{ -61.13~(-31\%)}$ & $221.80$ & $^{+9.77~(+4\%)}_{-25.24~(-11\%)}$ & $^{+4.98~(+2\%)}_{-5.42~(-2\%)}$ & $1.13$\\
       & $120$ & $153.47$ & $^{+ 77.05~(+50\%)}_{ -47.83~(-31\%)}$ & $174.28$ & $^{+8.01~(+5\%)}_{-20.08~(-12\%)}$ & $^{+3.94~(+2\%)}_{-4.40~(-2\%)}$ & $1.14$\\
&&&&&&&\\
  \hline\hline
&&&&&&&\\
  NNPDF& $40$  & $460.80$ & $^{+221.93~(+48\%)}_{-139.68~(-30\%)}$ & $572.18$ & $^{+11.14~(+2\%)}_{-56.23~(-10\%)}$ & $^{+11.31~(+2\%)}_{-11.31~(-2\%)}$ & $1.24$\\
       & $60$  & $312.34$ & $^{+150.81~(+48\%)}_{- 94.83~(-30\%)}$ & $399.61$ & $^{+12.74~(+3\%)}_{-42.42~(-11\%)}$ & $^{+9.15~(+2\%)}_{-9.15~(-2\%)}$ & $1.28$\\
       & $80$  & $227.37$ & $^{+109.97~(+48\%)}_{- 69.10~(-30\%)}$ & $297.64$ & $^{+11.92~(+4\%)}_{-33.13~(-11\%)}$ & $^{+7.40~(+2\%)}_{-7.40~(-2\%)}$ & $1.31$\\
       & $100$ & $172.51$ & $^{+ 83.53~(+48\%)}_{-52.47~(-30\%)}$ & $229.19$ & $^{+10.13~(+4\%)}_{-26.15~(-11\%)}$ & $^{+6.01~(+3\%)}_{-6.01~(-3\%)}$ & $1.33$\\
       & $120$ & $134.57$ & $^{+ 65.20~(+48\%)}_{-40.95~(-30\%)}$ & $180.15$ & $^{+ 8.31~(+5\%)}_{-20.82~(-12\%)}$ & $^{+5.06~(+3\%)}_{-5.06~(-3\%)}$ & $1.34$\\
&&&&&&&\\
\hline\hline
\end{tabular}
\end{center}
\end{table}
%

In the following we examine two choices for a dynamical factorisation
and renormalisation scale. As a first scale we adopt
$\mu_R=\mu_F=\mu_0=E_T/2$, where $E_T$ is defined in
Eq.~\eqref{scaledef}. Our second choice is $\mu_R=\mu_F=\mu_0=H_T/2$,
where $H_T$ is the sum of the transverse momenta of all final state
objects (bottom and light jets, missing transverse momentum and
charged leptons) according to Eq.~\eqref{scaledef}.  We repeat the
same analysis performed in the previous case, where we considered the
fixed scale $\mu_0 = m_t$. We start with results for $\mu_R=\mu_F=
\mu_0=E_T/2$ and the CT14 PDF set, which are as follows
\begin{equation}
\begin{split}
\sigma^{\rm LO}_{e^+\nu_e\mu^-\bar{\nu}_\mu b\bar{b}j}  ({\rm CT14},
\mu_0=E_T/2) &= 493.54^{+230.40~(+47\%)}_{-147.02~(-30\%)}
~[{\rm scales}] ~{\rm
  fb}\,,\\
\sigma^{\rm NLO}_{e^+\nu_e\mu^-\bar{\nu}_\mu b\bar{b}j} ({\rm CT14},
\mu_0=E_T/2) &= 
544.64^{~~\,+2.95~(~+1\%)}_{-117.47~(-22\%)} ~[{\rm scales}] ~
{}^{+18.10~(+3\%)}_{-18.92~(-3\%)} ~[{\rm PDF}] ~{\rm
  fb}\,.
\end{split}
\end{equation}
For  the MMHT14 PDF set we have the following integrated
cross sections
\begin{equation}
\begin{split}
\sigma^{\rm LO}_{e^+\nu_e\mu^-\bar{\nu}_\mu b\bar{b}j}  ({\rm MMHT14},
\mu_0=E_T/2) &= 536.43^{+268.93~(+50\%)}_{-166.94~(-31\%)}
 ~[{\rm scales}] ~{\rm
  fb}\,,\\
\sigma^{\rm NLO}_{e^+\nu_e\mu^-\bar{\nu}_\mu b\bar{b}j} ({\rm MMHT14},
\mu_0=E_T/2) &= 549.58^{~+3.11~(+1\%)}_{-49.90~(-9\%)} ~[{\rm scales}]
~ {}^{+12.74~(+2\%)}_{-11.61~(-2\%)} ~[{\rm PDF}] ~{\rm
  fb}\,,
\end{split}
\end{equation}
while for the NNPDF3.0 set we have 
\begin{equation}
\begin{split}
\sigma^{\rm LO}_{e^+\nu_e\mu^-\bar{\nu}_\mu b\bar{b}j}  ({\rm NNPDF3.0},
\mu_0=E_T/2) &=  473.88^{+223.00~(+47\%)}_{-144.34~(-30\%)}  ~[{\rm scales}] ~{\rm
  fb}\,,\\
\sigma^{\rm NLO}_{e^+\nu_e\mu^-\bar{\nu}_\mu b\bar{b}j} ({\rm NNPDF3.0},
\mu_0=E_T/2) &=  567.13^{~+3.15~(+1\%)}_{-51.53~(-9\%)} ~[{\rm
  scales}]~ {}^{+8.63~(+2\%)}_{-8.63~(-2\%)}  ~[{\rm PDF}] ~{\rm
  fb}\,.
\end{split}
\end{equation}
The use of the dynamical instead of the fixed scale hardly affects the
NLO integrated cross section. For each PDF set, a difference of only
$1.5\%$ is observed. On the other hand the LO cross sections are
lowered by more than $20\%$, which results in positive NLO
corrections. The size of the latter, however, remains the same,
i.e. it varies between $2\%-20\%$ depending on the PDF
set. Additionally, PDF uncertainties are of the same size. The
integrated NLO cross sections are shifted by maximally $4\%$ when
different PDF sets are used, which again remains  within the uncertainties
of the individual set.  Theoretical uncertainties at LO taken
conservatively (after symmetrisation) have been estimated to be around
$50\%$ $(40\%)$ and at NLO they are reduced down to $22\%$ $(11\%)$
for CT14 and to $10\%$ $(5\%)$ for MMHT14 and NNPDF3.0 sets. These
conclusions are not affected by the variation of the $p_{T,\,j_1}$
cut, that we move within the $40-120$ GeV range as can
been seen from Table \ref{tab:4}.  Lastly, for our third choice of
scale, $\mu_R=\mu_F=\mu_0=H_T/2$ and for the CT14 PDF set we can write
\begin{equation}
\begin{split}
\sigma^{\rm LO}_{e^+\nu_e\mu^-\bar{\nu}_\mu b\bar{b}j}  ({\rm CT14},
\mu_0=H_T/2) &=  479.38^{+221.91~(+46\%)}_{-142.05~(-30\%)}
~[{\rm scales}] ~{\rm
  fb}\,,\\
\sigma^{\rm NLO}_{e^+\nu_e\mu^-\bar{\nu}_\mu b\bar{b}j} ({\rm CT14},
\mu_0=H_T/2) &=  549.65^{+10.25~(~+2\%)}_{- 53.42~(-10\%)}
~[{\rm scales}] ~
{}^{+18.00~(+3\%)}_{-19.15~(-3\%)} ~[{\rm PDF}] ~{\rm
  fb}\,.
\end{split}
\end{equation}
For  the MMHT14 PDF set we obtain 
\begin{equation}
\begin{split}
\sigma^{\rm LO}_{e^+\nu_e\mu^-\bar{\nu}_\mu b\bar{b}j}  ({\rm MMHT14},
\mu_0=H_T/2) &= 521.08^{+259.12~(+50\%)}_{-161.36~(-31\%)}
 ~[{\rm scales}] ~{\rm
  fb}\,,\\
\sigma^{\rm NLO}_{e^+\nu_e\mu^-\bar{\nu}_\mu b\bar{b}j} ({\rm MMHT14},
\mu_0=H_T/2) &= 554.61^{+10.85~(~+2\%)}_{- 54.51~(-10\%)} ~[{\rm scales}]
~ {}^{+12.06~(+2\%)}_{-12.22~(-2\%)} ~[{\rm PDF}] ~{\rm
  fb}\,,
\end{split}
\end{equation}
and for the NNPDF3.0 PDF set our integrated cross section are as
follows
\begin{equation}
\begin{split}
\sigma^{\rm LO}_{e^+\nu_e\mu^-\bar{\nu}_\mu b\bar{b}j}  ({\rm NNPDF3.0},
\mu_0=H_T/2) &=  460.80^{+221.93~(+48\%)}_{-139.68~(-30\%)}
 ~[{\rm scales}] ~{\rm
  fb}\,,\\
\sigma^{\rm NLO}_{e^+\nu_e\mu^-\bar{\nu}_\mu b\bar{b}j} ({\rm NNPDF3.0},
\mu_0=H_T/2) &=   572.18^{+11.14~(~+2\%)}_{-56.23~(-10\%)} ~[{\rm scales}]
~ {}^{+11.31~(+2\%)}_{-11.31~(-2\%)} ~ [{\rm PDF}] ~{\rm
  fb}\,.
\end{split}
\end{equation}
The behaviour of the integrated cross section with the $\mu_0= H_T/2$
scale choice is similar to $\mu_0=E_T/2$. At the central value of the
scale positive and moderate NLO QCD corrections have been
obtained. Specifically, we observe $15\%$ corrections for the CT14 PDF
set, $6\%$ for MMHT and $24\%$ for the NNPDF set. Moreover PDF
uncertainties are of the same size, $2\%-3\%$ only. The only visible
difference is the magnitude of theoretical uncertainties due to the
scale variation. For the last choice, i.e. $H_T/2$, we have not only
obtained the smallest theoretical error, but this error remains the
same independently of the PDF set used.  Namely, LO uncertainties that
are of the order of $50\% ~(40\%)$ are cut down to $10\%$ $(6\%)$ at
NLO, independently of the PDF set, where for values in the brackets
the symmetrisation of errors is performed.  Results are also quite
stable when shifting the $p_{T,\,j_1}$ cut from $40$ GeV up to $120$
GeV as presented in Table \ref{tab:5}.  For $120$ GeV $p_{T, \,
j_1}-$cut, the NLO scale dependence increases by $2\%$ only up to
$12\%$ $(8\%)$ respectively.  Even if the scale choices $E_T /2$ and
$H_T /2$ have similar features, the latter leads to the smallest
theoretical errors and is therefore best suited for the calculation of
cross sections within the scope of our analysis.

\begin{figure}[t!]
\caption{\it Differential cross section distribution as a function of
$E_T$ and $H_T$ at LO (left panel) and at NLO (right panel) for the
$pp\to e^+ \nu_e \mu^- \bar{\nu}_\mu b\bar{b}j +X$ process at the LHC
run II with $\sqrt{s} = 13$ TeV.  Renormalisation and factorisation
scales are set to the common value $\mu_R=\mu_F=\mu_0=m_t$. The LO and
the NLO CT14 PDF sets are employed. Also shown is the 
$E_T/H_T$ ratio. }
\begin{center}
\includegraphics[width=0.49\textwidth]{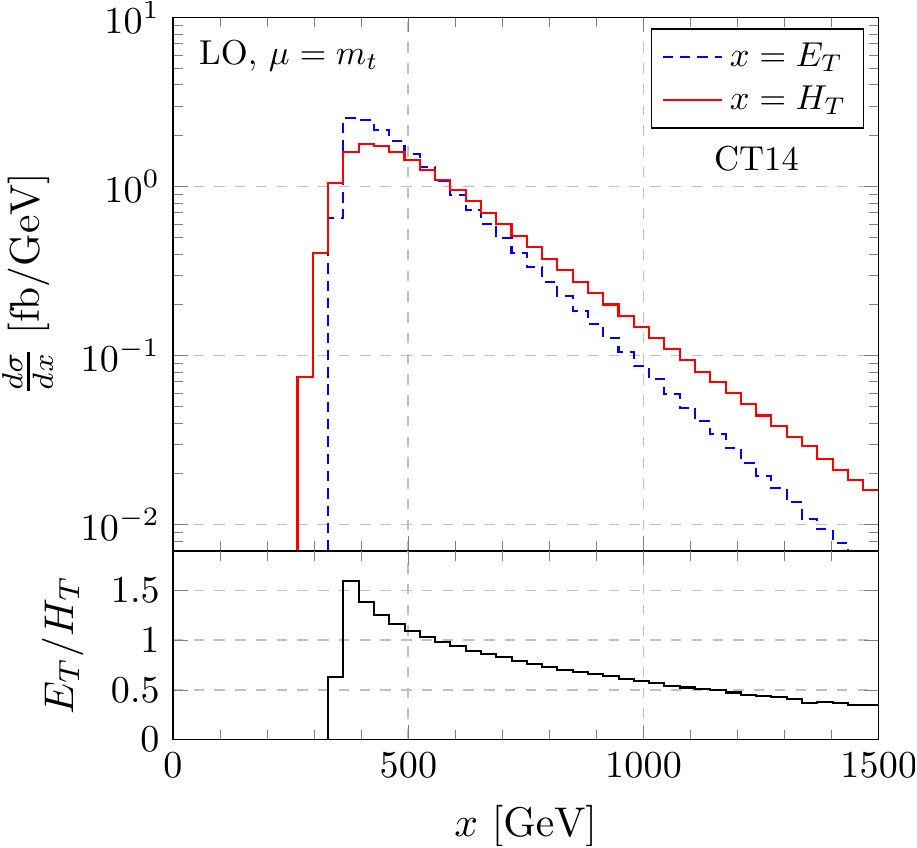} 
\includegraphics[width=0.49\textwidth]{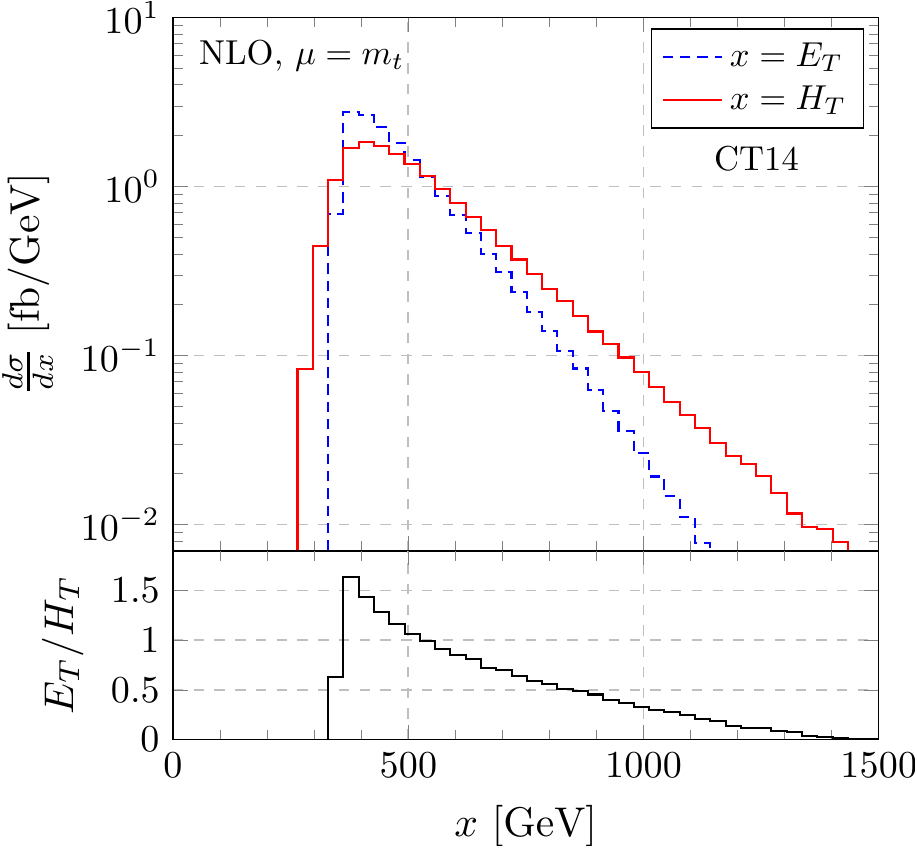} 
 \end{center}
\label{fig:4} 
\end{figure}

To illustrate why the two dynamical scale choices give similar results
we plot in Figure \ref{fig:4} differential cross section distributions
as a function of $E_T$ and $H_T$.  The left panel displays LO results
whereas the right panel NLO ones.  Renormalisation and factorisation
scales are set to the common (fixed) value $\mu_R=\mu_F=\mu_0=m_t$ and
the CT14 LO and NLO PDF sets have been employed. The upper panels
present observables while the lower panels display the $E_T/H_T$
ratio. With our selection cuts the $H_T$ distribution has its maximum 
around $2m_t$. Moreover, both observables are quite similar in
the region close to the $t\bar{t}$ threshold and up to about $750$
GeV, which influences the total integrated cross section. Above $750$
GeV the $H_T$ spectrum is much harder than the corresponding $E_T$
spectrum, which should be reflected in the high $p_T$ tails of various
differential cross sections that we are going to examine in the next
section.
%
\begin{figure}[t!]
\caption{\it Scale dependence of the LO cross section with the
individual contributions of the partonic channels (left panel) and
scale dependence of the LO and NLO cross sections (right panel) for
the $pp\to e^+ \nu_e \mu^- \bar{\nu}_\mu b\bar{b}j +X$ process at the
LHC run II with $\sqrt{s} = 13$ TeV.  Renormalisation and
factorisation scales are set to the common value $\mu_R=\mu_F=\mu_0$
with $\mu_0=m_t$, $\mu_0=H_T/2$ and $\mu_0=E_T/2$. The LO  and the 
NLO CT14 PDF sets are  employed.}
\begin{center}
\includegraphics[width=0.49\textwidth]{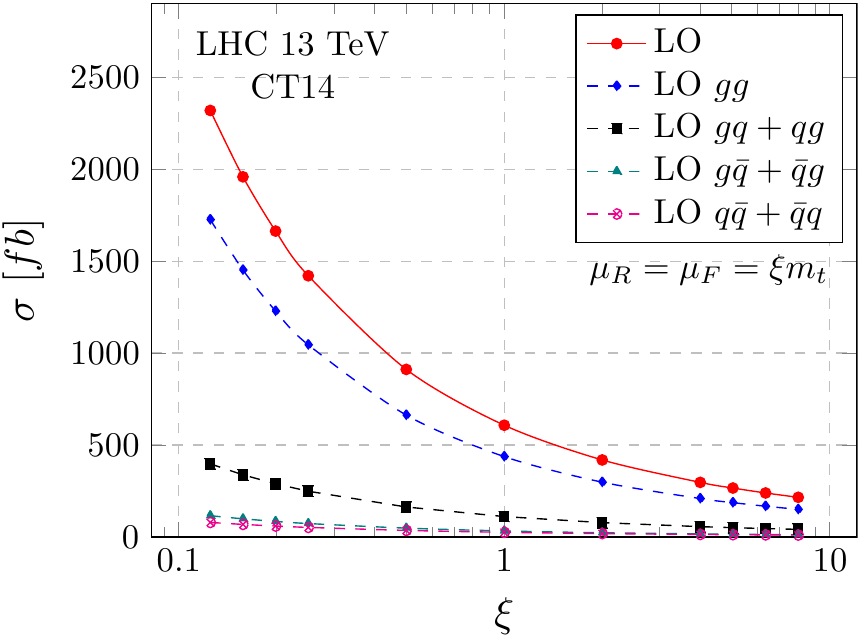} 
 \includegraphics[width=0.49\textwidth]{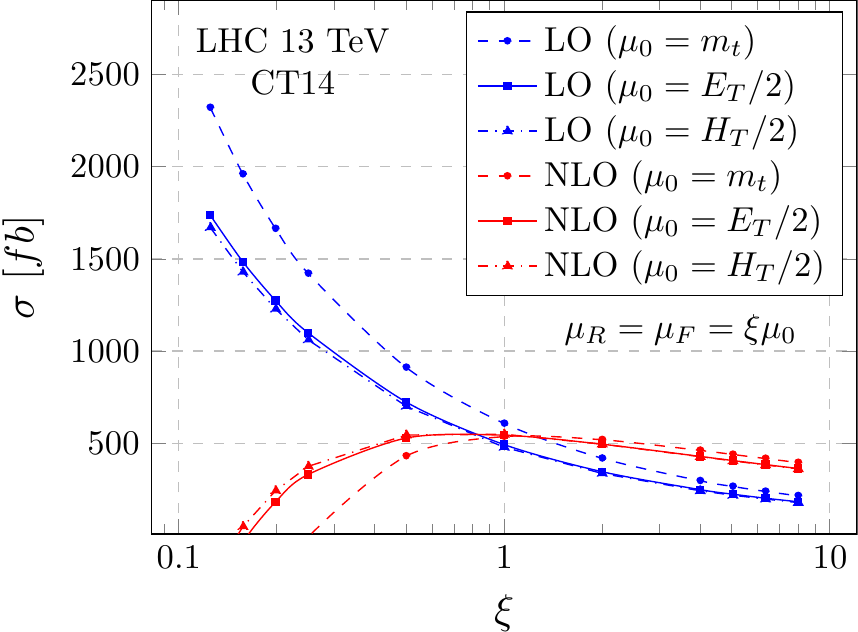}
\end{center}
\label{fig:5} 
\end{figure}
\begin{figure}[t!]
\caption{\it Scale dependence of the LO and NLO integrated cross
section for the $pp\to e^+ \nu_e \mu^- \bar{\nu}_\mu b\bar{b}j +X$
process at the LHC run II with $\sqrt{s} = 13$ TeV.  Renormalisation
and factorisation scales are set to the common value
$\mu_R=\mu_F=\mu_0$ with $\mu_0=m_t$, $\mu_0=E_T/2$ and
$\mu_0=H_T/2$. The LO and the NLO CT14 PDF sets are employed. For each
case of $\mu_0$ also shown is the variation of $\mu_R$ with fixed
$\mu_F$ and the variation of $\mu_F$ with fixed
$\mu_R$.}
\begin{center}
   \includegraphics[width=0.49\textwidth]{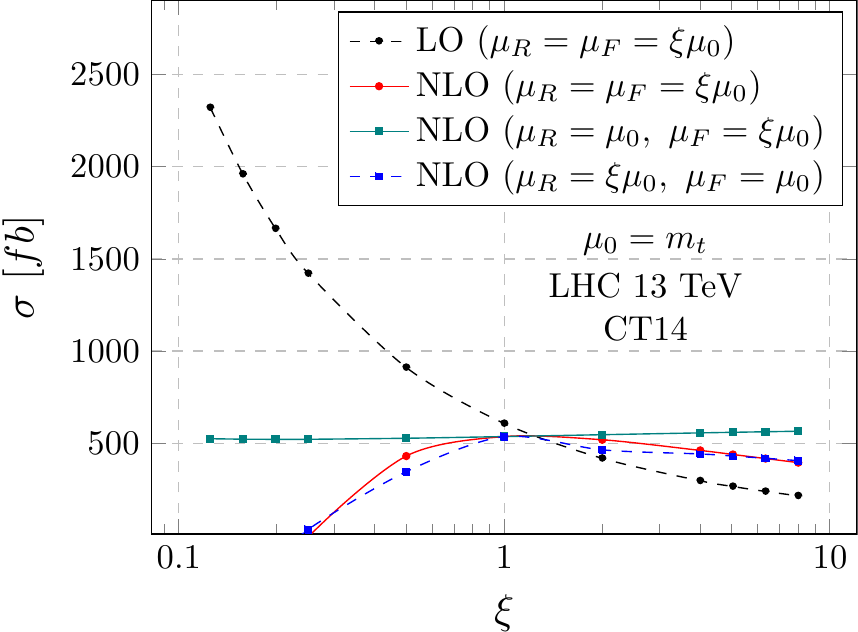} 
 \includegraphics[width=0.49\textwidth]{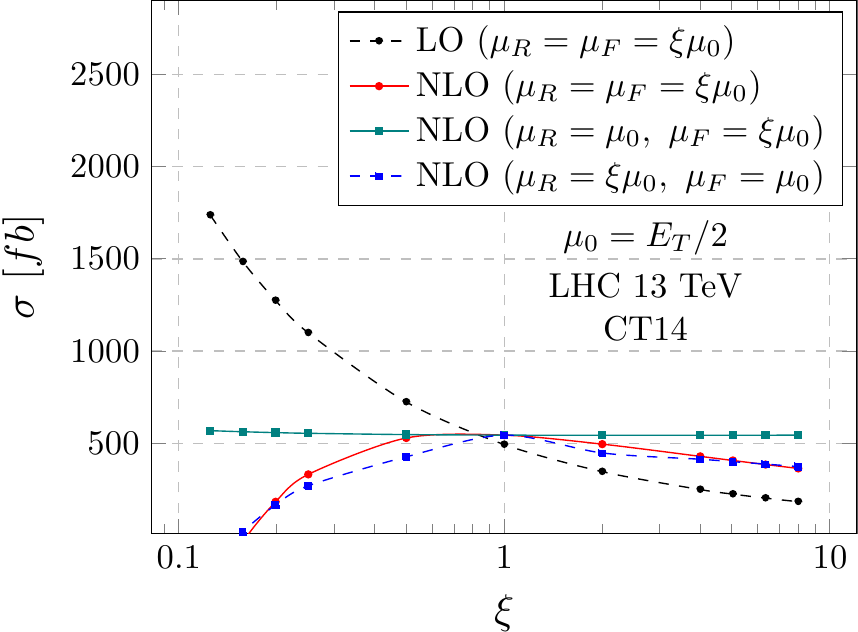} 
  \includegraphics[width=0.49\textwidth]{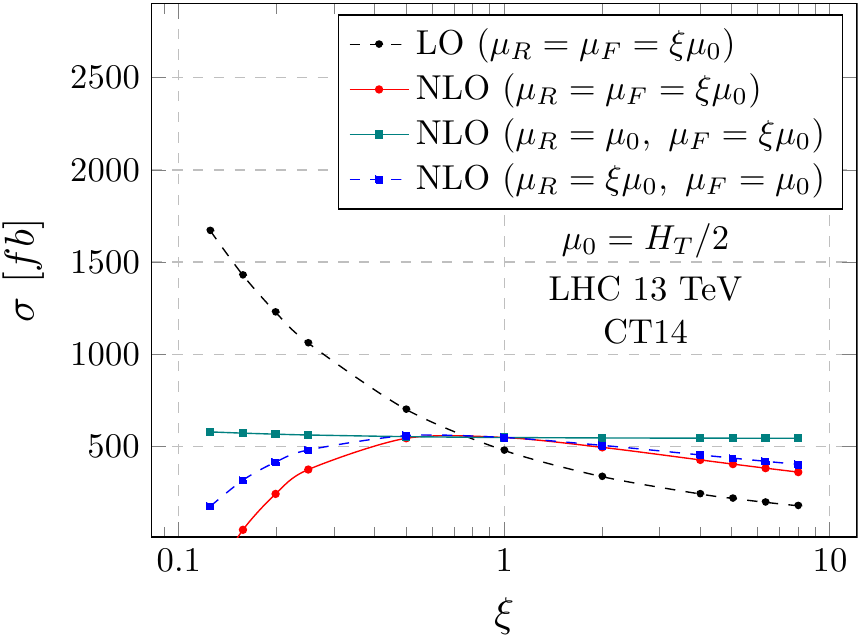}
\end{center}
\label{fig:6} 
\end{figure}
%

It is also instructive to present the scale dependence of our results
in a more graphical fashion. To this end, we show in Figure
\ref{fig:5} the total cross sections at LO and NLO, based on the CT14
PDF set. The scales $\mu_R$ and $\mu_F$ are varied simultaneously
according to the prescription $\mu_R = \mu_F = \mu_0 = \xi m_t$ with
$\xi \in (0.125,\dots,8)$. The dependence is large, illustrating the
well known fact that the LO prediction can only provide a rough
estimate. At the LHC with $\sqrt{s}=13$ TeV for  $\xi=1$ and for our
selection cut the $gg$ channel (blue dashed curve) dominates the total
$pp$ cross section by about $72\%$, followed by the $gq$ channel
(black dashed curve) with about $18\%$. The remaining $10\% $ comes
from two additional channels, $g\bar{q}$ (green dashed curve) and
$q\bar{q}$ (pink dashed curve) that contributes at the $6\%$ and
$4\%$ level respectively.  In the right panel of Figure \ref{fig:5}
the scale dependence of the NLO cross section (red curves) is shown
together with the LO one (blue curves). This time for three different
scale choices, namely $\mu_0=m_t$, $\mu_0 =E_T/2$ and $\mu_0
=H_T/2$. As already discussed, we observe a reduction of the scale
uncertainty while going from LO to NLO. Additionally, we confirm that
both $\mu_0=E_T/2$ and $\mu_0=H_T/2$ give similar results within the
whole plotted range. In Figure \ref{fig:6} we display again the
dependence of the integrated LO (black dashed curve) and NLO (red
solid curve) cross sections on the variation of the fixed and
dynamical scales for the CT14 PDF set. Here, however, we show
additionally results with individual variation of $\mu_R$ and
$\mu_F$. Each time we plot two additional curves, the first one (green
solid curve) corresponds to the case where $\mu_R$ is kept fix at the
central value, while $\mu_F$ is varied and the second one (blue dashed
curve) describes the opposite situation. We can observe that, 
independently of the scale choice, either it is $\mu_0=m_t$, $
\mu_0=E_T/2$ or $\mu_0=H_T/2$, the scale variation is driven by the
changes in $\mu_R$.

To summarise this part, for the total cross section, where effects of
the phase space regions close to the threshold for the $t\bar{t}$
production dominate, all three scales, $\mu_0=m_t$, $\mu_0=E_T/2$ and
$\mu_0=H_T/2$ describe the process under scrutiny very well. They all
agree within their respective theoretical errors, as it should be,
however, $\mu_0=H_T/2$ provides the smallest theoretical error, that
is independent of the PDF set and the $p_{T,\,j_1}$ cut applied. For
this reason, it can be recommended as the best one for the computation
of total cross sections for inclusive analyses at $\sqrt{s} = 13$
TeV. On the other hand, differential cross sections extend themselves
up to energy scales that are much larger than the $t\bar{t}$
threshold. Thus, in the next section we shall examine which scales are
also suitable for the description of differential cross sections.

%
\subsection{Differential Distributions}
%

In addition to the normalization of the integrated cross section, QCD
corrections can affect the shape of various kinematic
distributions. To quantify the size of these distortions we shall
examine differential distributions for various observables of
interest for the LHC. These distributions are
obtained with the CT14 PDF sets by applying the cuts and parameters
specified in the previous section. Also here we examine three
different scale choices, the fixed scale $\mu_0=m_t$ and two dynamical
scales $\mu_0=E_T/2$ and $\mu_0=H_T/2$. For each of the observables we
present three plots that correspond to the three scale choices. The
upper panel of each plot shows the absolute prediction at LO and NLO
together with their scale dependence bands obtained from the envelope
of results calculated according to Eq.~\eqref{scan}.  The lower panels
display the same LO and NLO predictions normalised to the LO result at
the central scale. Thus, the blue band provides the relative scale
uncertainty of the LO cross section, whereas the red band gives the
differential ${\cal K}$-factor together with the uncertainty band.

%
\begin{figure}[t!]
\caption{\it Averaged differential cross section distributions as a
function of the transverse momentum of the top quark for the $pp\to
e^+ \nu_e \mu^- \bar{\nu}_\mu b\bar{b}j +X$ process at the LHC run II
with $\sqrt{s} = 13$ TeV.  The upper plot shows absolute LO and NLO
predictions together with corresponding uncertainty bands resulting
from scale variations. The lower panel displays the differential
${\cal K}$ factor together with the uncertainty band (red band). Also
shown is the relative scale uncertainty of the LO cross section (blue
band).  Renormalisation and factorisation scales are set to the common
value $\mu_R=\mu_F=\mu_0$ where $\mu_0=m_t$, $\mu_0=E_T/2$ and
$\mu_0=H_T/2$. The CT14 PDF sets are employed.}
\begin{center}
\includegraphics[width=0.95\textwidth]{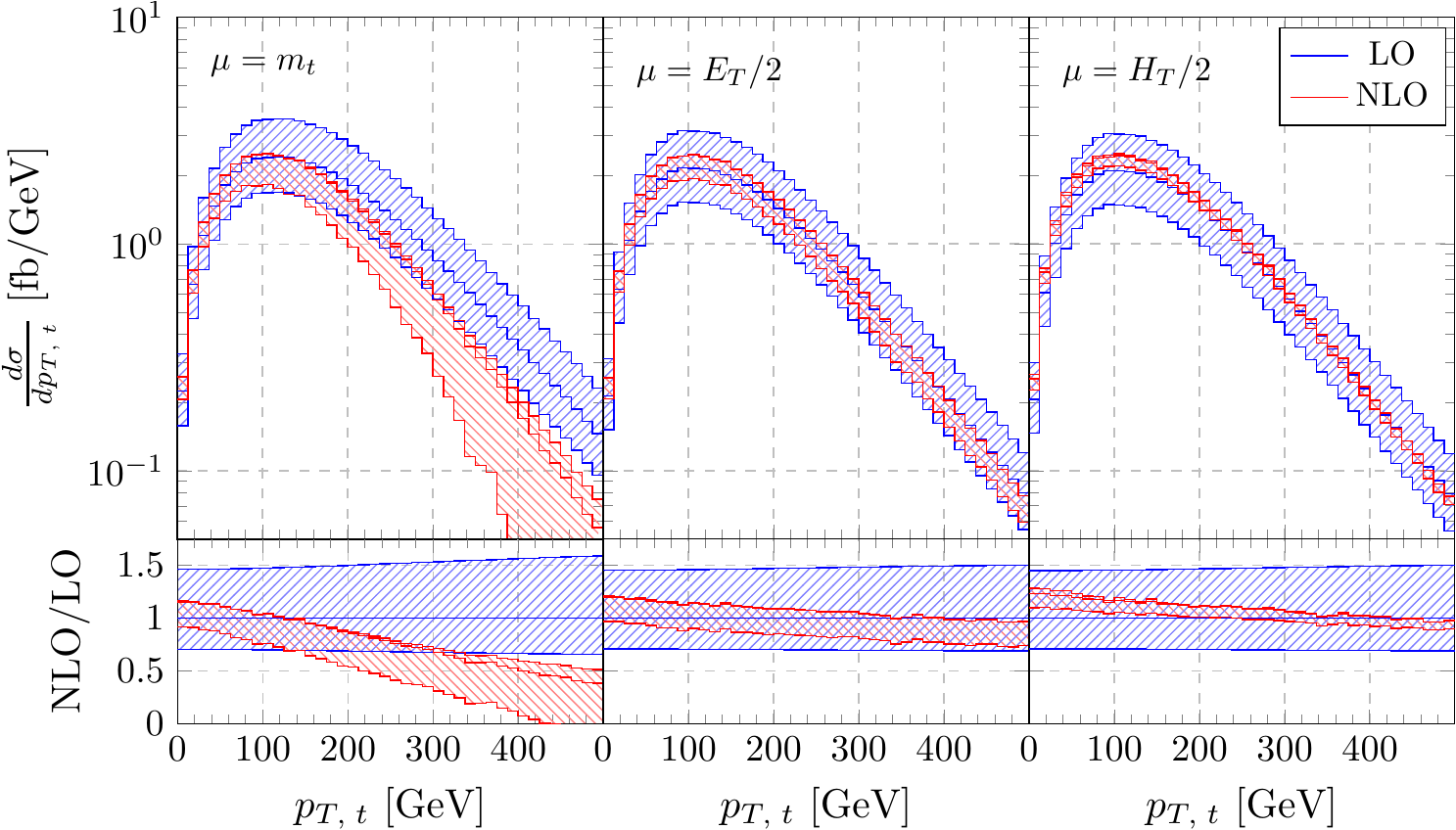}
\end{center}
\label{fig:dis:1a} 
\end{figure}
\begin{figure}[t!]
\caption{\it Averaged differential cross section distributions as a
function of the transverse momentum of the top quark at LO (left
panel) and at NLO (right panel). Results are given for the $pp\to e^+
\nu_e \mu^- \bar{\nu}_\mu b\bar{b}j +X$ process at the LHC run II with
$\sqrt{s} = 13$ TeV  with $\mu_R=\mu_F=\mu_0$ where $\mu_0=m_t$,
$\mu_0=E_T/2$ and $\mu_0=H_T/2$.  The lower panel displays a ratio to
the prediction with $\mu_0=m_t$.  The CT14 PDF sets are employed.}
\begin{center}
\includegraphics[width=0.45\textwidth]{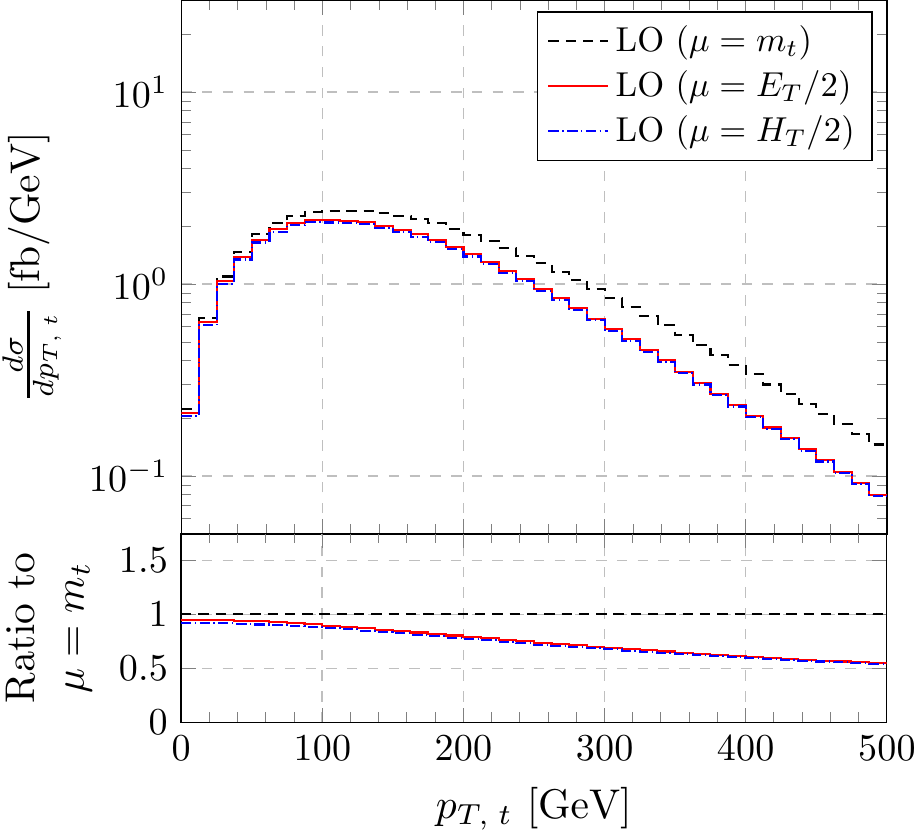}
\includegraphics[width=0.45\textwidth]{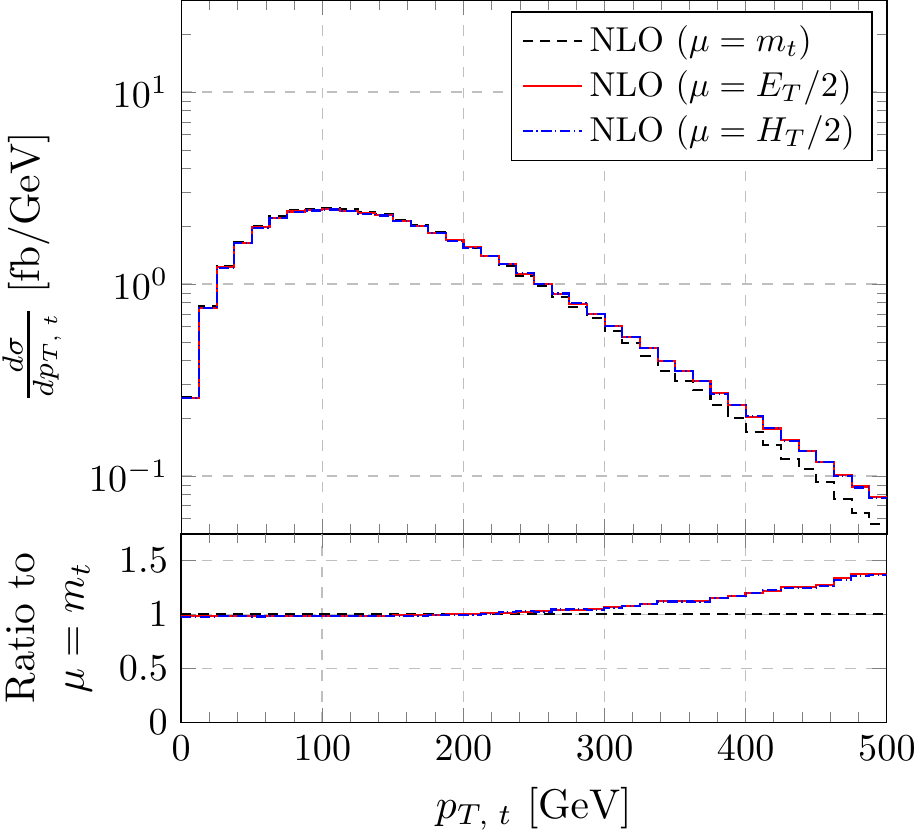}
\end{center}
\label{fig:dis:1aa} 
\end{figure}

\begin{figure}[t!]
\caption{\it Averaged differential cross section distributions as a
function of the rapidity of the top quark for
the $pp\to e^+ \nu_e \mu^- \bar{\nu}_\mu b\bar{b}j +X$ process at the
LHC run II with $\sqrt{s} = 13$ TeV.  }
\begin{center}
\includegraphics[width=0.95\textwidth]{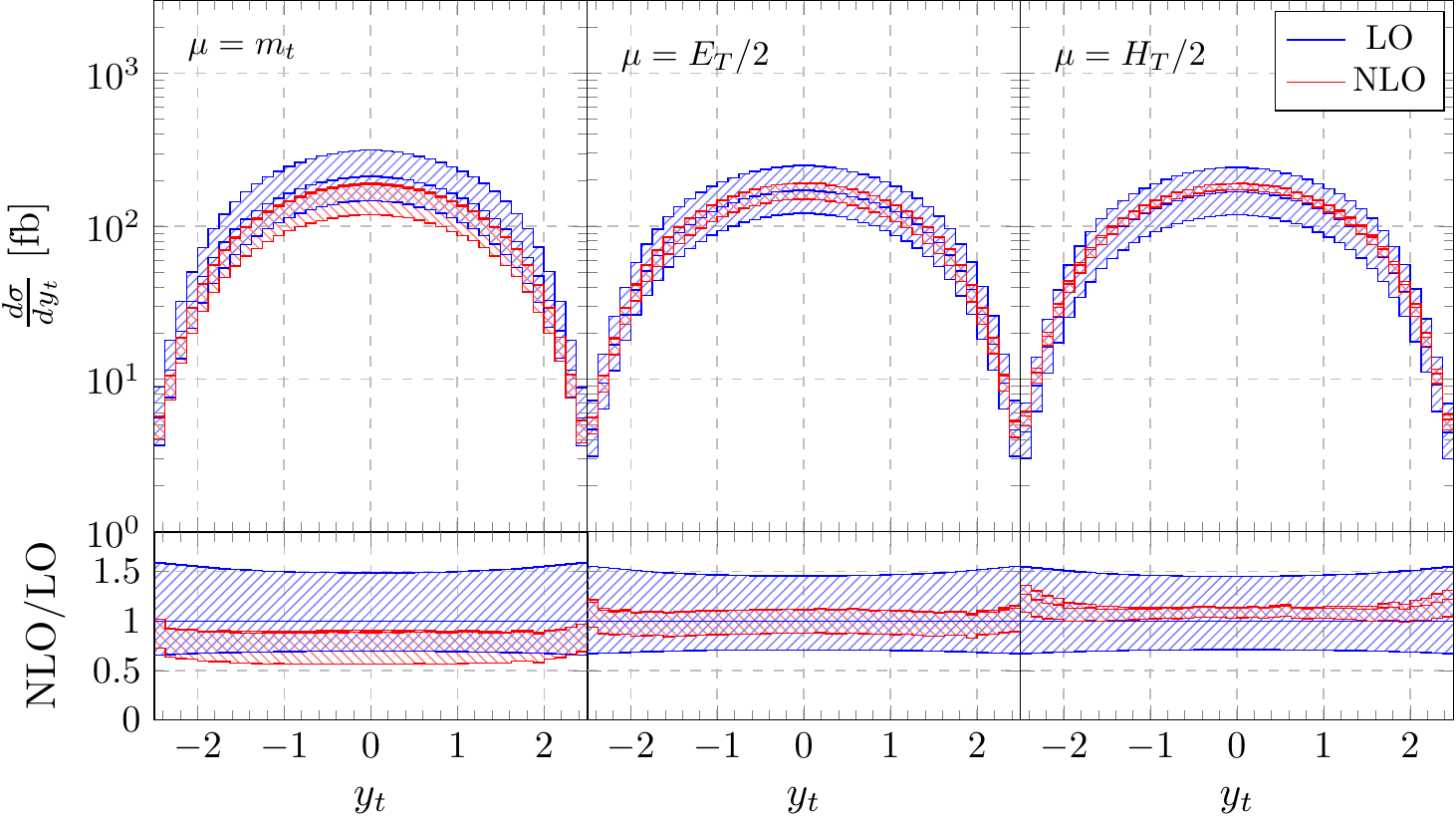} 
\end{center}
\label{fig:dis:1b} 
\end{figure}

\begin{figure}[t!]
\caption{\it Averaged differential cross section distributions as a
function of the rapidity of the top quark at LO (left
panel) and at NLO (right panel) for the $pp\to e^+ \nu_e \mu^-
\bar{\nu}_\mu b\bar{b}j +X$ process at the LHC run II with $\sqrt{s} =
13$ TeV.  }
\begin{center}
\includegraphics[width=0.45\textwidth]{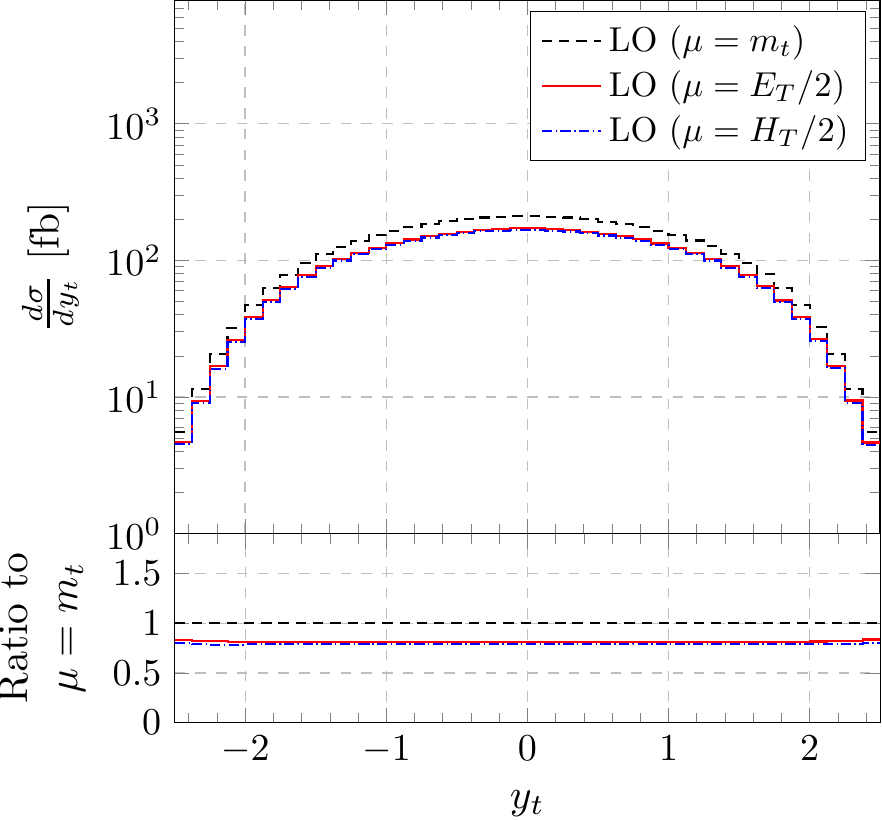}
\includegraphics[width=0.45\textwidth]{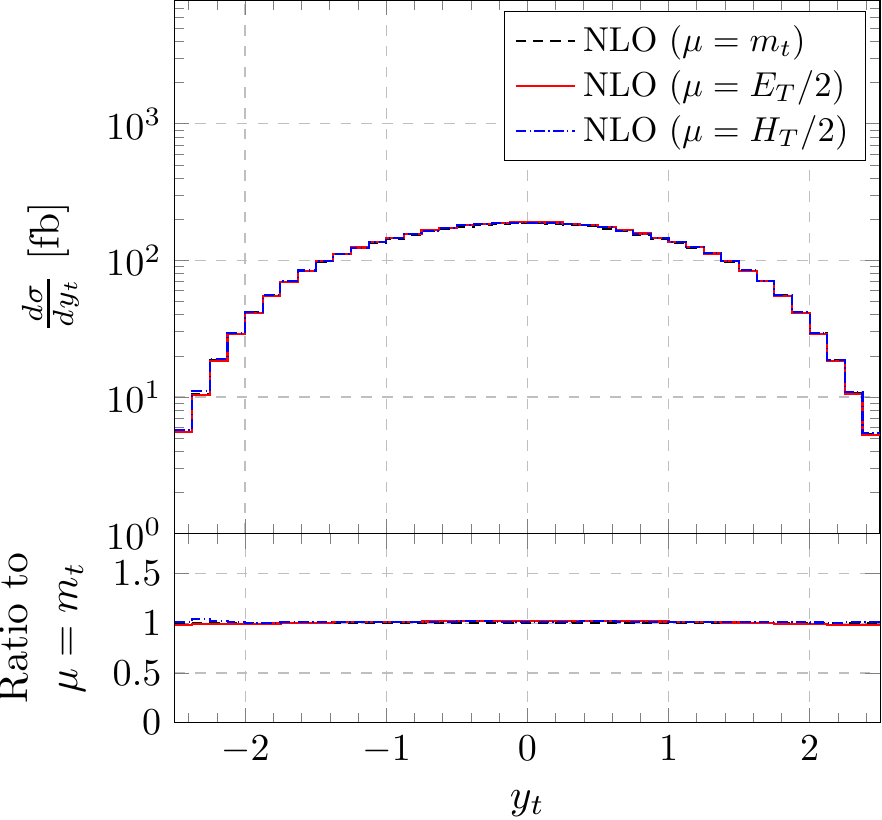}
\end{center}
\label{fig:dis:1bb} 
\end{figure}

\begin{figure}[t!]
\caption{\it Differential cross section distribution as a
function of the invariant mass of the $t\bar{t}$ system for
the $pp\to e^+ \nu_e \mu^- \bar{\nu}_\mu b\bar{b}j +X$ process at the
LHC run II with $\sqrt{s} = 13$ TeV. }
\begin{center}
\includegraphics[width=0.95\textwidth]{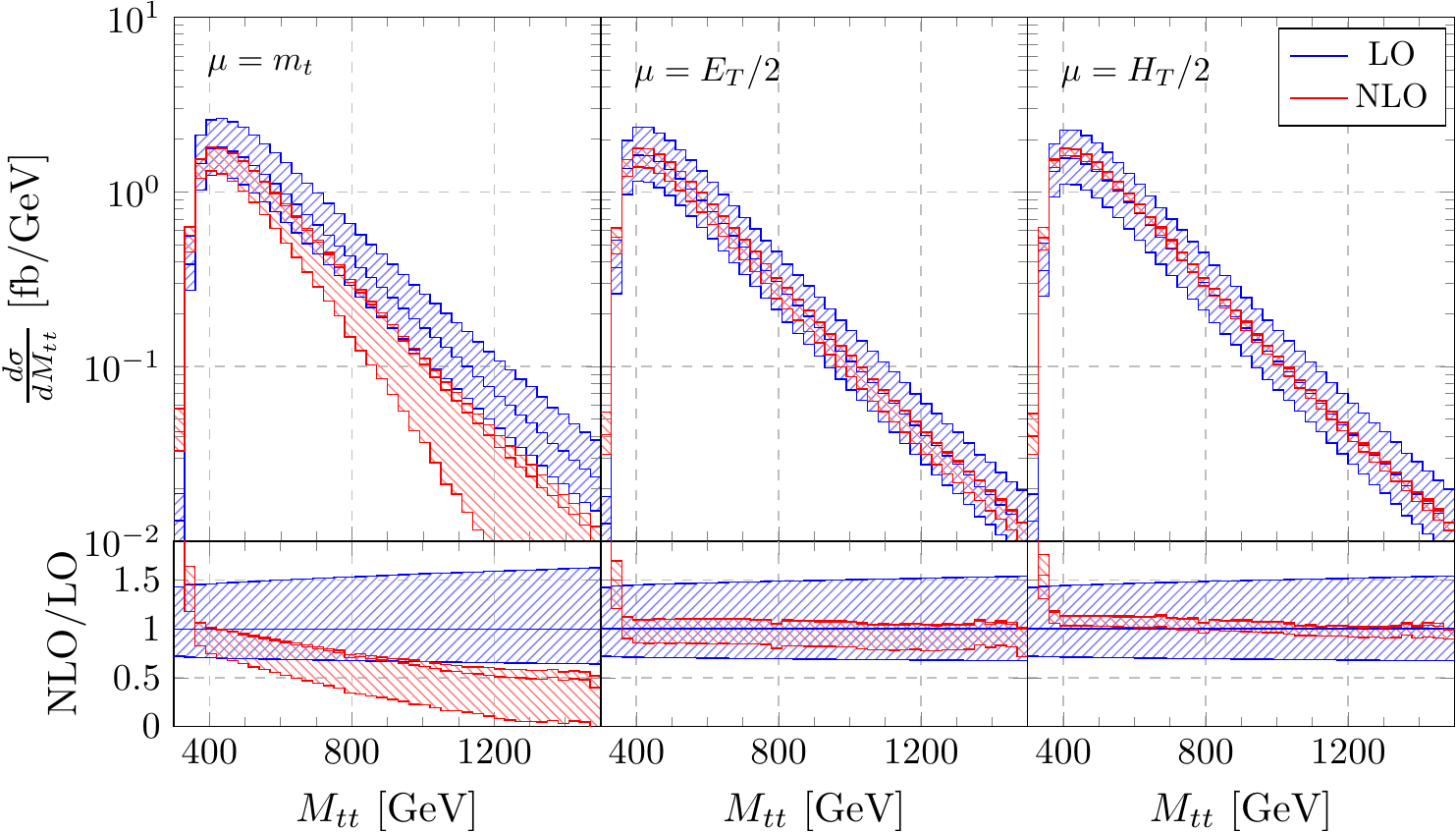} 
\end{center}
\label{fig:dis:1c} 
\end{figure}

\begin{figure}[t!]
\caption{\it Differential cross section distributions as a
function of the invariant mass of the  $t\bar{t}$ pair at LO (left
panel) and at NLO (right panel) for the $pp\to e^+ \nu_e \mu^-
\bar{\nu}_\mu b\bar{b}j +X$ process at the LHC run II with $\sqrt{s} =
13$ TeV.  }
\begin{center}
\includegraphics[width=0.45\textwidth]{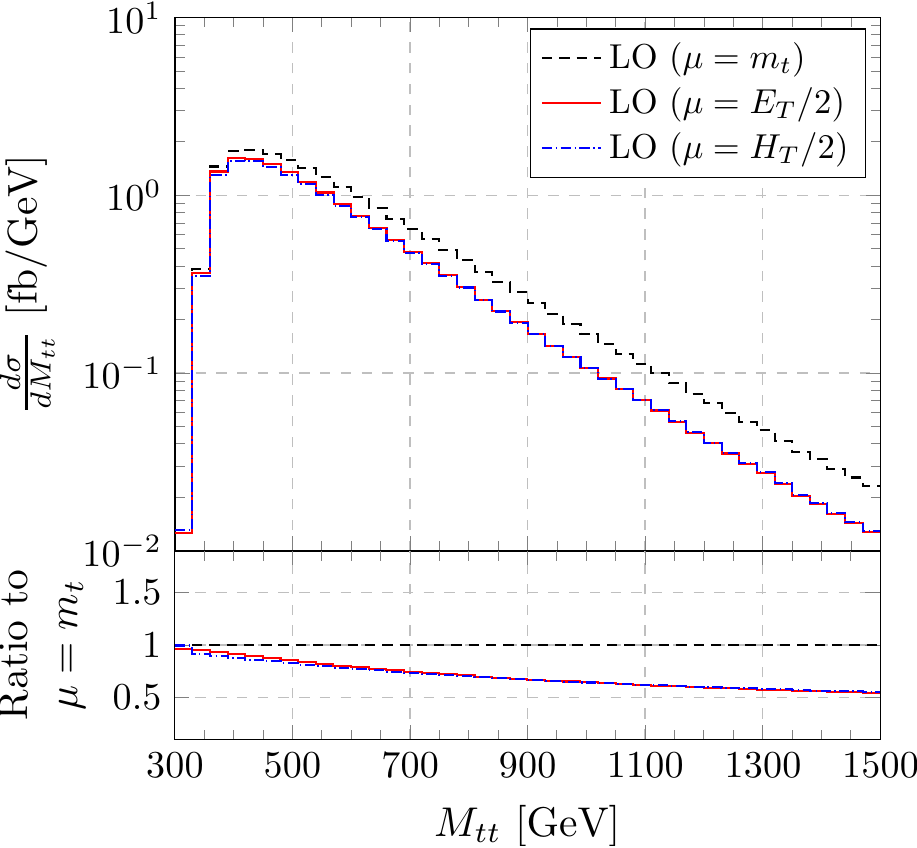}
\includegraphics[width=0.45\textwidth]{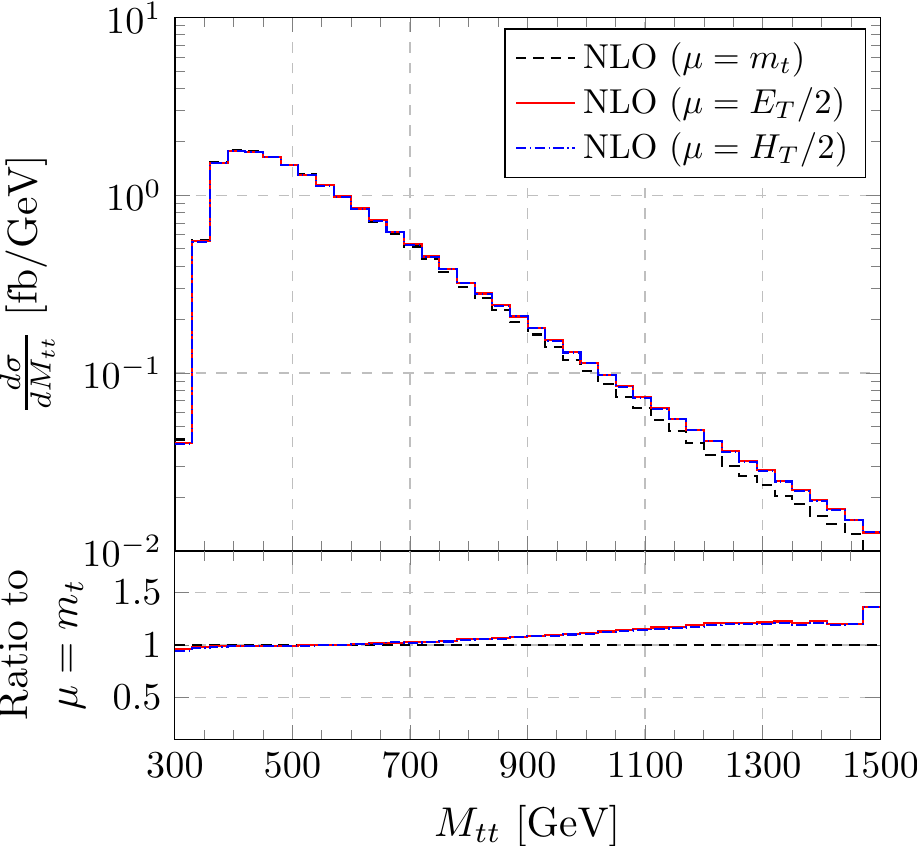}
\end{center}
\label{fig:dis:1cc} 
\end{figure}

We start with the top-quark kinematics.  In Figures~\ref{fig:dis:1a}
and \ref{fig:dis:1b} we present the averaged differential
cross section as a function of the transverse momentum and rapidity of
the top quark.  In Figure \ref{fig:dis:1c} the invariant mass of the
$t\bar{t}$ system, $M_{t\bar{t}}=\sqrt{(p_t+p_{\bar{t}})^2}$, is
plotted.  The kinematics of the top quark and top antiquark are
determined from the four-momenta of final state objects, i.e. leptons
and bottom-jets.  The $p_{T,\,t}$ and $y_t$ distributions are
especially useful to validate and tune a given parton shower model as
well as to check specific higher order QCD calculations.  They can
also be exploited in searches for physics beyond the SM.  On the other
hand, the invariant mass of the $t\bar{t}$ pair is the observable to
look for new $s$-channel resonances that may arise in the $t\bar{t}$
system.  It can be used to test new physics models, where top quark
pairs are produced through the exchange of new heavy particles, e.g.
heavy Higgs boson(s) from supersymmetric extensions of the
SM, a heavy neutral resonance $Z^\prime$ or Kaluza-Klein excitations
of gluons.  Any deviation from the SM shape and normalisation in these
observables could signal the presence of new physics. Thus, they
need to be described as precisely as possible without any
approximations.  In Figure \ref{fig:dis:1a} we can observe that for a
fixed scale, $\mu_0=m_t$, the NLO corrections to the top-quark
transverse momentum distribution do not simply rescale the LO shapes,
but induce distortions of the order of $80\%$. Clearly, substantial,
of the order of $60\%$, negative NLO corrections affect the high
$p_{T,\, t}$ region.  We also note that the NLO error bands do not fit
within the LO ones as one would expect from a well-behaved
perturbative expansion. Thus, the fixed scale choice does not ensure a
stable shape when going from LO to NLO for this observable.
Through the implementation of a dynamical scale, large discrepancies
between the shapes of these distributions at NLO and LO have
disappeared. Even though the resulting differential ${\cal K}$-factor
is not flat the NLO QCD corrections are substantially decreased in the
tails, which is mainly  due to large changes in the LO distributions in
that region. To be more specific,  at the central value of the scale, 
high $p_T$ tails received negative but tiny ($3\%$) NLO
corrections. Overall distortions are around $25\%$ independently of
the dynamical scale choice.  In general the LO curve is much more
sensitive to the variation of the scale and will change more rapidly
than the NLO one. In addition, one can observe that the NLO error
bands as calculated through scale variation nicely fit within the LO
error bands. Also in the case of the differential cross section
$\mu_0=H_T/2$ provides the smallest theoretical uncertainties in the
whole plotted range. In Figure~\ref{fig:dis:1aa} we plot $p_{T,\, t}$
again, this time, however, LO and NLO spectra are given separately
only for the central value of the given scale. On the other hand, in
the lower panel a ratio of both dynamical scale choices to $\mu_0=m_t$
is displayed. We can notice that at LO already around $100$ GeV curves
described by dynamical scales, $\mu_0=E_T/2$ and $\mu_0= H_T/2$ vary
substantially from the one given by $\mu_0=m_t$. The latter yields
a much harder spectrum.  At NLO, the difference between the fixed and the
dynamical scale is smaller, as it should be, because of the reduced
dependence of NLO results on the renormalisation and factorisation
scales. Up to $300$ GeV predictions for all three scales are in
agreement. Above $300$ GeV, however, $\mu_0=m_t$ gives a softer spectrum
as compared to $\mu_0=E_T/2$ and $\mu_0=H_T/2$.  Given the better
performance in terms of perturbative stability, we believe that the
dynamical scales are more appropriate to model the high-$p_T$
tails. For the rapidity distribution of the top quark, shown
in Figure~\ref{fig:dis:1b}, we observe a different pattern. QCD
corrections for $\mu_0 = m_t$  are negative, moderate (below $12\%$)
and quite stable in the whole rapidity range. This can be
easily understood since $y_t$ is a dimensionless observable, that
receives contributions from all scales, most notably from those that
are sensitive to the threshold for the $t\bar{t}$ production.
Dynamical scales do not alter this behaviour but rather affect only
the normalisation of the LO prediction, which can be observed in
Figure~\ref{fig:dis:1bb}. As a consequence for $\mu_0=E_T/2$ and
$\mu_0=H_T/2$ positive, moderate ($10\%-15\%$) and quite stable NLO
corrections are obtained.  Also for this observable
$\mu_0=H_T/2$ provides the smallest theoretical
uncertainties. Finally, for the invariant mass of the $t\bar{t}$ pair
we expect a similar behaviour as in case of $p_{T,\,t}$ due to
dimensionful nature of the observable. Indeed, we can see in
Figures~\ref{fig:dis:1c} and \ref{fig:dis:1cc} that our conclusions
remain qualitatively unaffected.  The large negative QCD corrections,
of the order $40\% - 60\%$, which characterize the TeV range in the
case of $\mu_0 = m_t$ are sensibly reduced to about $5\%$ using $\mu_0
= E_T /2$ or $\mu_0 = H_T /2$. The latter two choices are also
legitimate options to describe correctly the NLO spectrum in the
$M_{t\bar{t}} \in (0.7-1.5)$ TeV range as can be observed in
Figure~\ref{fig:dis:1cc}.
%
\begin{figure}[t!]
\caption{\it Differential cross section distribution as a
function of the transverse momentum and rapidity of the hardest jet  
for the $pp\to e^+ \nu_e \mu^- \bar{\nu}_\mu
b\bar{b}j +X$ process at the LHC run II with $\sqrt{s} = 13$ TeV.  }
\begin{center}
\includegraphics[width=0.95\textwidth]{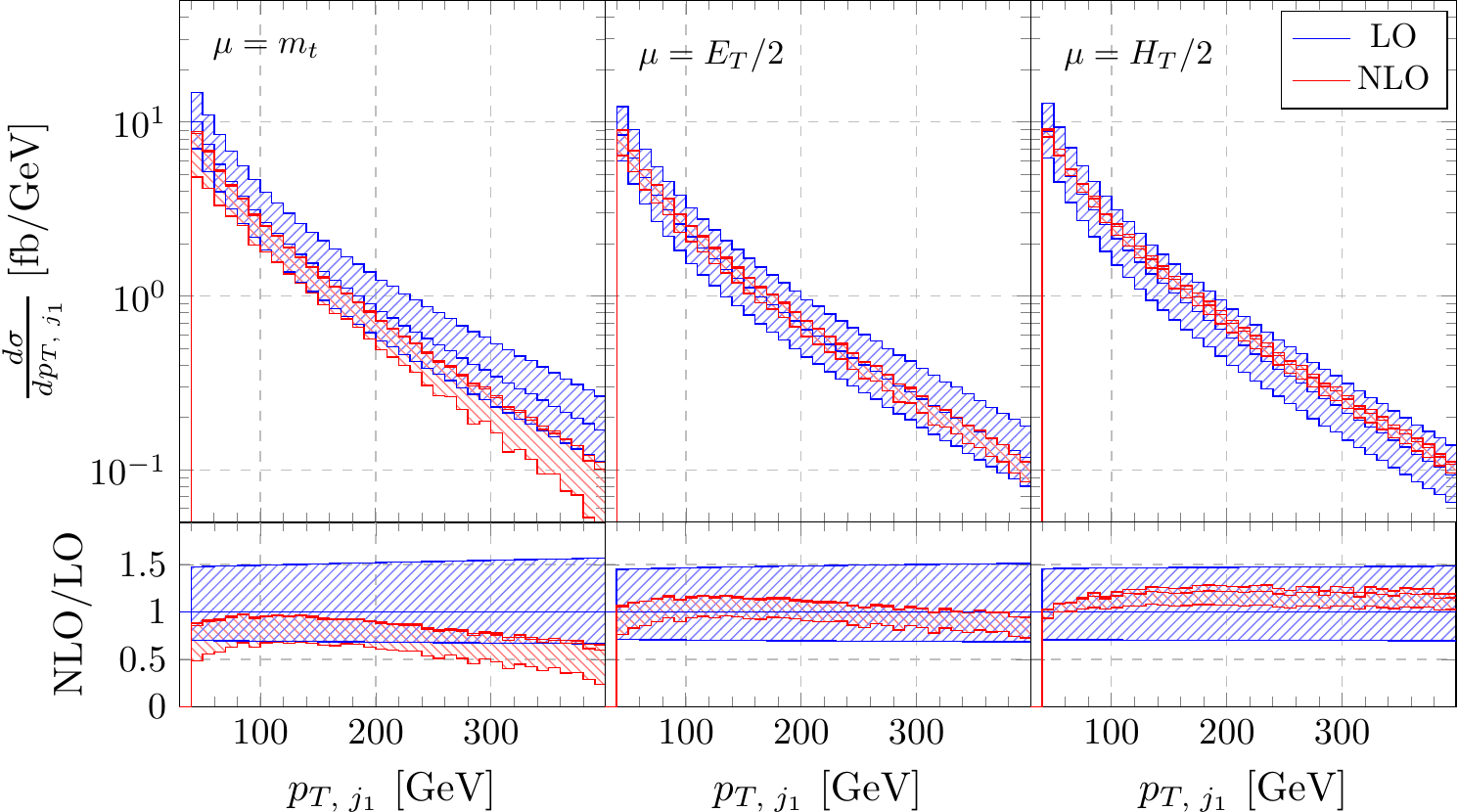} \\
\vspace{0.2cm}
\includegraphics[width=0.95\textwidth]{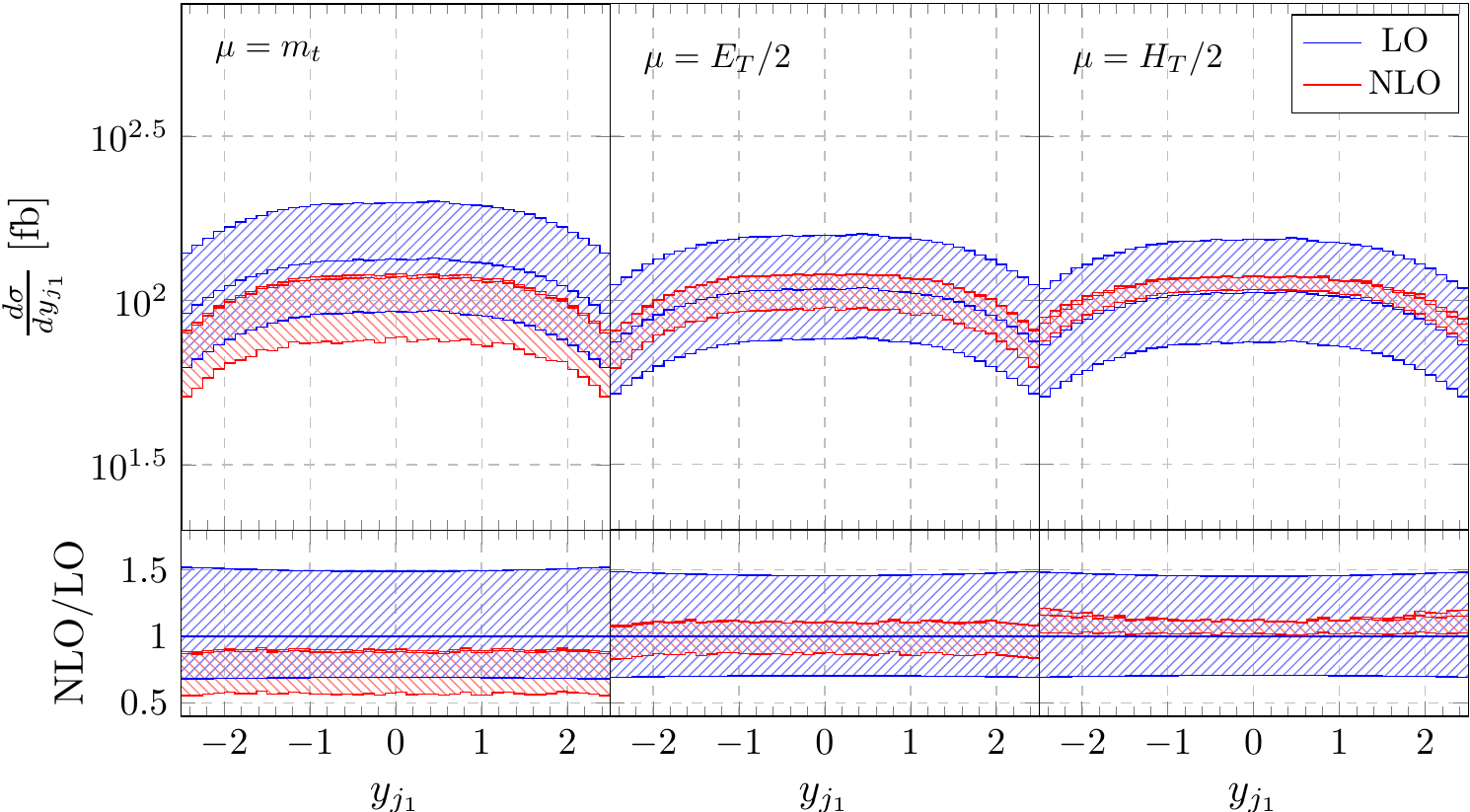} 
\end{center}
\label{fig:dis:2} 
\end{figure}
\begin{figure}[t!]
\caption{\it Differential cross section distributions as a function of
the transverse momentum of the first hardest jet at LO (left panel)
and at NLO (right panel) for the $pp\to e^+ \nu_e \mu^- \bar{\nu}_\mu
b\bar{b}j +X$ process at the LHC run II with $\sqrt{s} = 13$ TeV.  }
\begin{center}
\includegraphics[width=0.45\textwidth]{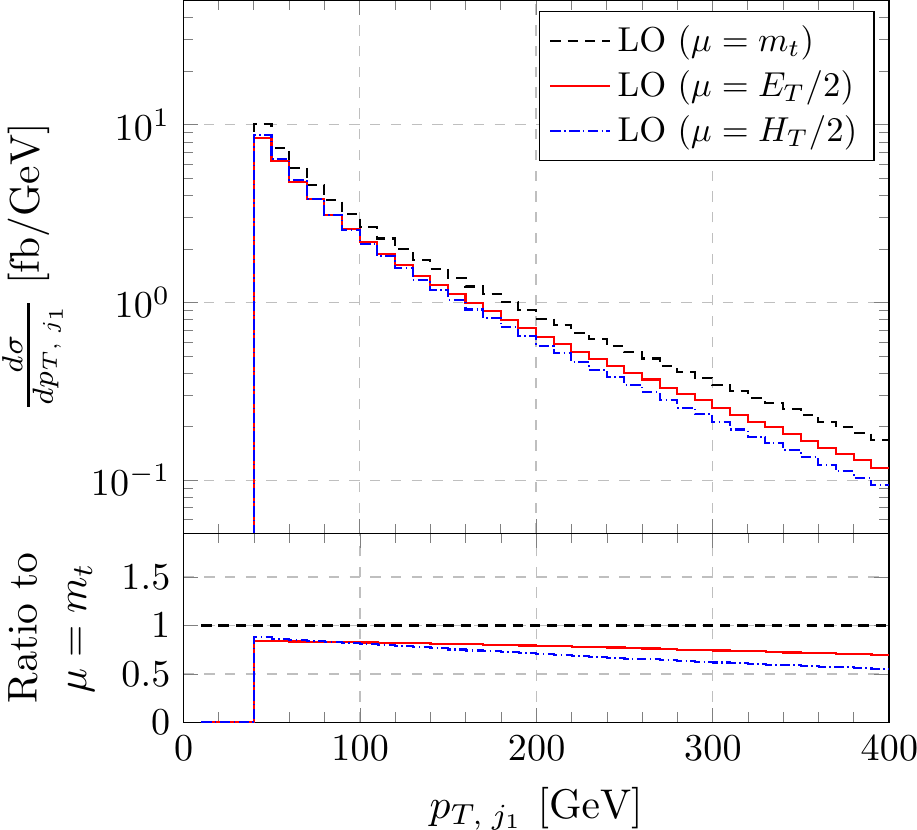}
\includegraphics[width=0.45\textwidth]{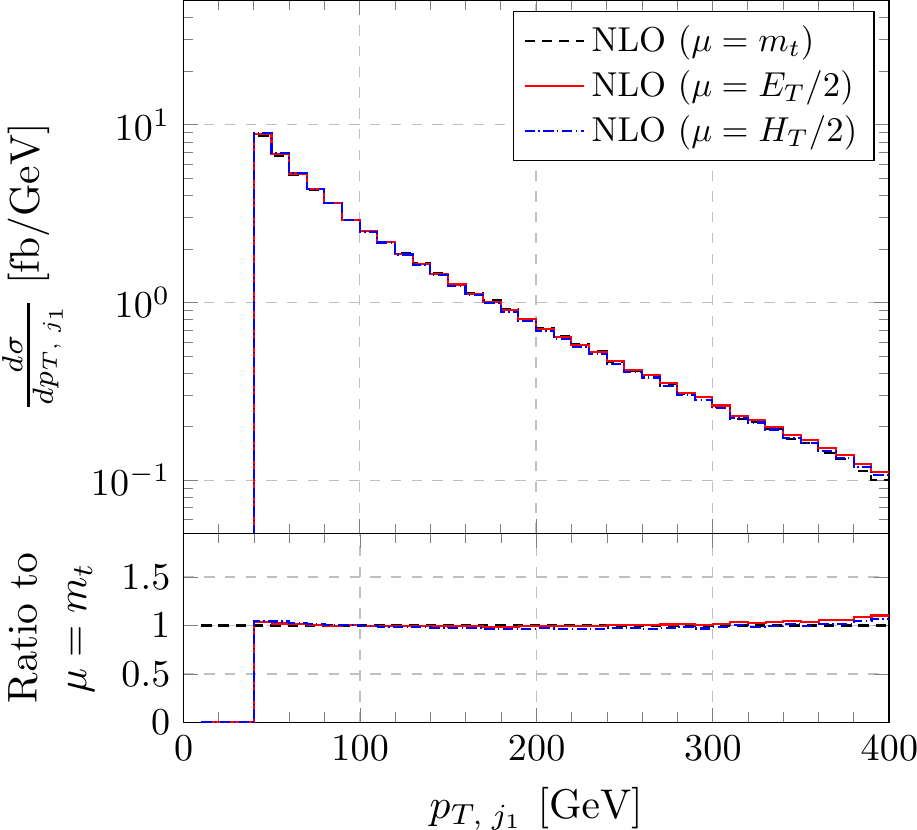}
\end{center}
\label{fig:dis:2a} 
\end{figure}
\begin{figure}[t!]
\caption{\it Differential cross section distributions as a function of
the rapidity of the first hardest jet at LO (left panel) and at NLO
(right panel) for the $pp\to e^+ \nu_e \mu^- \bar{\nu}_\mu b\bar{b}j
+X$ process at the LHC run II with $\sqrt{s} = 13$ TeV.}
\begin{center}
\includegraphics[width=0.45\textwidth]{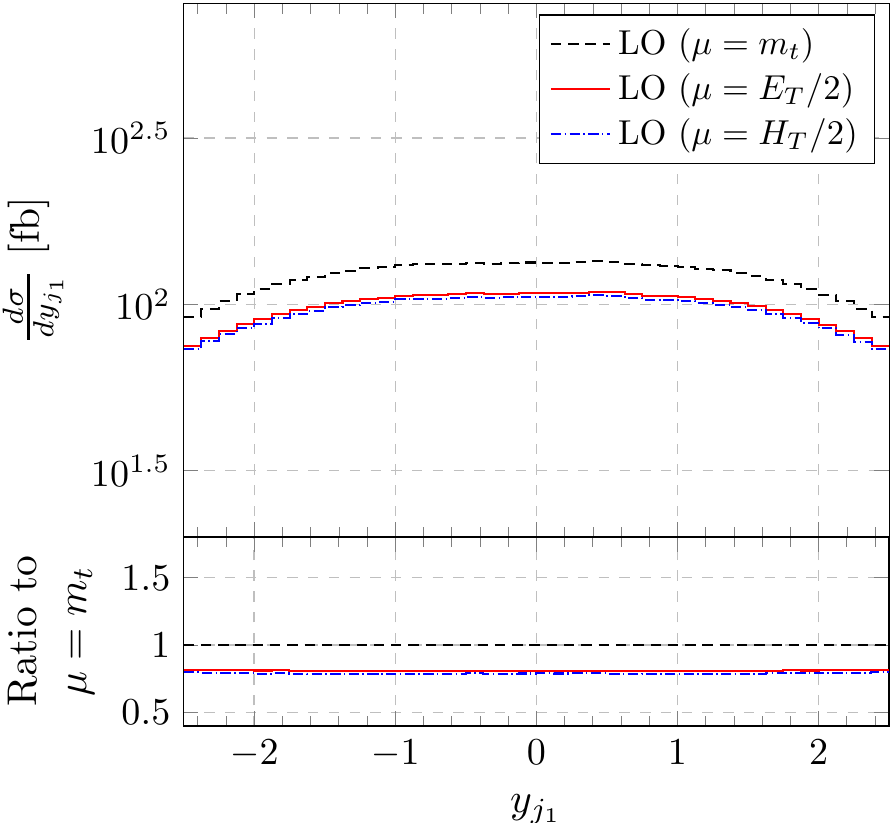}
\includegraphics[width=0.45\textwidth]{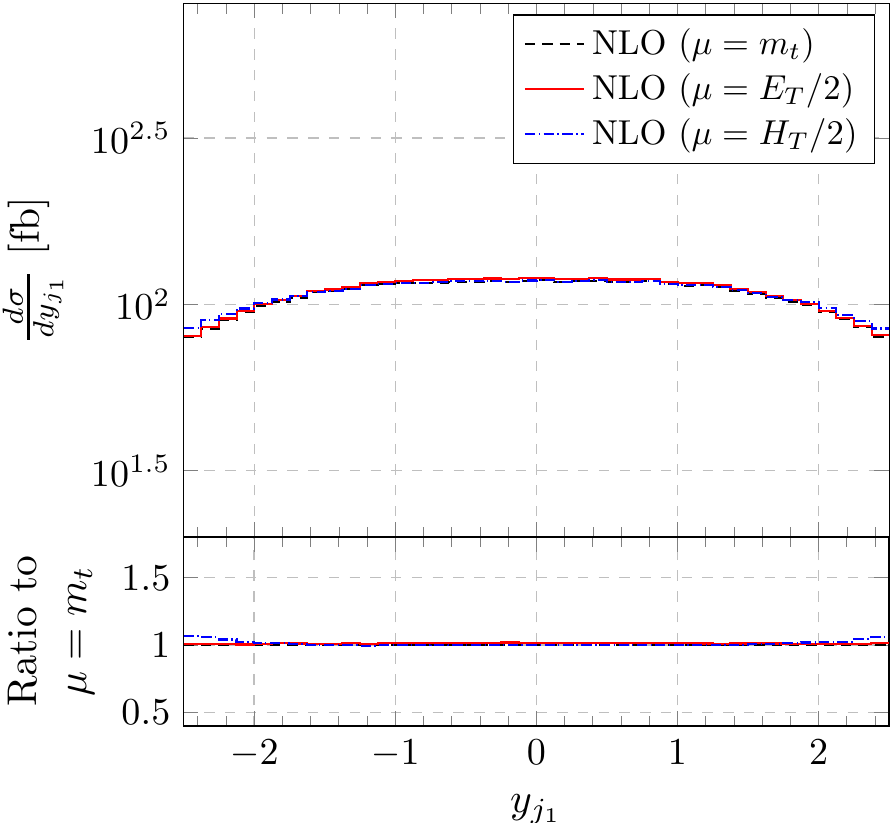}
\end{center}
\label{fig:dis:2b} 
\end{figure}
\begin{figure}[t!]
\caption{\it Averaged differential cross section distribution as a
function of the transverse momentum and rapidity of the bottom-jet for
the $pp\to e^+ \nu_e \mu^- \bar{\nu}_\mu b\bar{b}j +X$ process at the
LHC run II with $\sqrt{s} = 13$ TeV. }
\begin{center}
\includegraphics[width=0.95\textwidth]{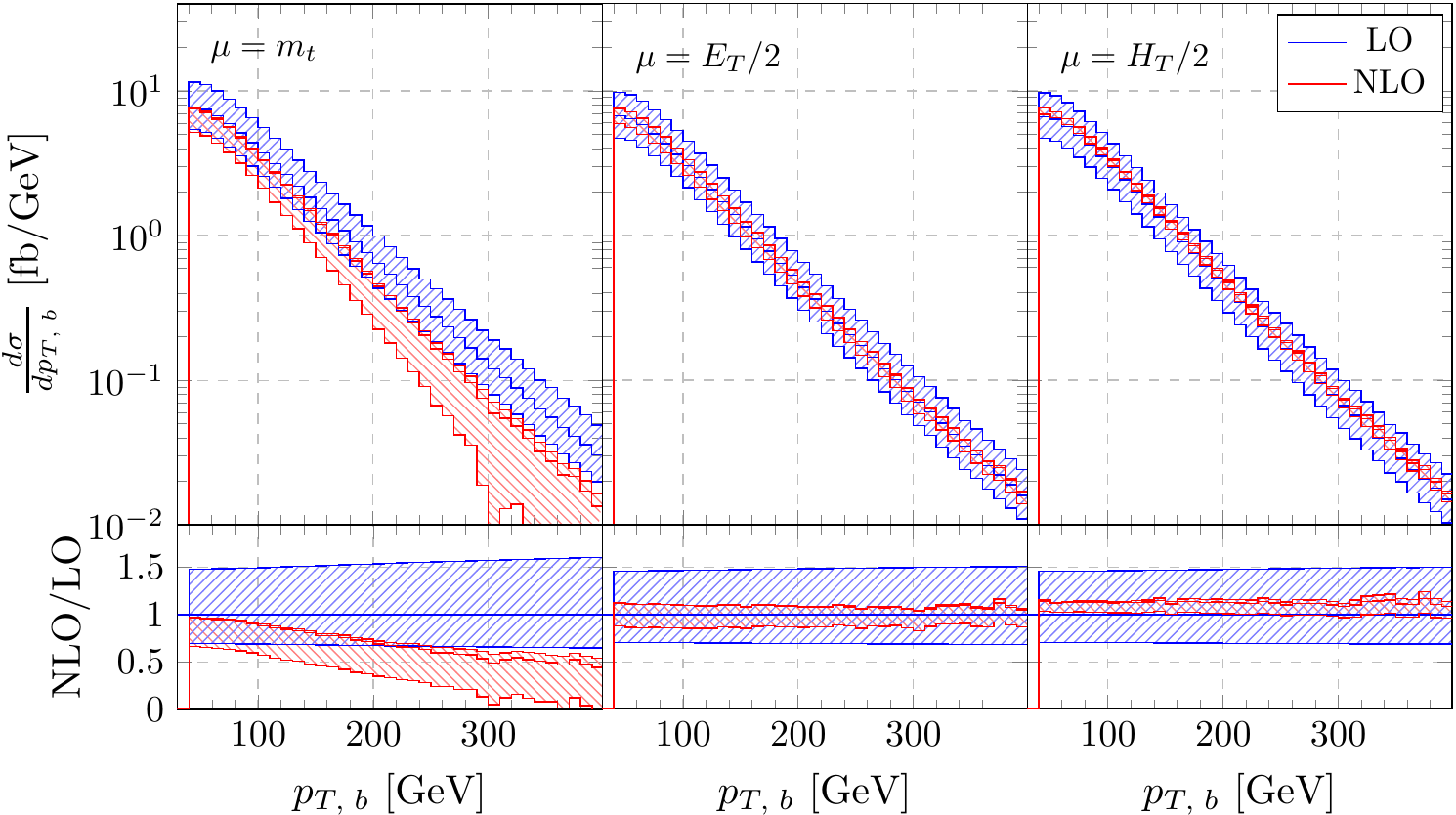} \\
\vspace{0.2cm}
\includegraphics[width=0.95\textwidth]{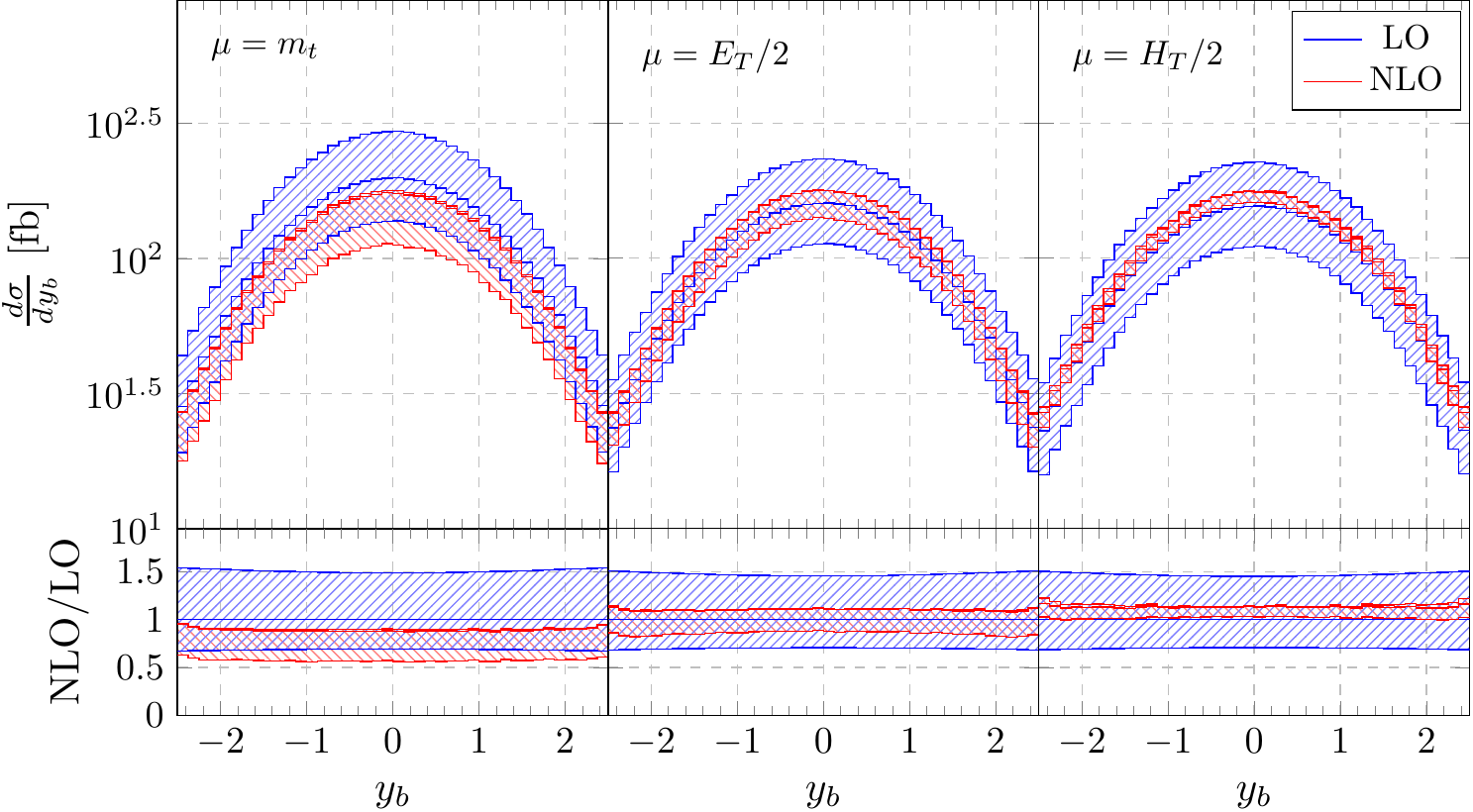} 
\end{center}
\label{fig:dis:3} 
\end{figure}
\begin{figure}[t!]
\caption{\it Averaged differential cross section distribution as a
function of the transverse momentum and rapidity of the charged lepton
for the $pp\to e^+ \nu_e \mu^- \bar{\nu}_\mu b\bar{b}j +X$ process at
the LHC run II with $\sqrt{s} = 13$ TeV.  }
\begin{center}
\includegraphics[width=0.95\textwidth]{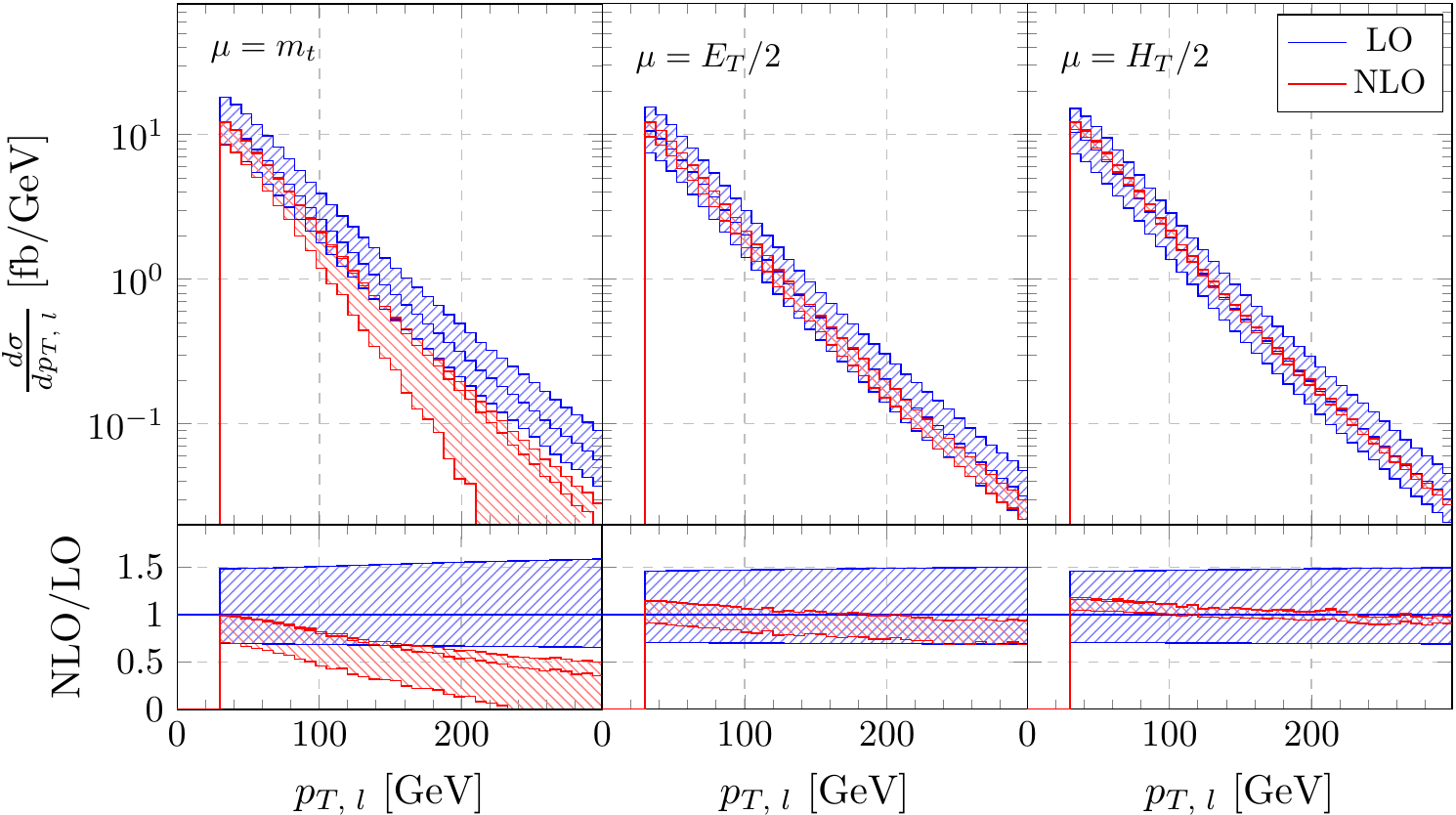} \\
\vspace{0.2cm}
\includegraphics[width=0.95\textwidth]{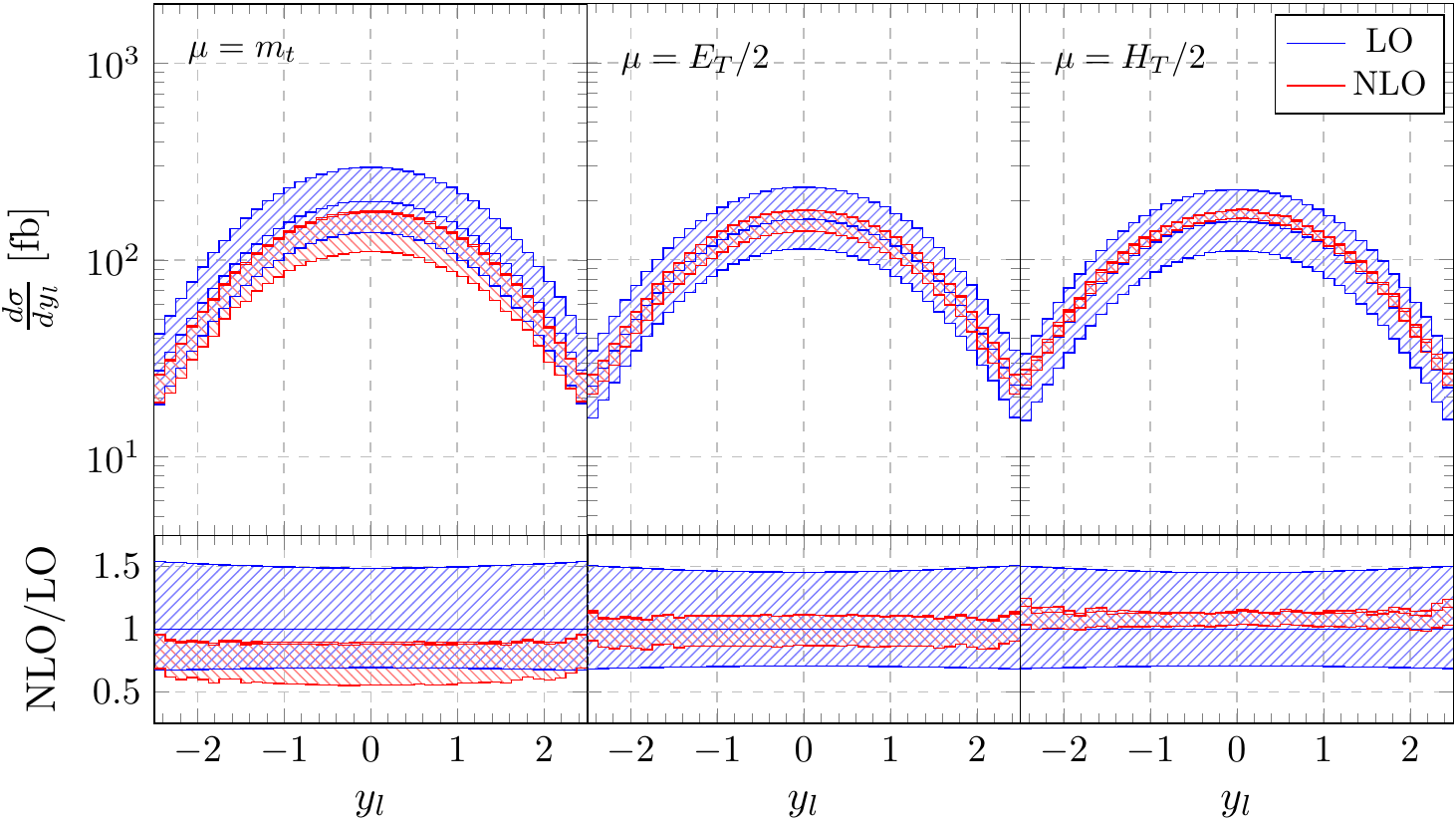} 
\end{center}
\label{fig:dis:4} 
\end{figure}
\begin{figure}[t!]
\caption{\it Averaged differential cross section distributions as a function of
the transverse momentum of the bottom-jet at LO (left panel) and at NLO
(right panel) for the $pp\to e^+ \nu_e \mu^- \bar{\nu}_\mu b\bar{b}j
+X$ process at the LHC run II with $\sqrt{s} = 13$ TeV.}
\begin{center}
\includegraphics[width=0.45\textwidth]{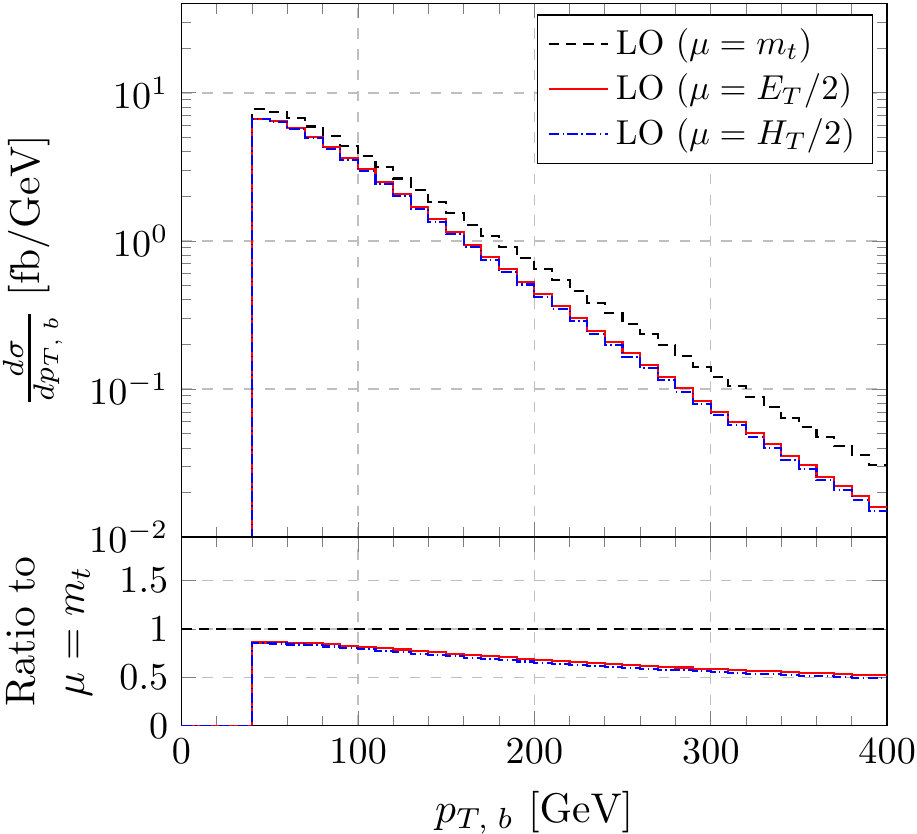}
\includegraphics[width=0.45\textwidth]{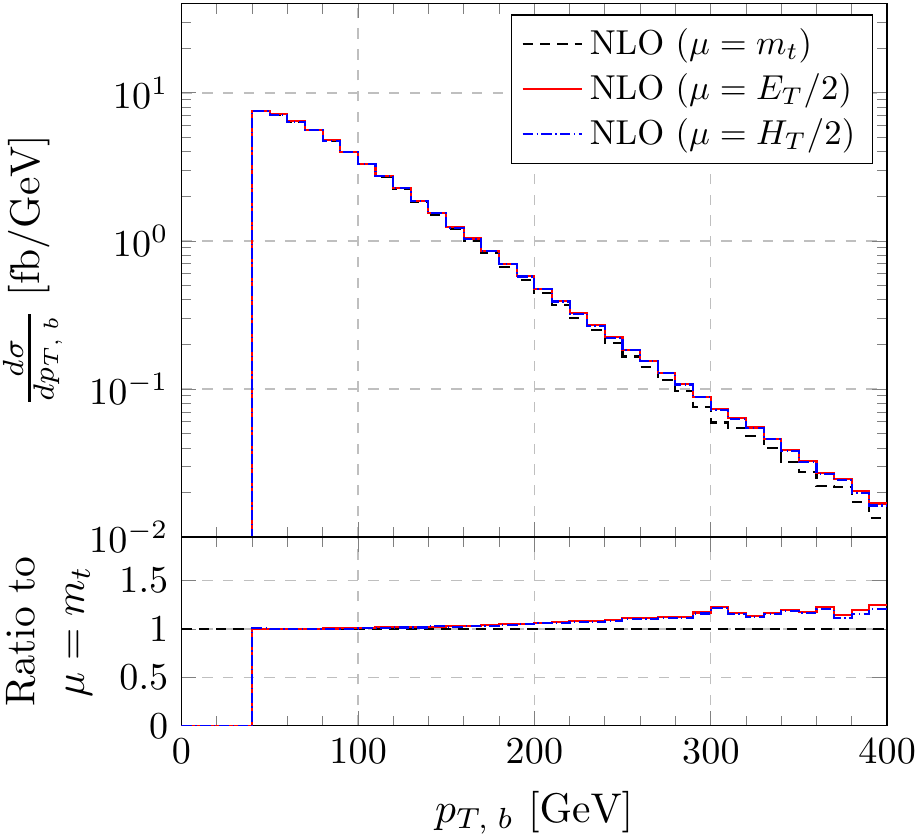}
\end{center}
\label{fig:dis:3a} 
\end{figure}
\begin{figure}[t!]
\caption{\it Averaged differential cross section distributions as a function of
the rapidity of the bottom-jet at LO (left panel) and at NLO
(right panel) for the $pp\to e^+ \nu_e \mu^- \bar{\nu}_\mu b\bar{b}j
+X$ process at the LHC run II with $\sqrt{s} = 13$ TeV.}
\begin{center}
\includegraphics[width=0.45\textwidth]{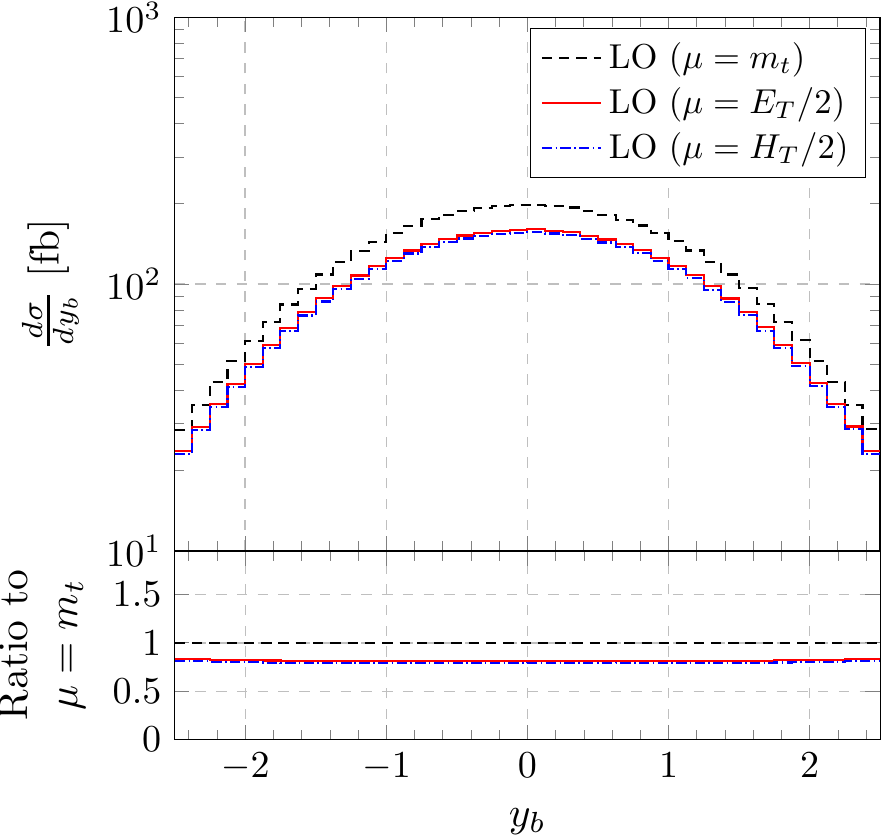}
\includegraphics[width=0.45\textwidth]{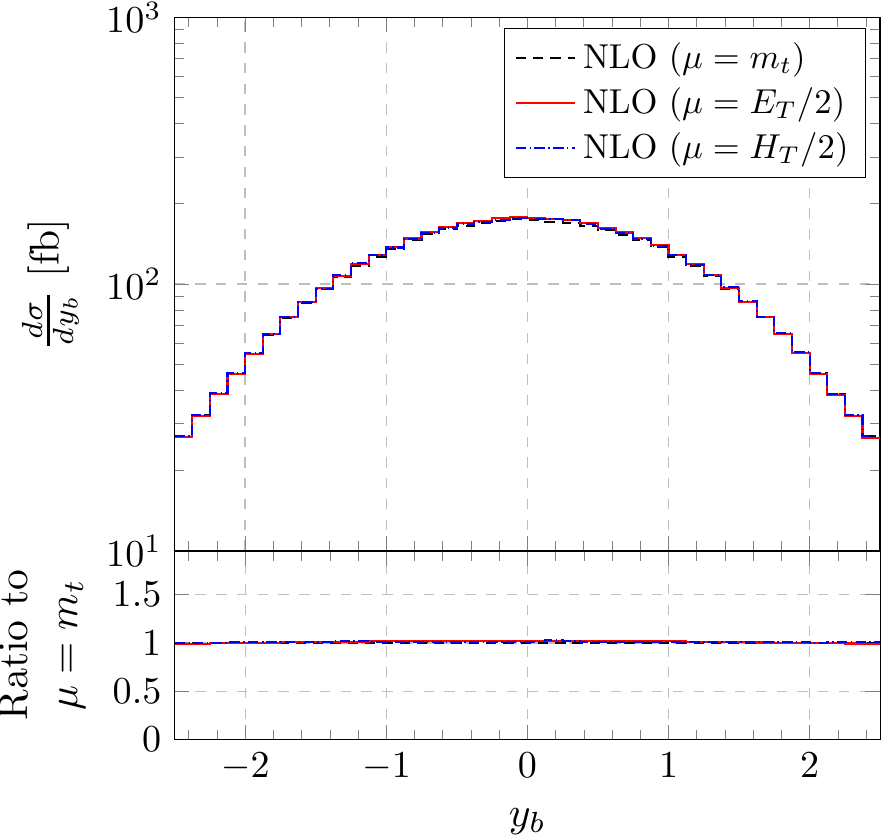}
\end{center}
\label{fig:dis:3b} 
\end{figure}
\begin{figure}[t!]
\caption{\it Averaged differential cross section distributions as a function of
the transverse momentum of  the charged lepton at LO (left panel) and at NLO
(right panel) for the $pp\to e^+ \nu_e \mu^- \bar{\nu}_\mu b\bar{b}j
+X$ process at the LHC run II with $\sqrt{s} = 13$ TeV.}
\begin{center}
\includegraphics[width=0.45\textwidth]{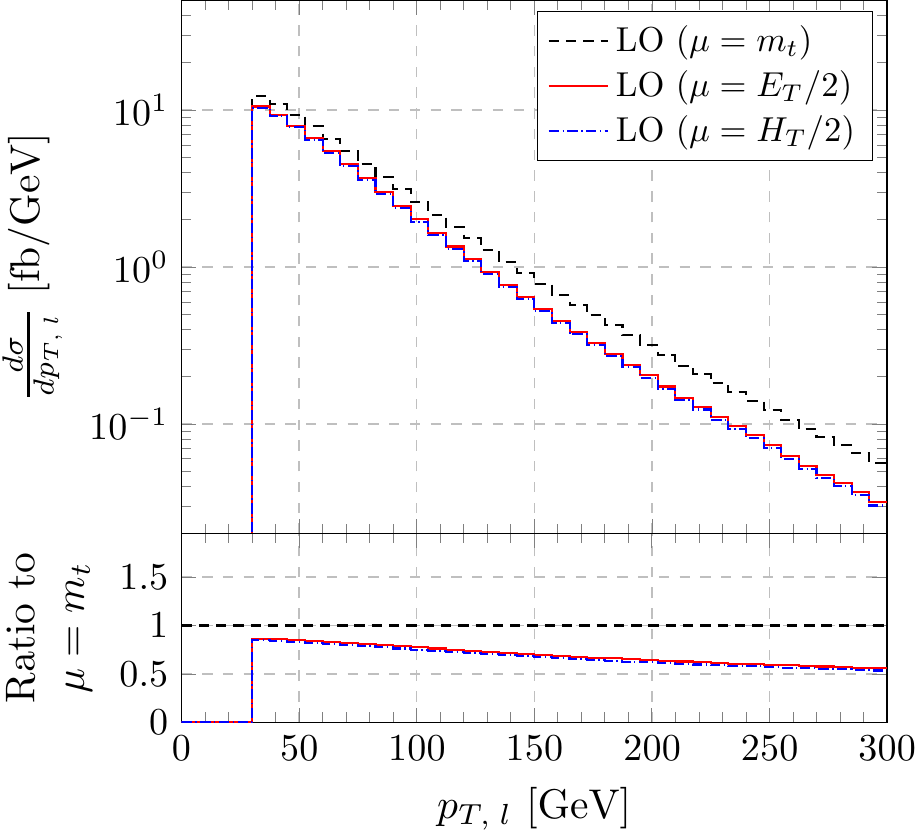}
\includegraphics[width=0.45\textwidth]{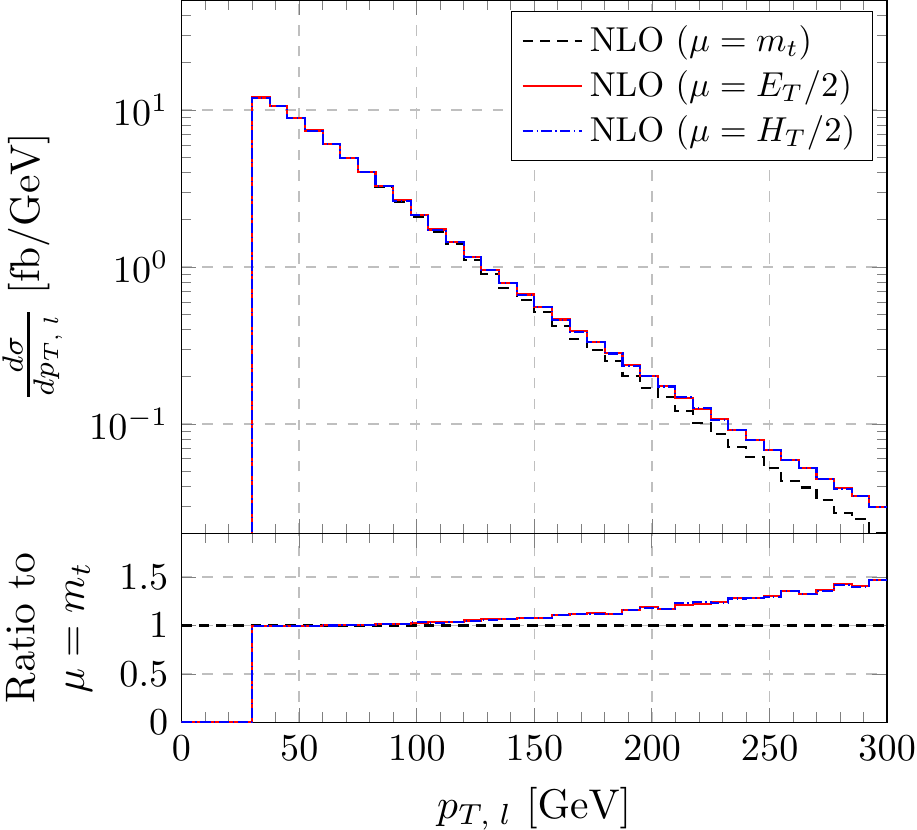}
\end{center}
\label{fig:dis:4a} 
\end{figure}
\begin{figure}[t!]
\caption{\it Averaged differential cross section distributions as a function of
the rapidity of the charged lepton at LO (left panel) and at NLO
(right panel) for the $pp\to e^+ \nu_e \mu^- \bar{\nu}_\mu b\bar{b}j
+X$ process at the LHC run II with $\sqrt{s} = 13$ TeV.}
\begin{center}
\includegraphics[width=0.45\textwidth]{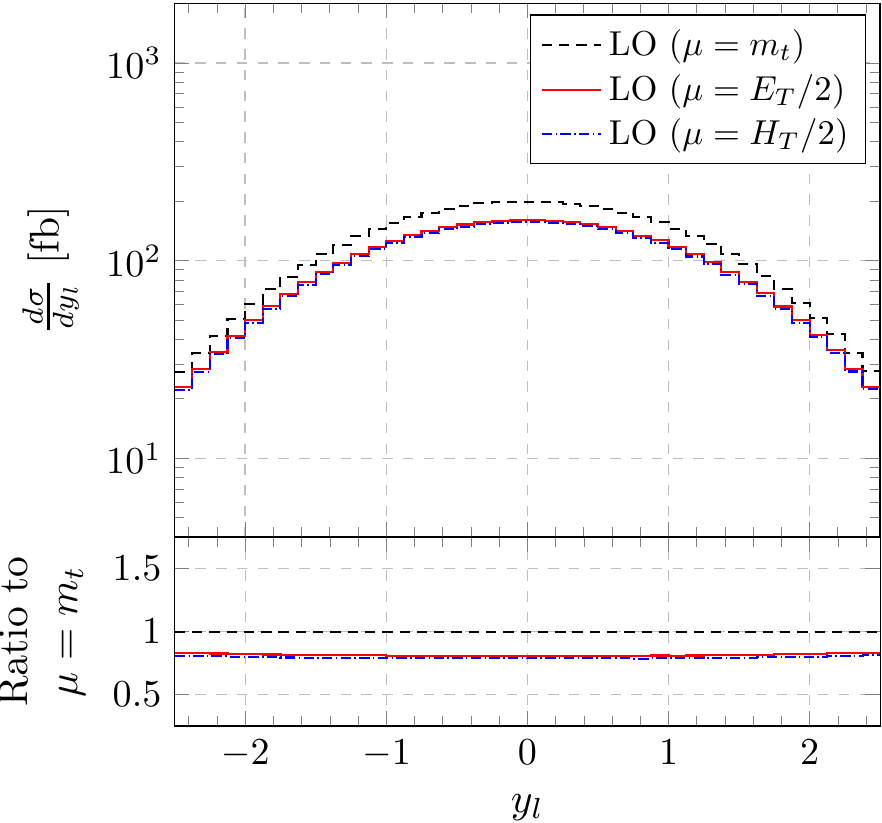}
\includegraphics[width=0.45\textwidth]{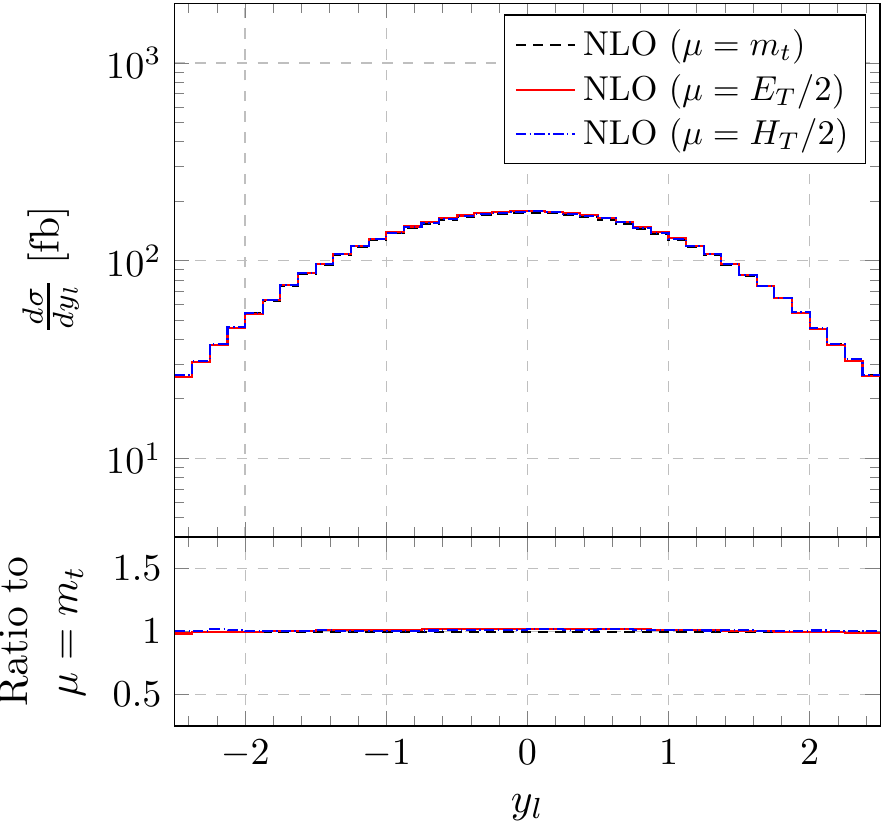}
\end{center}
\label{fig:dis:4b} 
\end{figure}
\begin{figure}[t!]
\caption{\it Differential cross section distribution as a function of
the missing transverse momentum for the $pp\to e^+ \nu_e \mu^-
\bar{\nu}_\mu b\bar{b}j +X$ process at the LHC run II with $\sqrt{s} =
13$ TeV. }
\begin{center}
\includegraphics[width=0.95\textwidth]{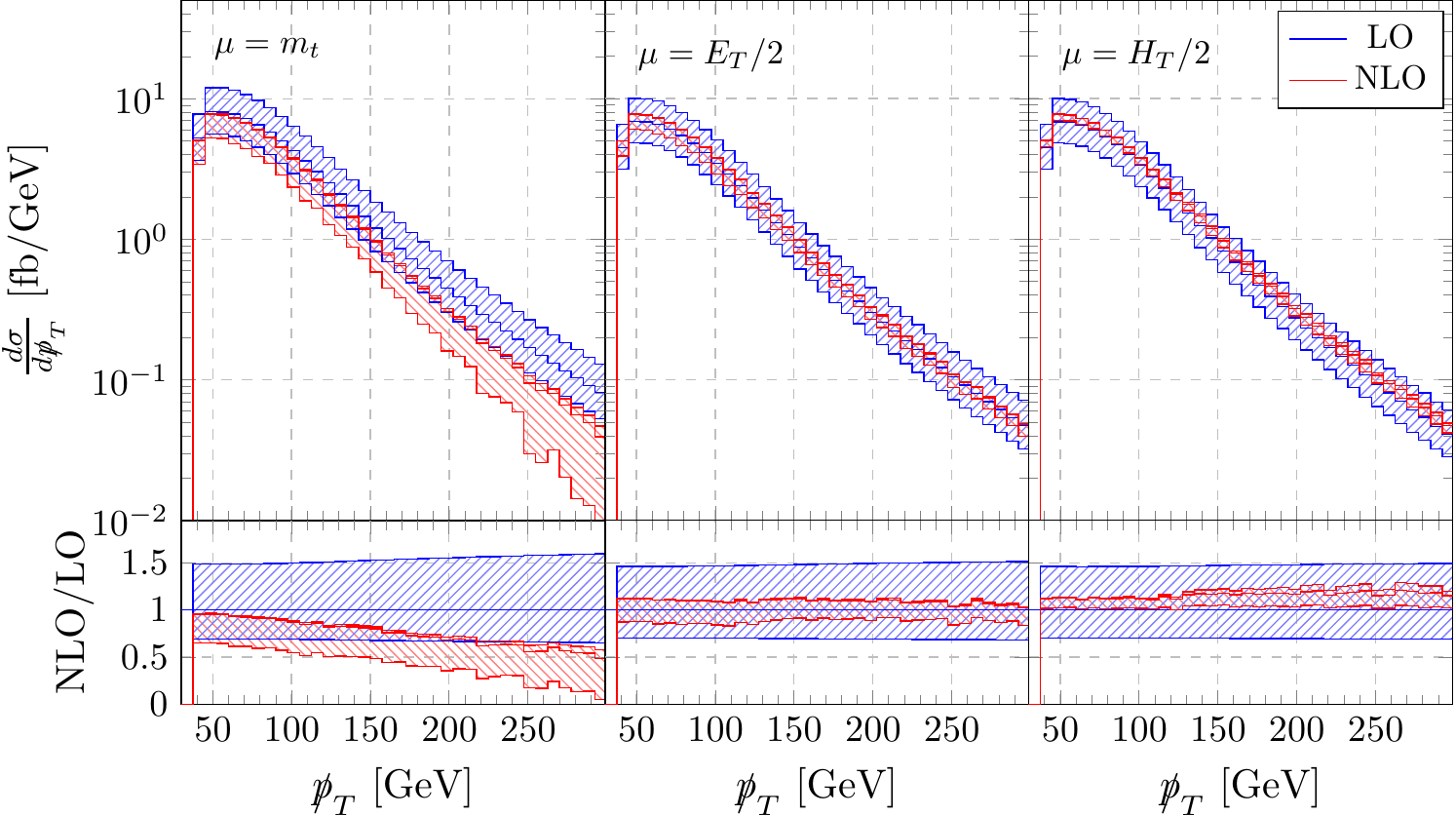} 
\end{center}
\label{fig:dis:5} 
\end{figure}
\begin{figure}[t!]
\caption{\it Differential cross section distribution as a function of
the total transverse momentum of the system, $H_T$ and the invariant
mass of two charged leptons for the $pp\to e^+ \nu_e \mu^-
\bar{\nu}_\mu b\bar{b}j +X$ process at the LHC run II with $\sqrt{s} =
13$ TeV. }
\begin{center}
\includegraphics[width=0.95\textwidth]{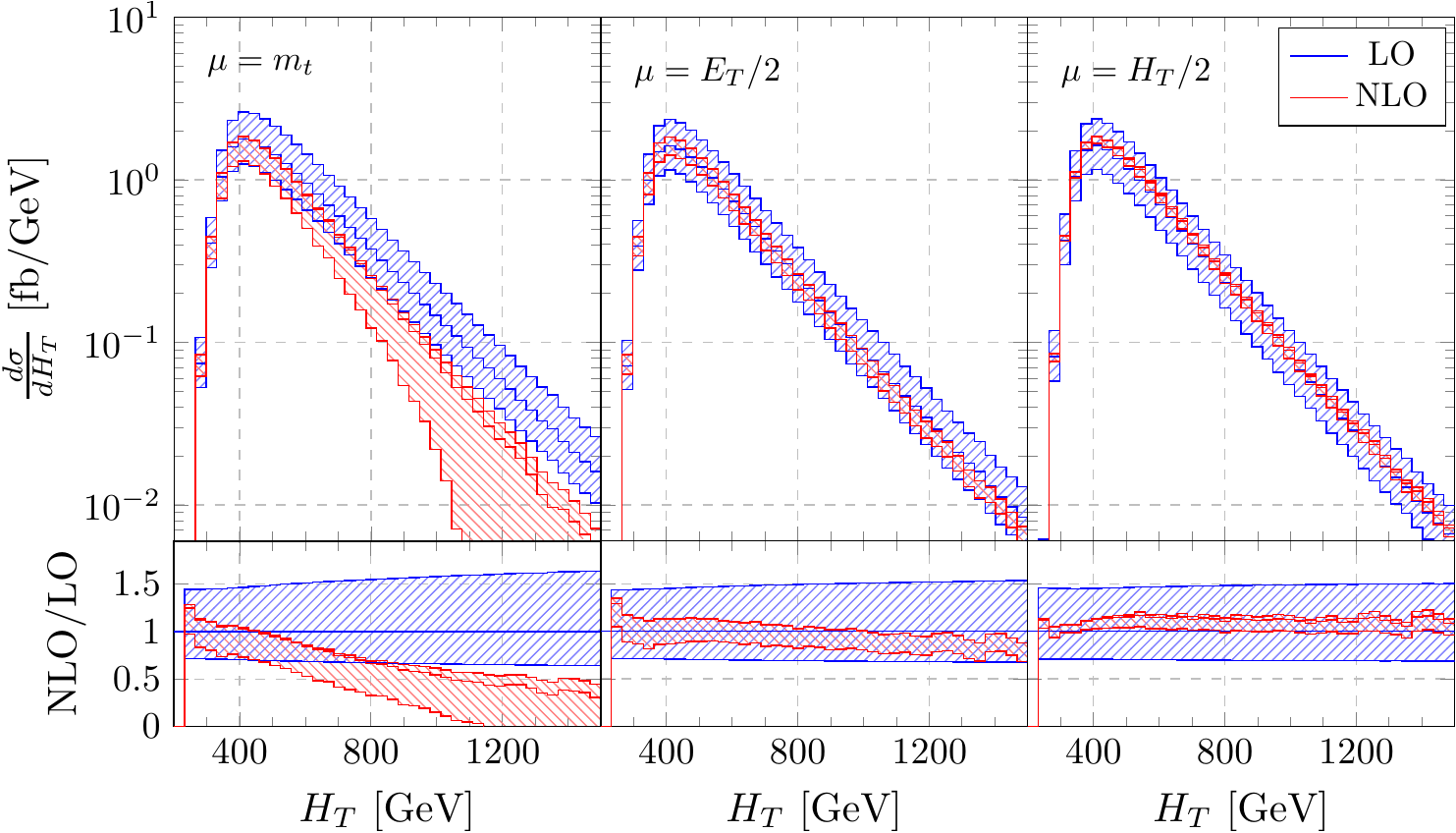} \\
\vspace{0.2cm}
\includegraphics[width=0.95\textwidth]{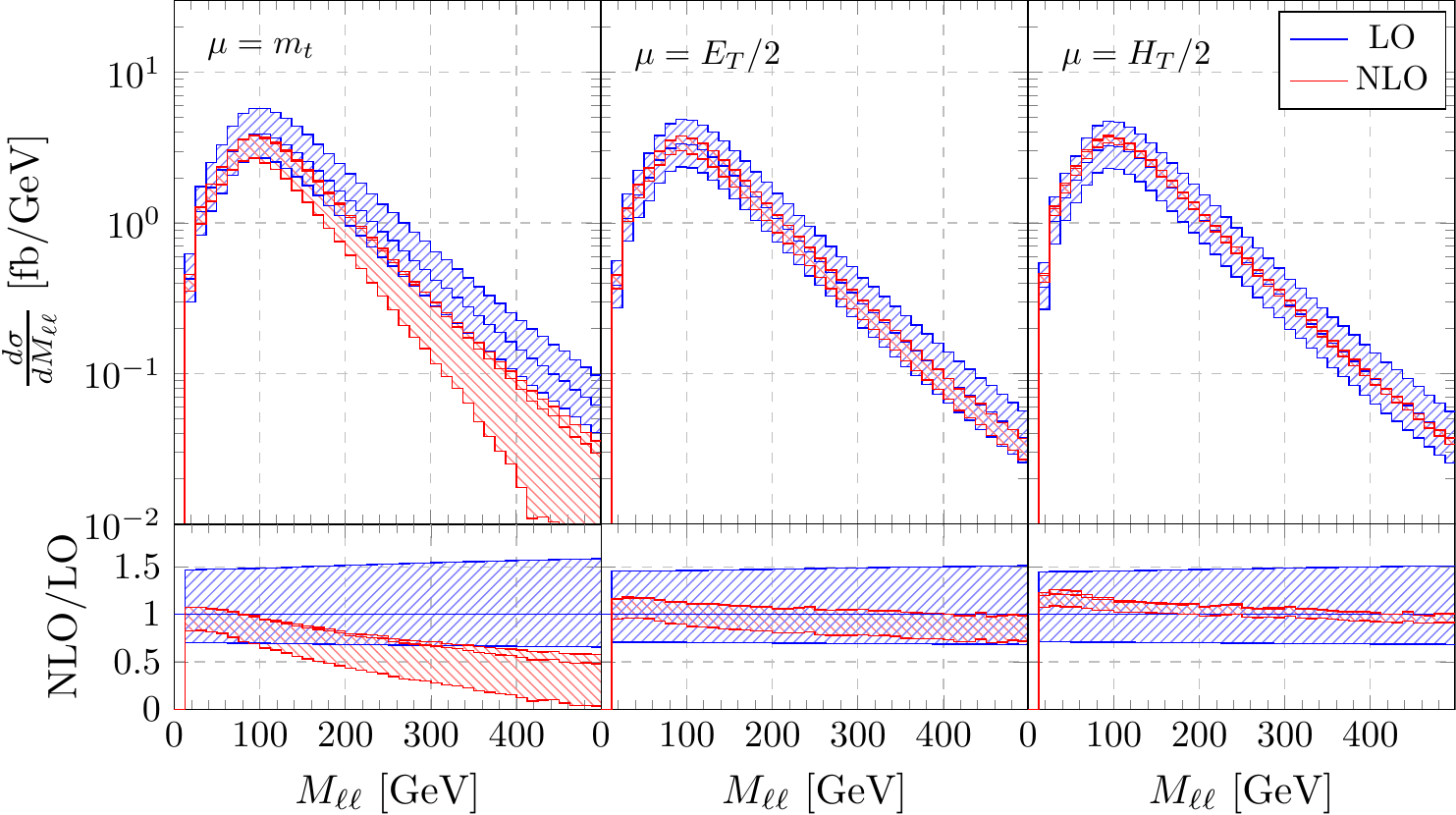} 
\end{center}
\label{fig:dis:6} 
\end{figure}
\begin{figure}[t!]
\caption{\it Differential cross section distributions as a function of
the missing transverse momentum at LO (left panel) and at NLO (right
panel) for the $pp\to e^+ \nu_e \mu^- \bar{\nu}_\mu b\bar{b}j +X$
process at the LHC run II with $\sqrt{s} = 13$ TeV. }
\begin{center}
\includegraphics[width=0.45\textwidth]{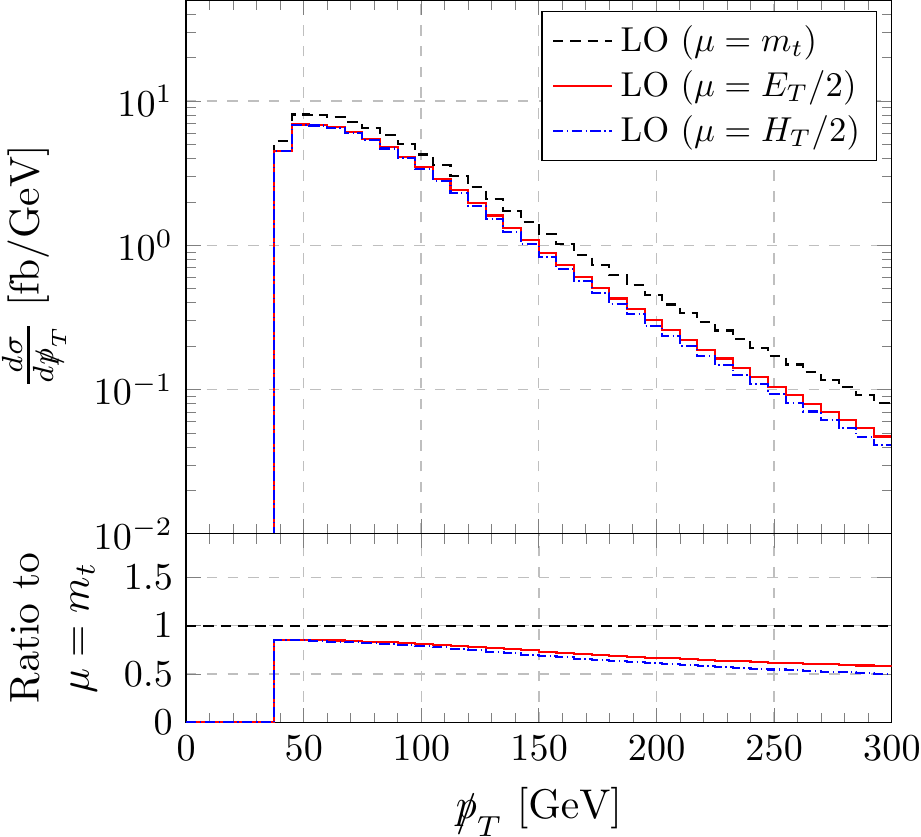}
\includegraphics[width=0.45\textwidth]{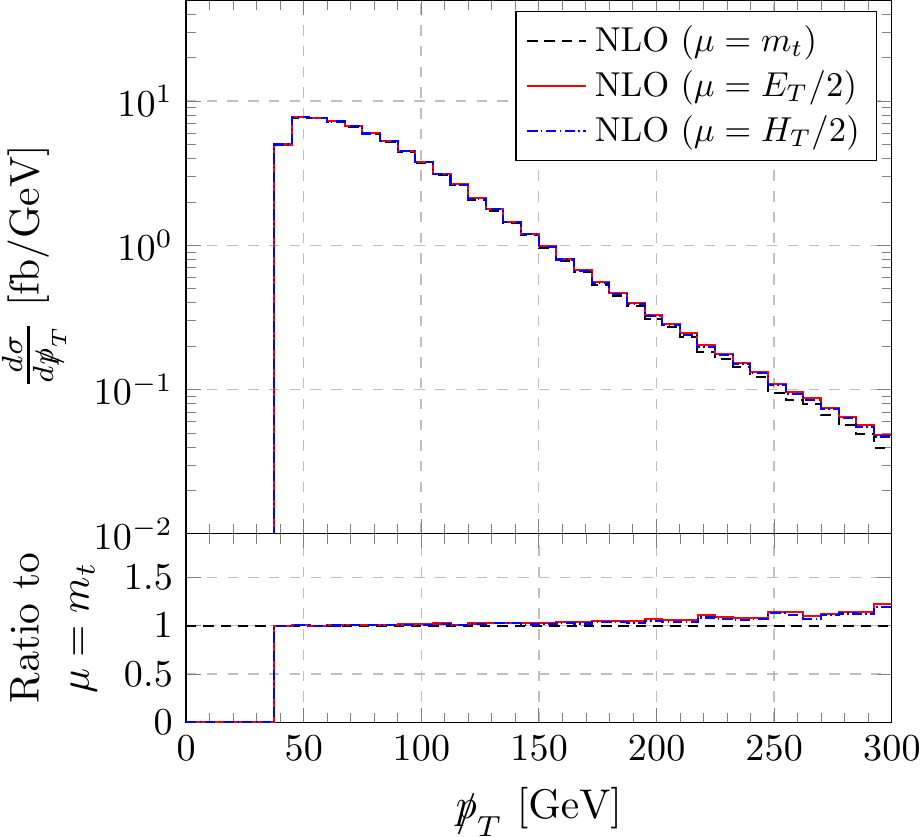}
\end{center}
\label{fig:dis:5a} 
\end{figure}
\begin{figure}[t!]
\caption{\it Differential cross section distributions as a function of
the total transverse momentum of the system, $H_T$, at LO (left panel)
and at NLO (right panel) for the $pp\to e^+ \nu_e \mu^- \bar{\nu}_\mu
b\bar{b}j +X$ process at the LHC run II with $\sqrt{s} = 13$ TeV. }
\begin{center}
\includegraphics[width=0.45\textwidth]{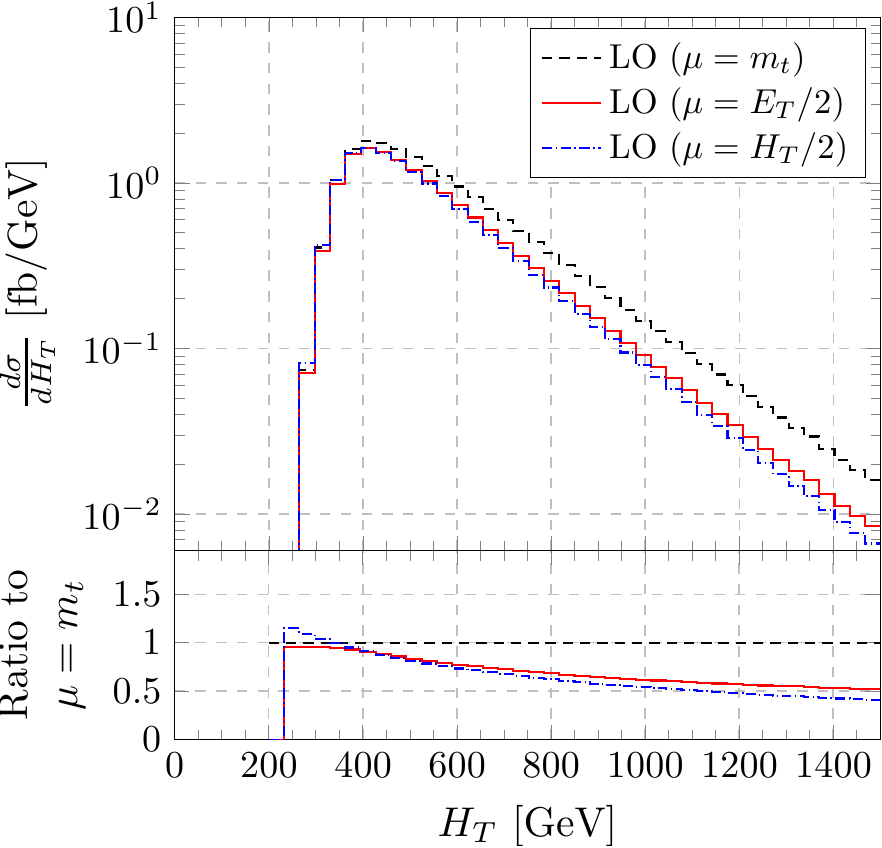}
\includegraphics[width=0.45\textwidth]{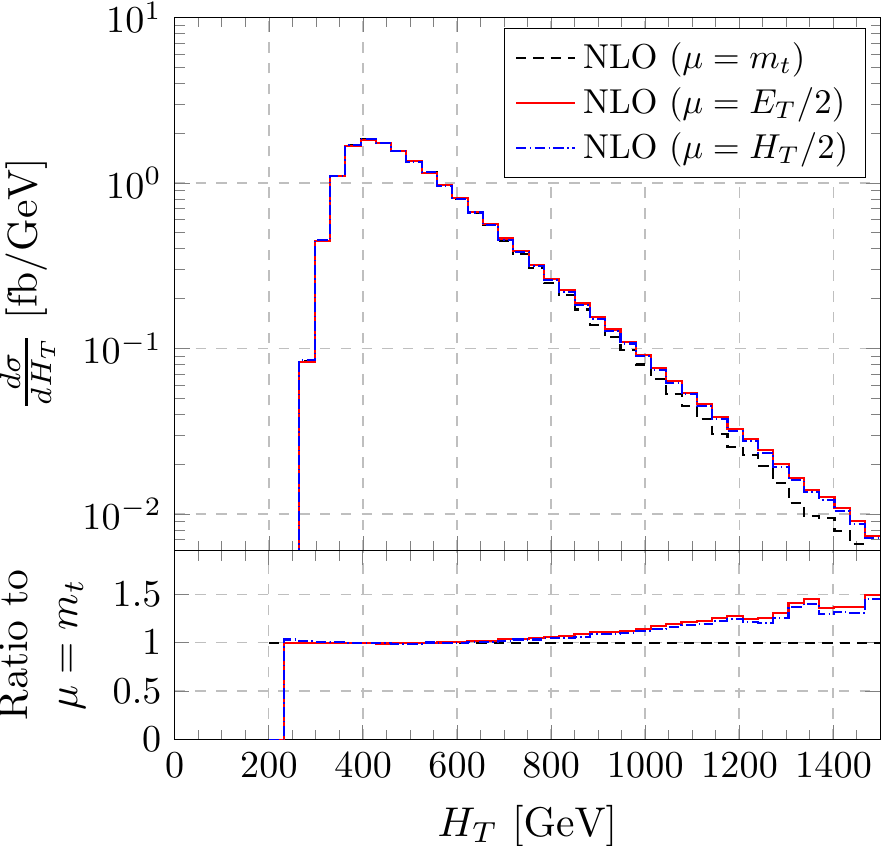}
\end{center}
\label{fig:dis:6a} 
\end{figure}
\begin{figure}[t!]
\caption{\it Differential cross section distributions as a function of
the invariant mass of two charged leptons at LO (left panel) and at
NLO (right panel) for the $pp\to e^+ \nu_e \mu^- \bar{\nu}_\mu
b\bar{b}j +X$ process at the LHC run II with $\sqrt{s} = 13$ TeV. }
\begin{center}
\includegraphics[width=0.45\textwidth]{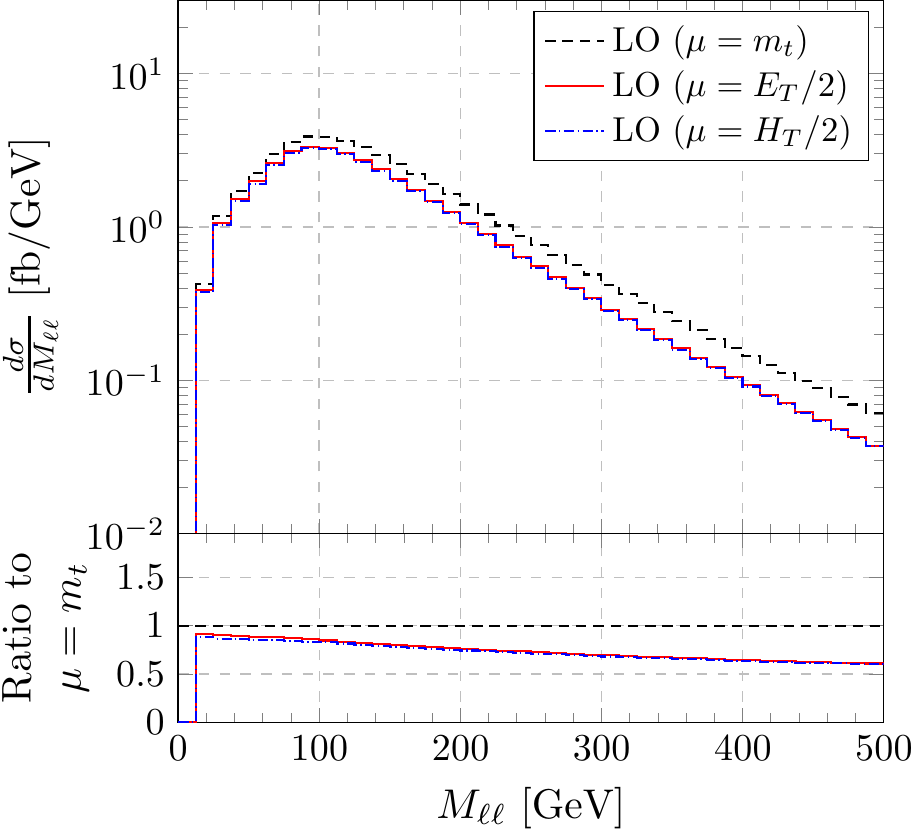}
\includegraphics[width=0.45\textwidth]{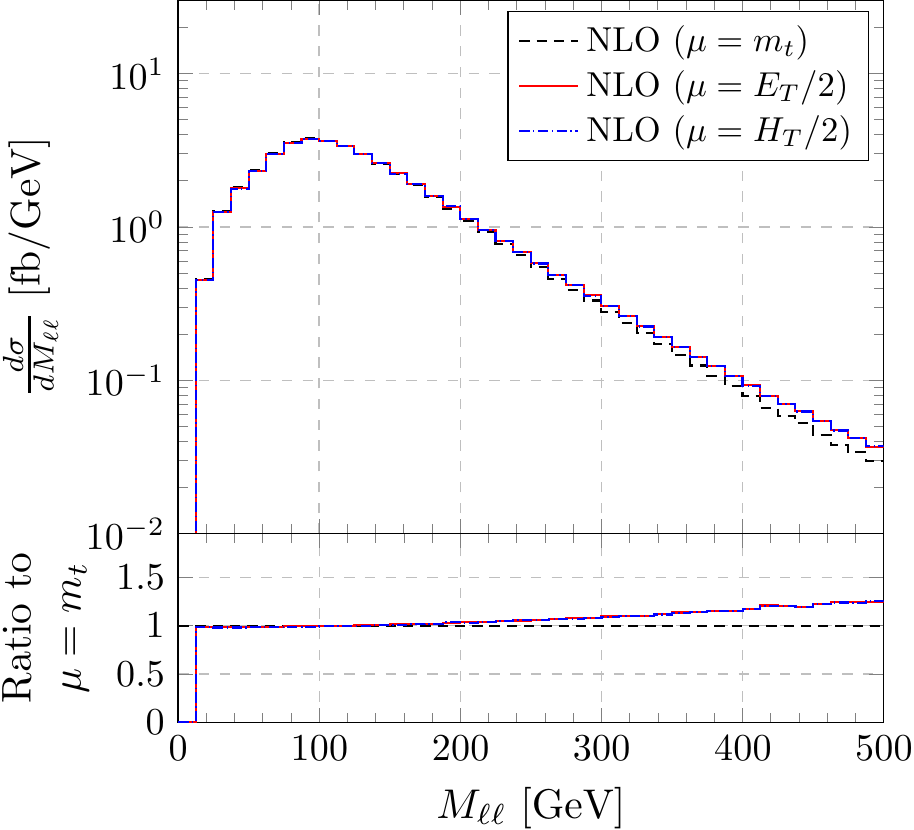}
\end{center}
\label{fig:dis:6b} 
\end{figure}
%

In the next step we shall present observables that are constructed
from visible top-quark decay products, i.e. light- and bottom-jets,
charged leptons and missing transverse momentum.  Therefore, they are
directly accessible without any need for the convoluted
reconstruction. We start with the transverse momentum and rapidity of
the hardest light-jet, depicted in Figures~\ref{fig:dis:2},
\ref{fig:dis:2a} and \ref{fig:dis:2b}.  The kinematics of the hardest
light-jet is particularly important when additional jet activity in 
 $t\bar{t}$ events is studied.  It is greatly sensitive to
higher-order perturbative QCD effects and several theoretical
approaches are available to model it.  The detailed description of
$p_{T,\,j_1}$ and $y_{j_1}$ can be used for example to test various
parton-shower models and different methods for matching fixed-order
QCD calculations with parton shower frameworks. We note that for
$p_{T, \, j_1}$ the fixed scale choice yields negative corrections
within the $4\%-40\%$ range.  Once dynamical scales are employed
positive NLO QCD corrections below $20\%$ are obtained. Moreover, at
NLO up to $400$ GeV all scale choices can be applied to describe the
$p_T$ spectrum of the hardest light-jet. A similar reduction can be
observed for the rapidity distribution where negative corrections of
the order of $10\%$ for $\mu_0=m_t$ are replaced by positive
corrections of the similar size for both $\mu_0=E_T/2$ and
$\mu_0=H_T/2$. Again theoretical uncertainties as obtained with the
$\mu_0=H_T/2$ are the smallest for both observables.  We have drawn
qualitatively similar conclusions for bottom-jet and charged lepton
kinematics that are displayed for completeness in
Figures~\ref{fig:dis:3} and \ref{fig:dis:4} as well as in
Figures~\ref{fig:dis:3a} $-$ \ref{fig:dis:4b}.  For the averaged $p_T$
distributions of the bottom-jet and charged leptons, we observe even
more pronounced NLO corrections, reaching $55\%$ and $65\%$
respectively in the tails. They have been replaced by positive
corrections below $20\%$ when $\mu_0=E_T/2$ or $\mu_0=H_T/2$ has been
used instead. For both rapidity distributions negative corrections of
the order of $10\%$ with $\mu_0=m_t$ have been substituted by positive
ones, which are of the same order for $\mu_0=E_T/2$ and below $20\%$
with the $\mu_0=H_T/2$ scale choice. Additionally, already around
$250$ GeV and $150$ GeV respectively for $p_{T, \, b}$ and $p_{T,\,
\ell}$, NLO distributions are properly described only by the dynamical
scale choice, either $\mu_0=E_T/2$ or $\mu_0=H_T/2$.  Let us also note
here that looking at the rapidity distributions of the light- and the
bottom-jet as well as the charged lepton we can observe a very well
known fact, namely, that bottom-jets and positrons or muons are
distributed centrally in the rapidity, while the light-jet spans a
broader  range.  This information is used for example to 
develop dedicated cuts to reduce top-quark backgrounds for various
signal processes.

Other observables, that are crucial in new physics searches in the
dilepton decay channel of the top quark, are the missing transverse
momentum, denoted here as $\slashed{p}_T$, the total transverse
momentum of the system, $H_T$, and the invariant mass of two charged
leptons, $M_{\ell\ell}$. They are presented in Figures~\ref{fig:dis:5}
and \ref{fig:dis:6}. Various new physics models postulate the
existence of new particles that might decay into a $t\bar{t}$ pair
plus other final states.  The most prominent example is pair
production of top-quark partners decaying to a top-quark pair and a
long-lived neutral particle, which escapes undetected. This weakly
interacting particle would manifest itself as a large missing energy
in the ATLAS and CMS detectors and would lead to the $pp \to
T\overline{T} \to t\bar{t} + \slashed{p}_T$ signature, where $T$
generically denotes the top-quark partner.  The above signature
appears in numerous new physics scenarios, see
e.g. \cite{Plehn:2012pr,Cao:2012rz,Chen:2012uw,
Boughezal:2013pja}. Since these three observables constitute a very
powerful tool in the BSM physics searches we also plot them separately
in Figures~\ref{fig:dis:5a}, \ref{fig:dis:6a} and \ref{fig:dis:6b},
for the central scale only and for three different scale choices.  
From the latter plots we can see that only $\mu_0=E_T/2$ and
$\mu_0=H_T/2$ describe these observables properly in the hight $p_T$
tails and that both dynamical scales give the same prediction in
the whole plotted ranges. However, the former distributions tell us
that $\mu_0=H_T/2$ grants the smallest theoretical uncertainties for
each observable. Overall, for $\slashed{p}_T$ we obtained large and
negative NLO corrections, which reach $50\%$ around $300$ GeV when
$\mu_0=m_t$ is applied. As long as dynamical scales are used instead,
these corrections are replaced by positive and moderate ones, which
are up to  $10\%$ for $\mu_0=E_T/2$ and of the order of
$10\%-20\%$ for $\mu_0=H_T/2$. For the total transverse momentum of the
$t\bar{t}j$ system we have noticed a comparable
performance. Specifically, around $1500$ GeV $-70\%$ corrections at
$\mu_0=m_t$ have been downsized to about $-10\%$ and $+10\%$ for
$\mu_0=E_T/2$ and $\mu_0=H_T/2$ respectively. Lastly, for the
invariant mass of two charged leptons $-50\%$ NLO corrections around
$500$ GeV for $\mu_0=m_t$ have been converted to $-1.5\%$ for
$\mu_0=E_T/2$ and almost to zero corrections  for $\mu_0=H_T/2$.

\begin{figure}[t!]
\caption{\it Differential cross section distribution as a
function of  $\Delta R_{\ell\ell}$ and $\Delta \phi_{\ell\ell}$
for the $pp\to e^+ \nu_e \mu^- \bar{\nu}_\mu b\bar{b}j +X$ process at
the LHC run II with $\sqrt{s} = 13$ TeV.  }
\begin{center}
\includegraphics[width=0.95\textwidth]{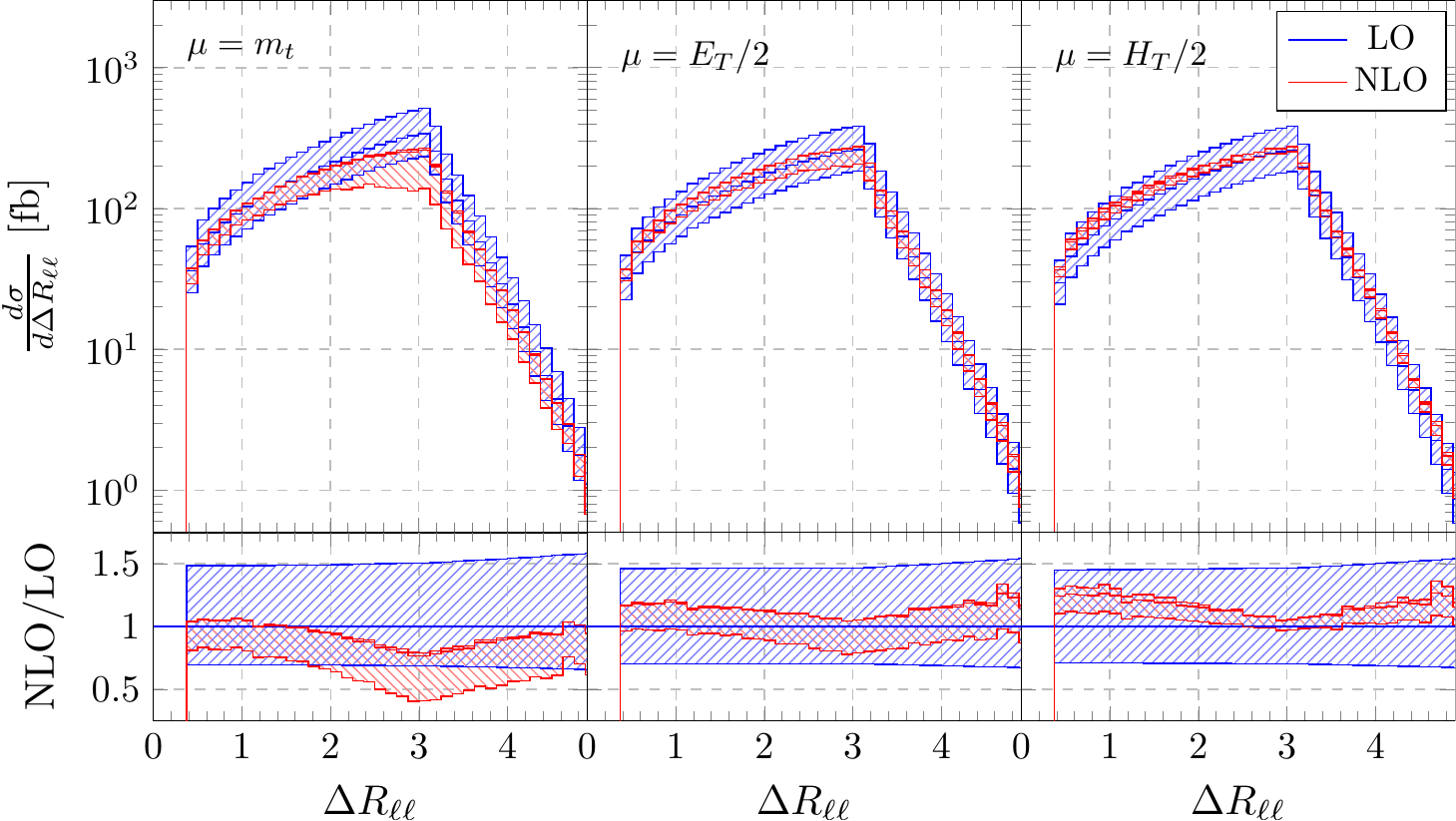} \\
\vspace{0.2cm}
\includegraphics[width=0.95\textwidth]{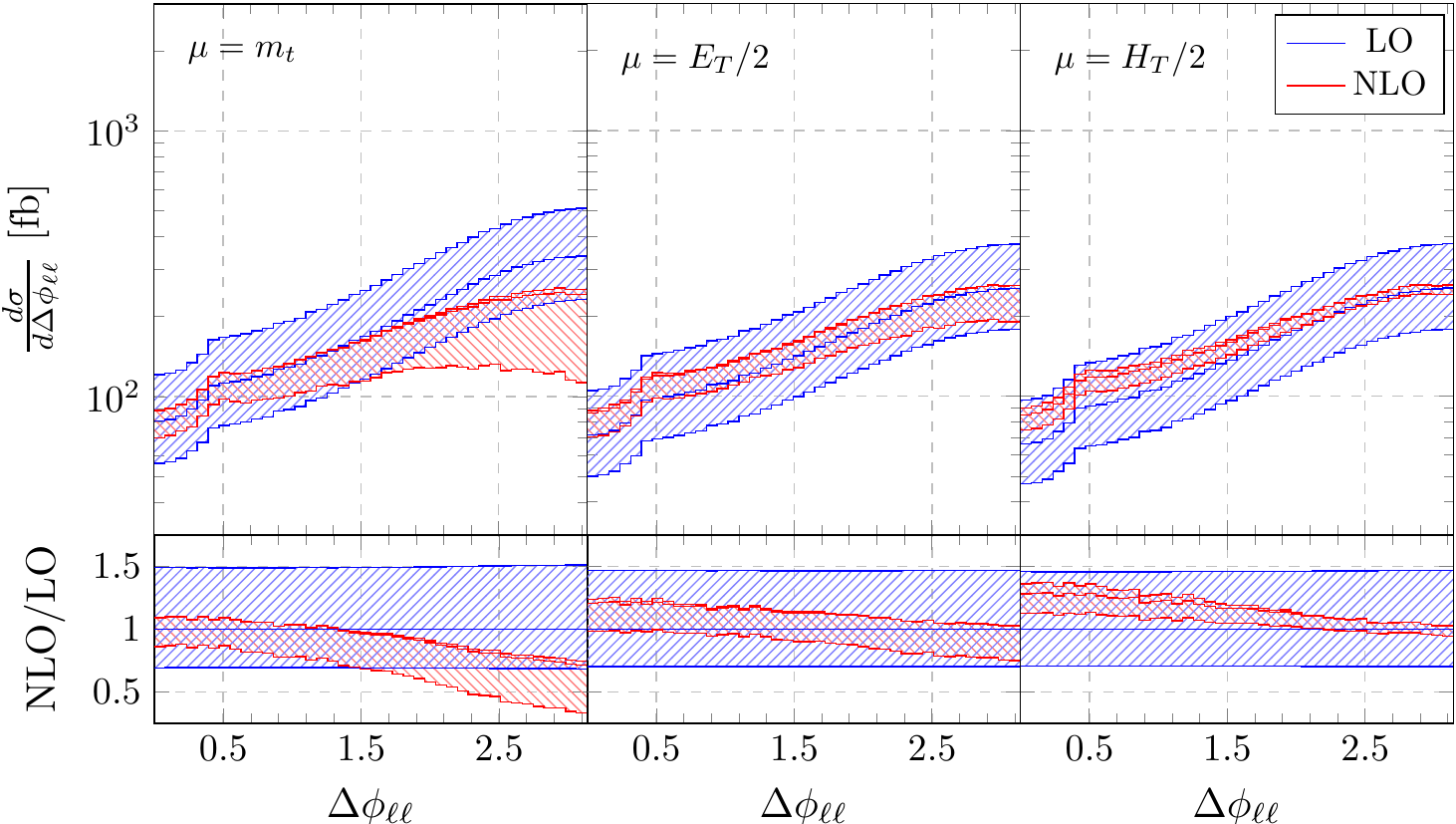} 
\end{center}
\label{fig:dis:7} 
\end{figure}

In Figure~\ref{fig:dis:7} we present the differential cross section as
a function of the separation of charged leptons in the
rapidity-azimuthal angle plane, $\Delta R_{\ell \ell}=\sqrt{\Delta
\phi_{\ell \ell}^2+\Delta y_{\ell \ell}^2}$, and the azimuthal angle
between the charged leptons
$\Delta\phi_{\ell\ell}=|\phi_{\ell_1}-\phi_{\ell_2}|$.  They are
measured very precisely at the LHC by both ATLAS and CMS
collaborations and do not require the reconstruction of the top
quarks. In general, angular distributions of charged leptons are of
huge importance since they reflect spin correlations of the top-quark
pair. Because of its large mass, the top quark is extremely
short-lived. As a result, top quarks do not have time to form hadrons
before they decay. Thus, the spin of the top-quark pair at production
is transferred to the decay products and can be measured directly via
their angular distributions \cite{Bernreuther:2010ny}. Many models of
new physics predict vastly different spin correlations while keeping
similar production cross sections, an example being the production of
heavy spin-zero states with undefined CP parity and mass below $400$
GeV that are resonantly produced in the $t\bar{t}$ channel
\cite{Bernreuther:2015fts}.  Therefore, in practice top-quark pair
spin correlations can be used by experimental collaborations at the
LHC to provide a handle on the determination of the nature of the new
particle that decays as $pp\to h_{\rm new} \to t\bar{t} +X \to
W^+W^-b\bar{b} +X \to \ell^+ \ell^- \nu_\ell \bar{\nu}_\ell b\bar{b}
+X$, where $h_{\rm new}$ is the heavy spin-zero state.
%
\begin{figure}[t!]
\caption{\it Differential cross section distributions as a function of
$\Delta R_{\ell \ell}$ at LO (left panel) and at NLO (right panel) for
the $pp\to e^+ \nu_e \mu^- \bar{\nu}_\mu b\bar{b}j +X$ process at the
LHC run II with $\sqrt{s} = 13$ TeV.  }
\begin{center}
\includegraphics[width=0.45\textwidth]{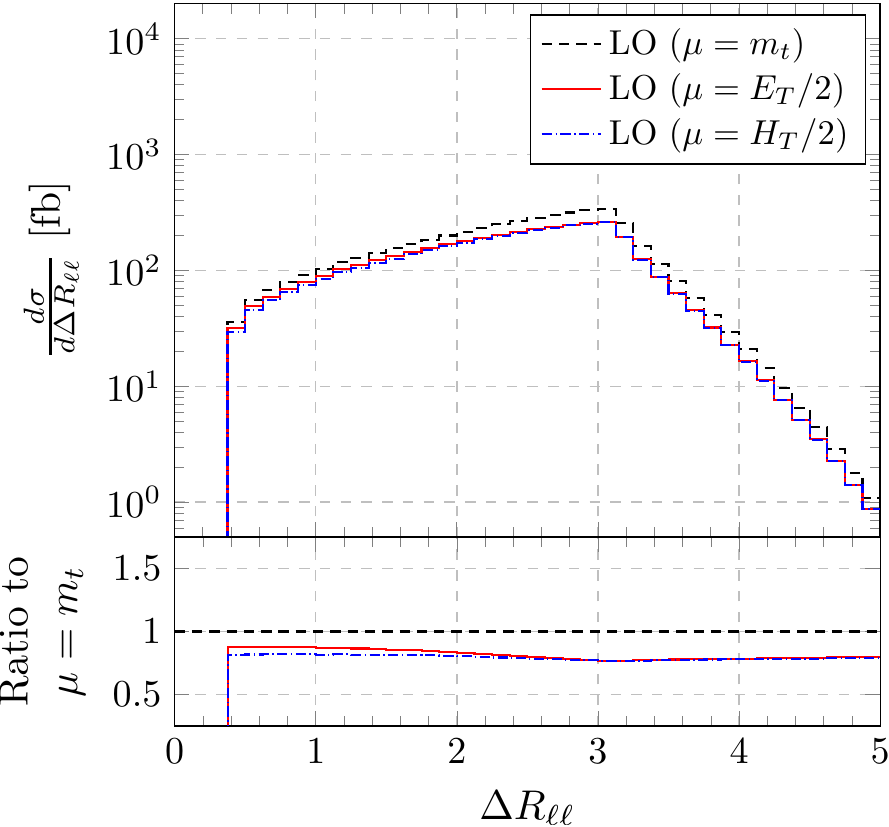}
\includegraphics[width=0.45\textwidth]{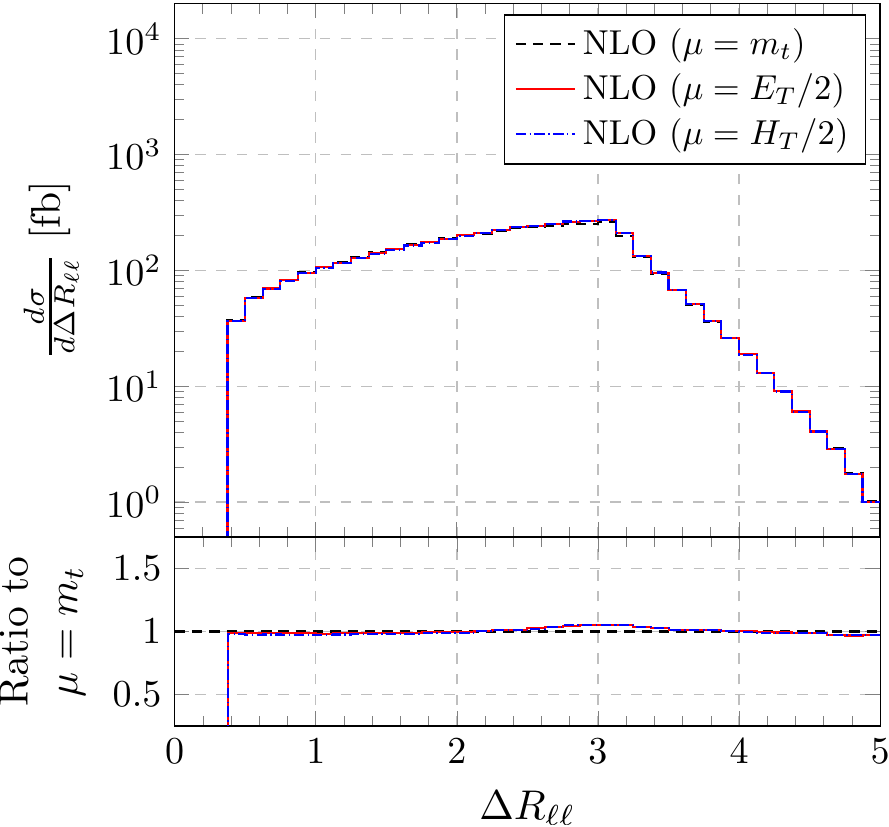}
\end{center}
\label{fig:dis:7a} 
\end{figure}
\begin{figure}[t!]
\caption{\it Differential cross section distributions as a function of
$\Delta \phi_{\ell \ell}$ at LO (left panel) and at NLO (right panel)
for the $pp\to e^+ \nu_e \mu^- \bar{\nu}_\mu b\bar{b}j +X$ process at
the LHC run II with $\sqrt{s} = 13$ TeV.  }
\begin{center}
\includegraphics[width=0.45\textwidth]{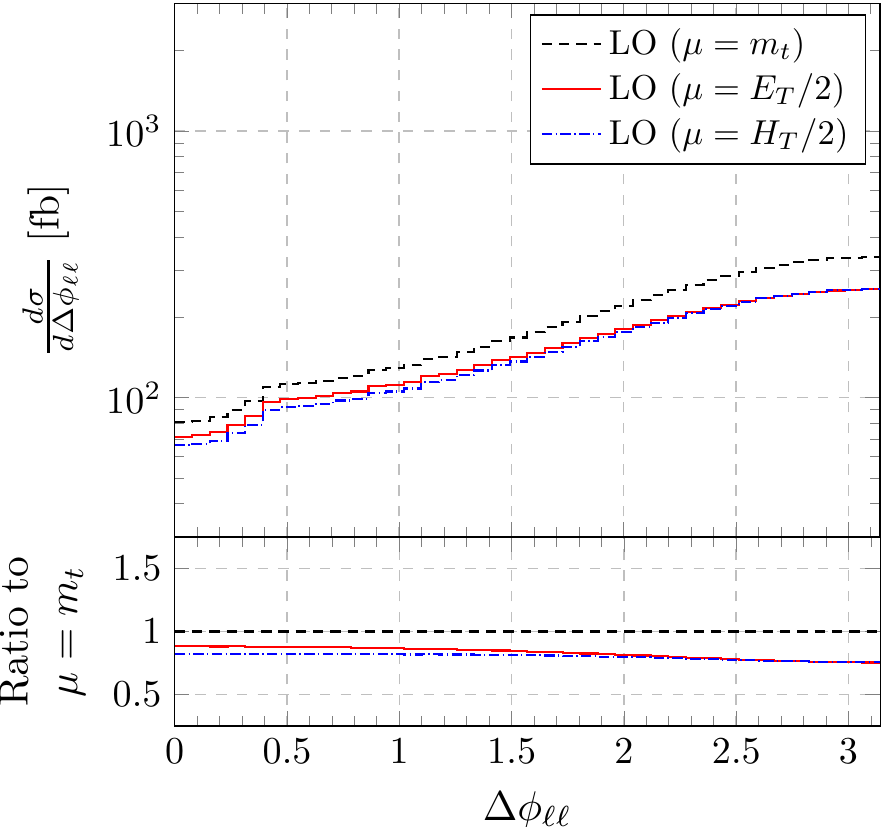}
\includegraphics[width=0.45\textwidth]{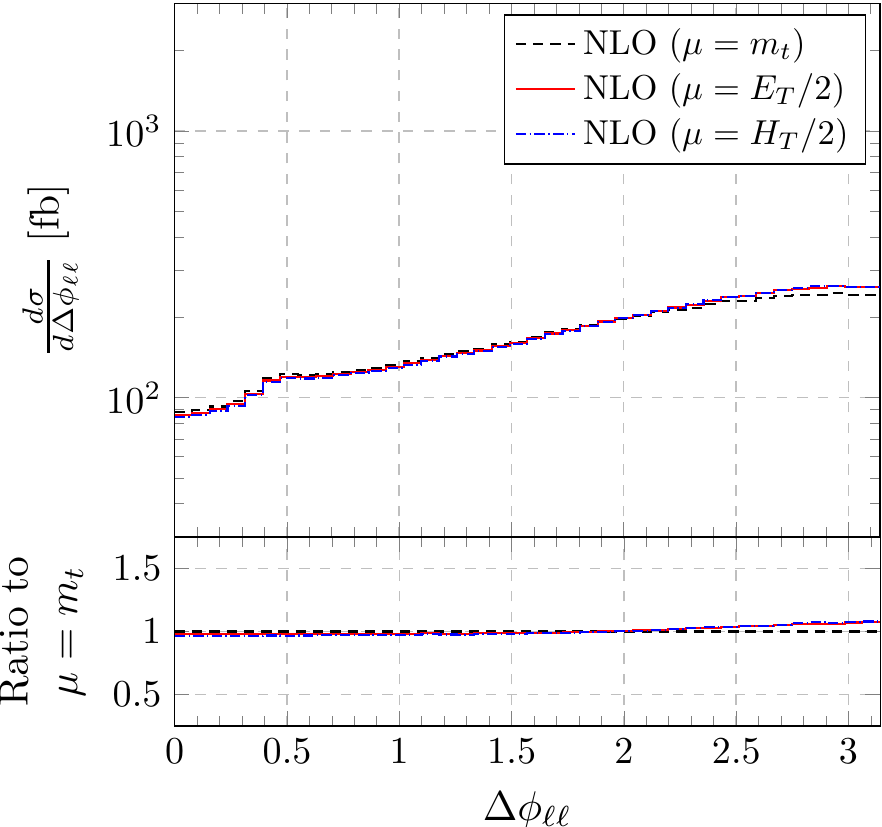}
\end{center}
\label{fig:dis:7b} 
\end{figure}
\begin{figure}[t!]
\caption{\it Differential cross section distribution as a function of
the invariant mass of the positron and bottom-jet and the averaged 
differential cross section distribution as a function of invariant
mass of the top quark for the $pp\to e^+ \nu_e \mu^- \bar{\nu}_\mu
b\bar{b}j +X$ process at the LHC run II with $\sqrt{s} = 13$ TeV.  }
\begin{center}
\includegraphics[width=0.95\textwidth]{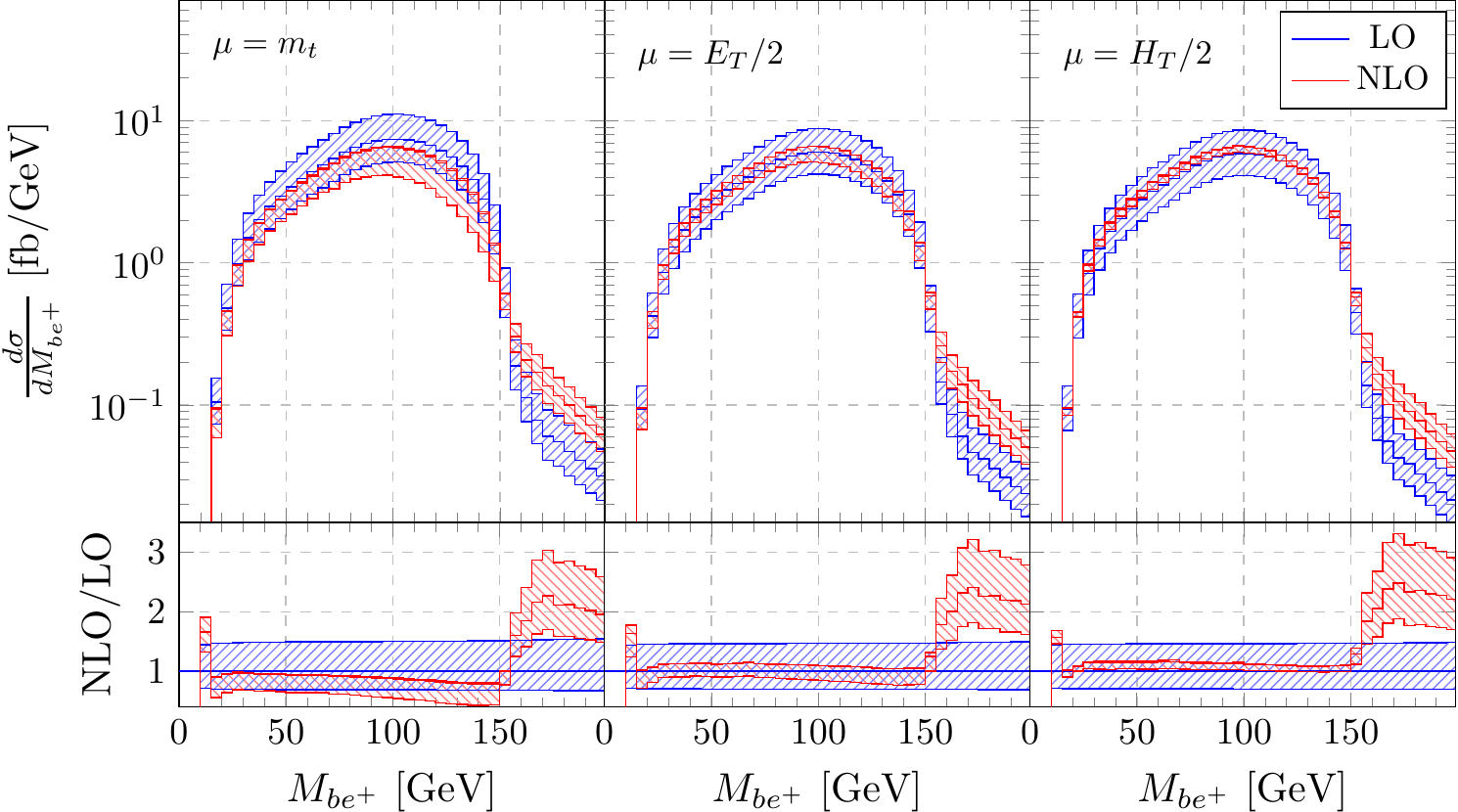} \\
\vspace{0.2cm}
\includegraphics[width=0.95\textwidth]{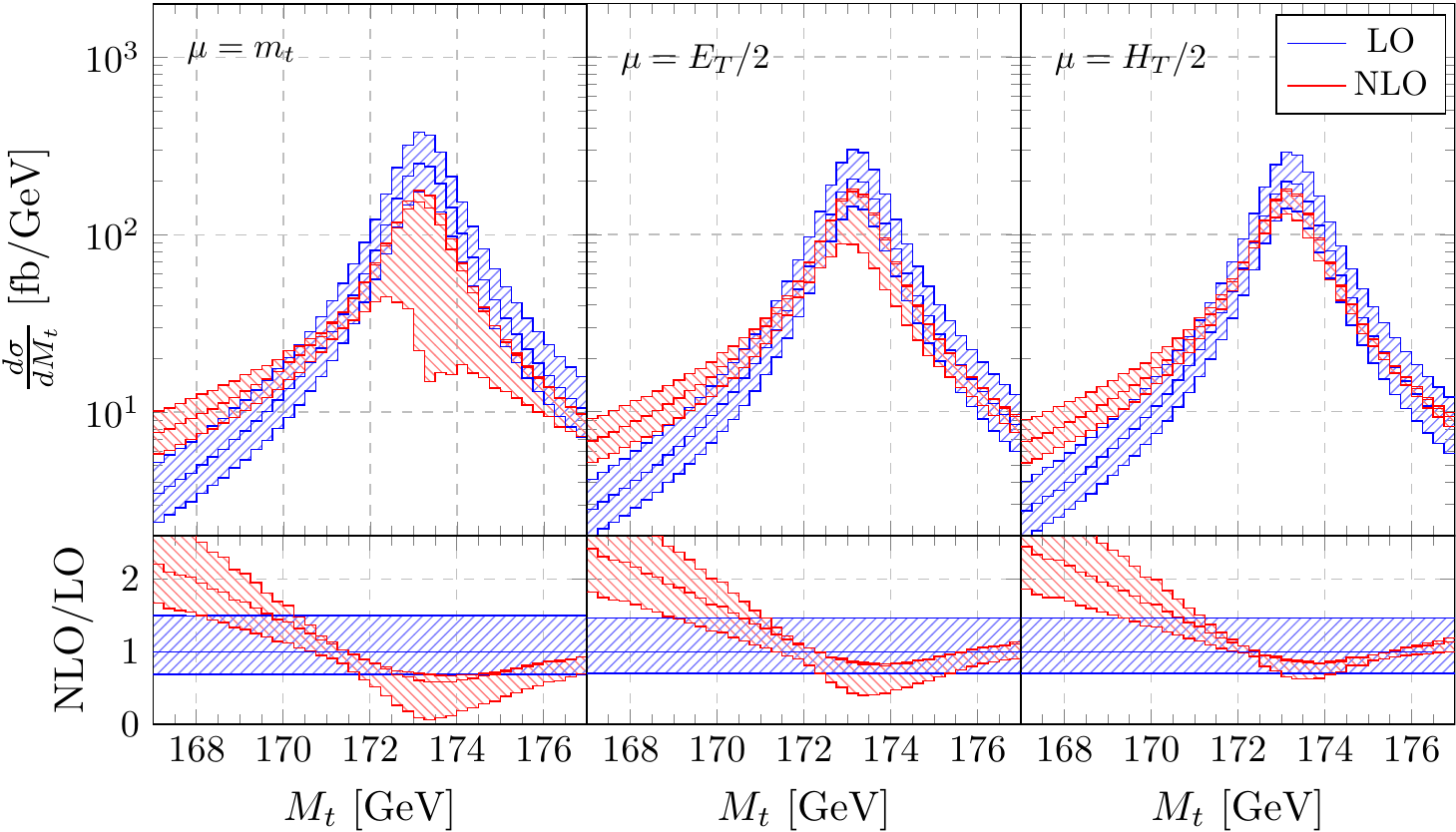} 
\end{center}
\label{fig:dis:8} 
\end{figure}
\begin{figure}[t!]
\caption{\it Differential cross section distributions as a function of
the invariant mass of the bottom-jet and positron at LO (left panel)
and at NLO (right panel) for the $pp\to e^+ \nu_e \mu^- \bar{\nu}_\mu
b\bar{b}j +X$ process at the LHC run II with $\sqrt{s} = 13$ TeV.  }
\begin{center}
\includegraphics[width=0.45\textwidth]{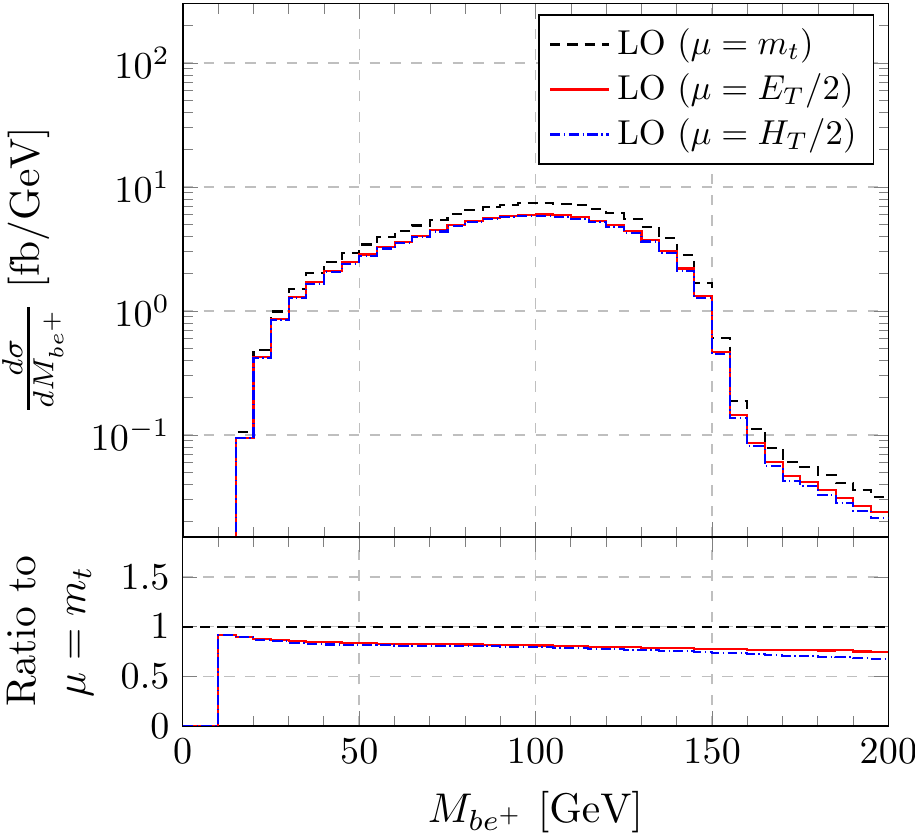}
\includegraphics[width=0.45\textwidth]{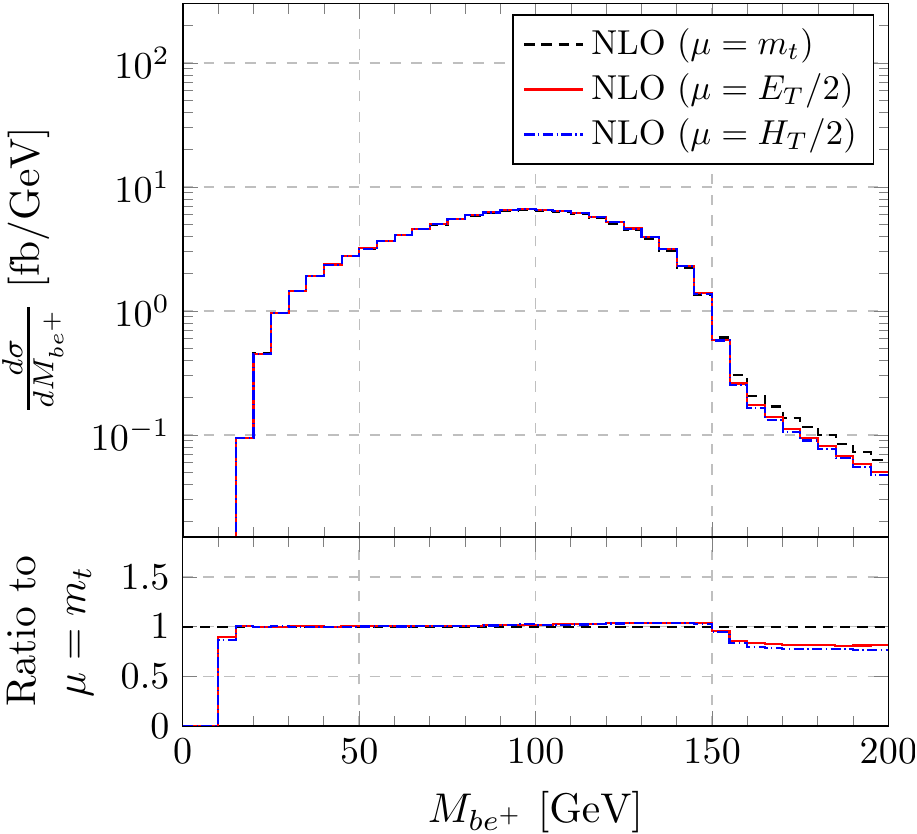}
\end{center}
\label{fig:dis:8a} 
\end{figure}
\begin{figure}[t!]
\caption{\it Averaged differential cross section distributions as a
function of the invariant mass of the top quark at LO (left panel) and
at NLO (right panel) for the $pp\to e^+ \nu_e \mu^- \bar{\nu}_\mu
b\bar{b}j +X$ process at the LHC run II with $\sqrt{s} = 13$ TeV.  }
\begin{center}
\includegraphics[width=0.45\textwidth]{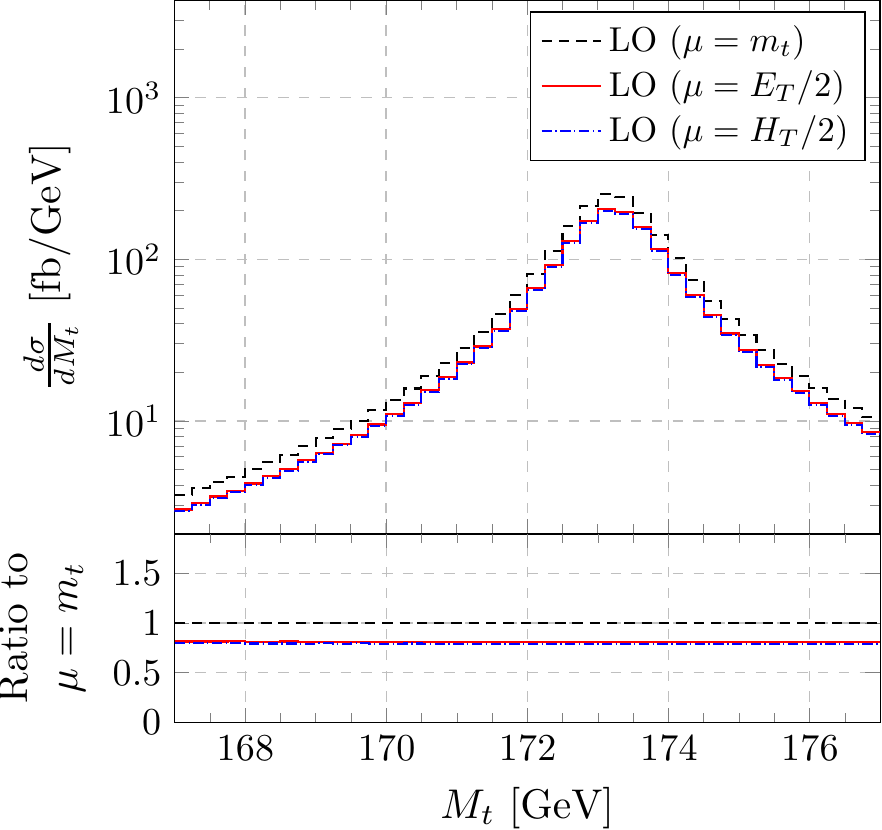}
\includegraphics[width=0.45\textwidth]{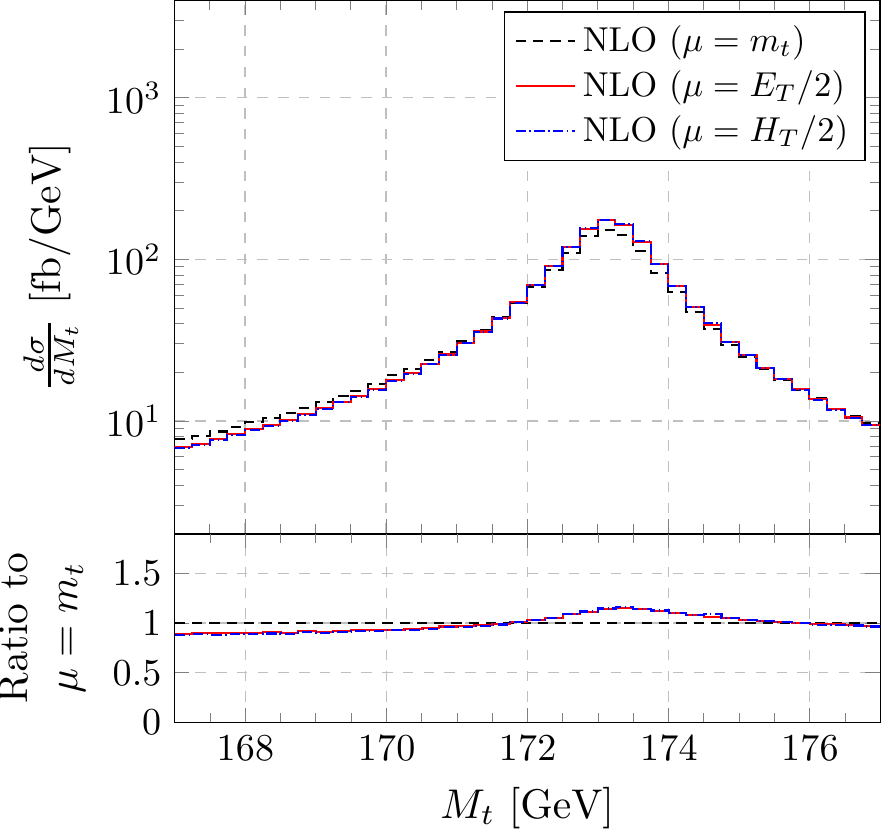}
\end{center}
\label{fig:dis:8b} 
\end{figure}
%
Even though $\Delta R_{\ell\ell}$ and $\Delta \phi_{\ell \ell}$ are
dimensionless observables, we observe in Figure~\ref{fig:dis:7} that
they receive quite large NLO corrections with $\mu_0=m_t$, which
vary  within the plotted ranges. To be more specific for $\Delta
R_{\ell\ell}$ we obtained a variation between $+5\%$ and $-25\%$, for
$\Delta \phi_{\ell \ell}$ we reached a change between $+10\%$ and
$-30\%$. On the other hand, for our best scale choice, $\mu_0=H_T/2$,
positive corrections in the whole shown range are realised for both
observables. Specifically, we have noticed $5\%-25\%$ corrections for
$\Delta R_{\ell\ell}$ and $2\%-30\%$ for $\Delta \phi_{\ell
\ell}$. Similar results have been reached with $\mu_0=E_T/2$. From
Figures~\ref{fig:dis:7a} and \ref{fig:dis:7b} we can further see
that at NLO dependence on the scale choice is practically non existing
in both cases, unlike at LO where the $\mu_0=m_t$ choice always
predicts higher spectra.

The last observables that we present are the invariant mass of the positron
and bottom-jet, $M_{be^+}$ and the  mass of the reconstructed  top quark,
$M_t$.  The latter is given only in the vicinity of the resonance.
They are both plotted in Figures~\ref{fig:dis:8}, 
\ref{fig:dis:8a} and \ref{fig:dis:8b}. These observables
are crucial for the top-quark mass extraction. In the case of
$M_{be^+}$ one cannot determine, which b-jet should be paired with the
positron. To increase the probability that both final states come from
the decay cascade initiated by the same top quark we select the $be^+$
pair, that returns the smallest invariant mass
\cite{Beneke:2000hk}. Thus, $M_{be^+}$ is defined as $M_{be^+}=\min
\left\{ \sqrt{(p_{b_1} +p_{e^+})^2} \, ,\, \sqrt{(p_{b_2} +p_{e^+})^2}
\right\}$ and contains a kinematic endpoint that can be derived from
the on-shell top-quark decay into $t \to W^+ b \to e^+ \nu_e b$.
Neglecting the masses of all decay products we can write
\begin{equation} 
m_t^2=p_t^2= m^2_W+2p_bp_{e^+}
+2p_bp_{\nu_e} \,.
\end{equation} 
As a result $M^2_{be^+}= 2p_bp_{e^+}\le m_t^2-m_W^2$. At
lowest order when both top quarks and $W$ gauge bosons are treated as
on-shell particles there is a strict kinematic limit for the invariant
mass of the bottom quark and the positron given by
\begin{equation}
M_{be^+}^{\rm max} = \sqrt{m^2_t -
m^2_W }\approx 153 ~{\rm GeV}\,. 
\end{equation}
For off-shell top quarks this kinematic limit is smeared, also
additional NLO radiation affects this region, nevertheless there is a
sharp fall of the cross section in the fixed order prediction. The two
bottom-jets stemming from each $t\bar{t}$ decay give rise to a
matching ambiguity. Pairings in which the bottom-jet and positron
emerge from different top quarks do not necessarily obey the upper
bound $M_{be^+}^{\rm max}$ and, thus, do not have a clean kinematic
endpoint. Although a priori it is impossible to distinguish between
correct and incorrect pairing, the easiest solution is to select the
smallest $M_{be^+}$ value in each event as we have done. In this
fashion the kinematic endpoint of the distribution is always preserved
simply because $M_{be^+} \le M^{\rm correct}_{be^+} $.  Strong
sensitivity of the kinematic endpoint to $m_t$ causes this
distribution to be extremely useful for the top-quark mass
extraction. In the same manner the mass of the reconstructed top quark,
defined as $M_t=M_{be^+\nu_e}=\sqrt{p_t^2}$ is susceptible to the
modelling of the top-quark decays. Off-shell effects and additional
gluon radiation further smear the peak resulting from the NWA. NLO QCD
corrections affect both distributions greatly. For $M_{be^+}$ above
$150$ GeV corrections above $100\%$ have been obtained. In more details we
have attained NLO QCD corrections of the order of $125\%$, $140\%$ and
$150\%$ correspondingly for $\mu_0=m_t$, $\mu_0=E_T/2$ and
$\mu_0=H_T/2$. In this region, the theoretical uncertainties are also 
immense independently of the scale used in the calculation. On the
other hand, below the kinematical endpoint, moderate negative
(positive) corrections in the range $5\%-20\%$ ($5\%-15\%$ and
$10\%-15\%$) are visible for $\mu_0=m_t$ ($\mu_0=E_T/2$ and
$\mu_0=H_T/2$ respectively). The remarkably different behaviour
between the two regions can be understood if one considers that the
phase space above $M^{\rm max}_{be^+}$ is populated at LO by genuine
off-shell contributions only. As Figure \ref{fig:dis:8a} suggests, a
more proper modeling of the NLO distribution for $M_{be^+} > 153$ GeV
is expected by the  use of the dynamical scales. Also, for the $M_t$
observable, shown in Figure~\ref{fig:dis:8b}, both at LO and
NLO, the NLO shape is accurately given only with $\mu_0=E_T/2$ and
$\mu_0=H_T/2$.  Moreover, from Figure~\ref{fig:dis:8} we can read that
at the beginning of the spectrum NLO corrections are large up to
$120\%$, $140\%$ and $145\%$ for $\mu_0=m_t$, $\mu_0=E_T/2$ and
$\mu_0=H_T/2$ respectively. Additionally, large distortions are
observed for this observable independently of the scale choice.

To summarise this part, we have studied the size of NLO QCD
corrections to numerous differential cross sections. For many
observables we have found substantial variations in their magnitude,
which depend on the observable itself, the scale choice and the
considered phase-space regions.  Overall we confirm the validity of
the proposed dynamical scales $\mu_0=E_T /2$ and $\mu_0=H_T /2$, where
the latter provides the smallest scale uncertainties. The fixed scale
choice $\mu_0 = m_t$, on the contrary, does not prove adequate in our
analysis for the modeling of differential cross sections.

%
\subsection{Theoretical Uncertainties for Differential Cross 
Sections}
%

%
\begin{figure}[t!]
\caption{\it Averaged NLO differential cross section distributions as
a function of the transverse momentum of the top quark, bottom-jet and
charged lepton. Also given is the NLO differential cross section as a
function of the transverse momentum of the hardest light jet. Results
are shown for the $pp\to e^+ \nu_e \mu^- \bar{\nu}_\mu b\bar{b}j +X$
process at the LHC run II with $\sqrt{s} = 13$ TeV for  three
different PDF sets.  Lower panels display scale and PDF
uncertainties of the NLO cross section normalised to the central NLO
prediction with CT14.}
\begin{center}
\includegraphics[width=0.45\textwidth]{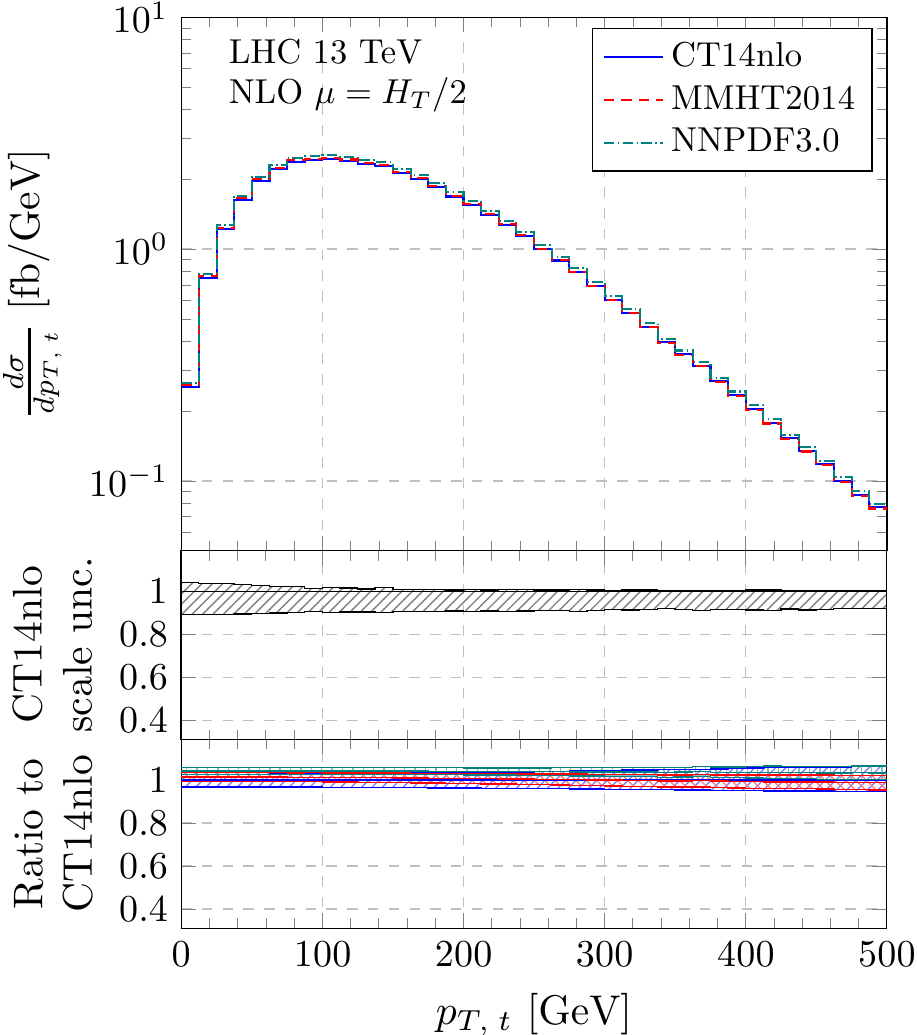}
\includegraphics[width=0.45\textwidth]{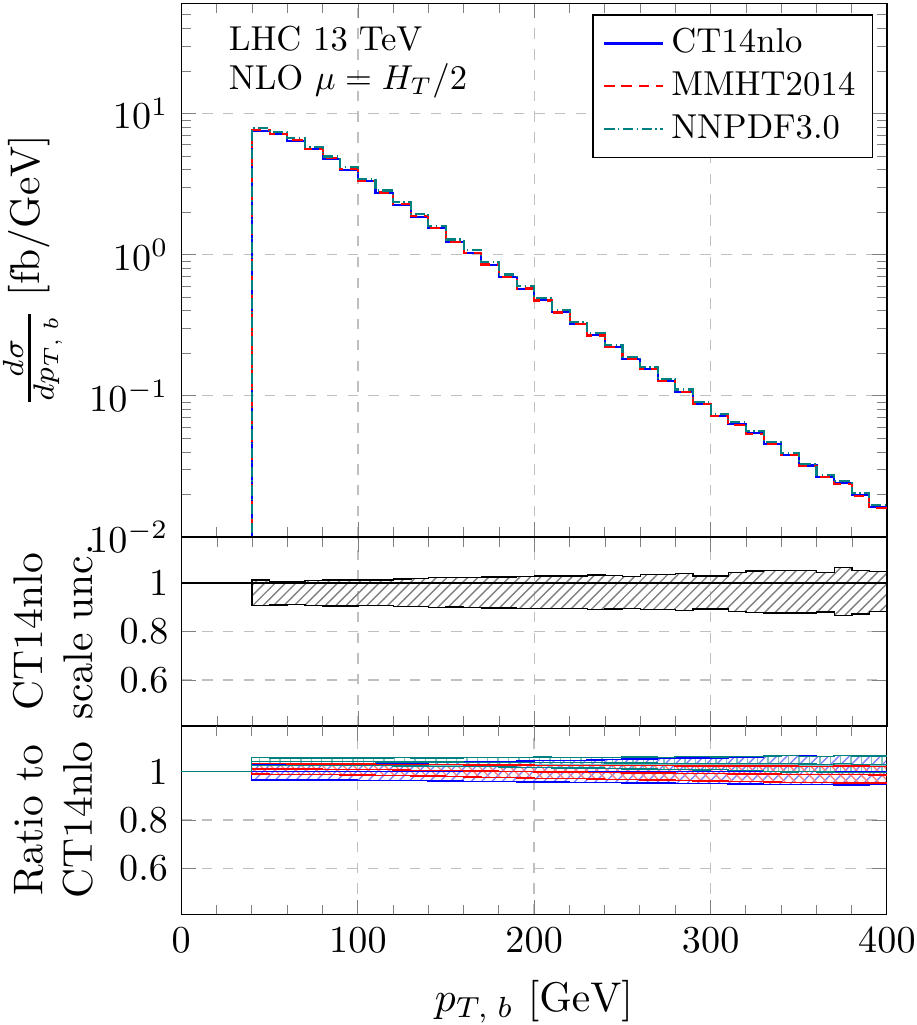}\\
\vspace{0.2cm}
\includegraphics[width=0.45\textwidth]{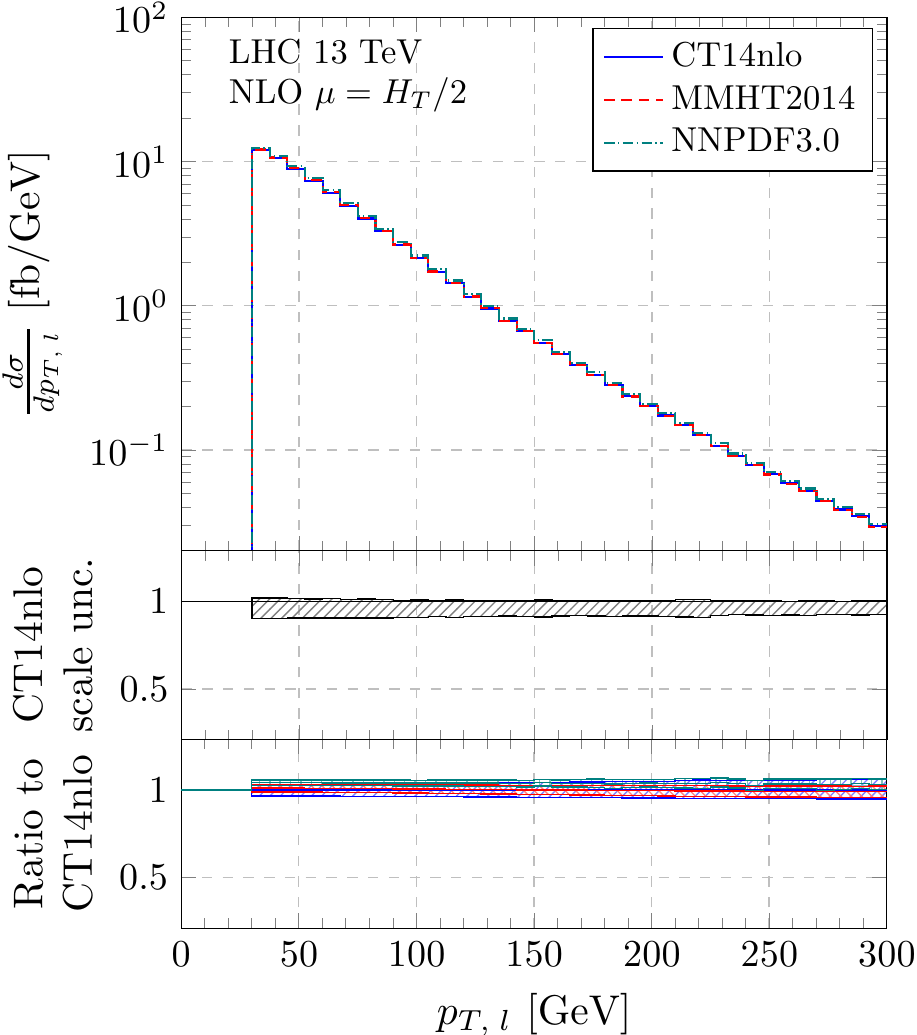}
\includegraphics[width=0.45\textwidth]{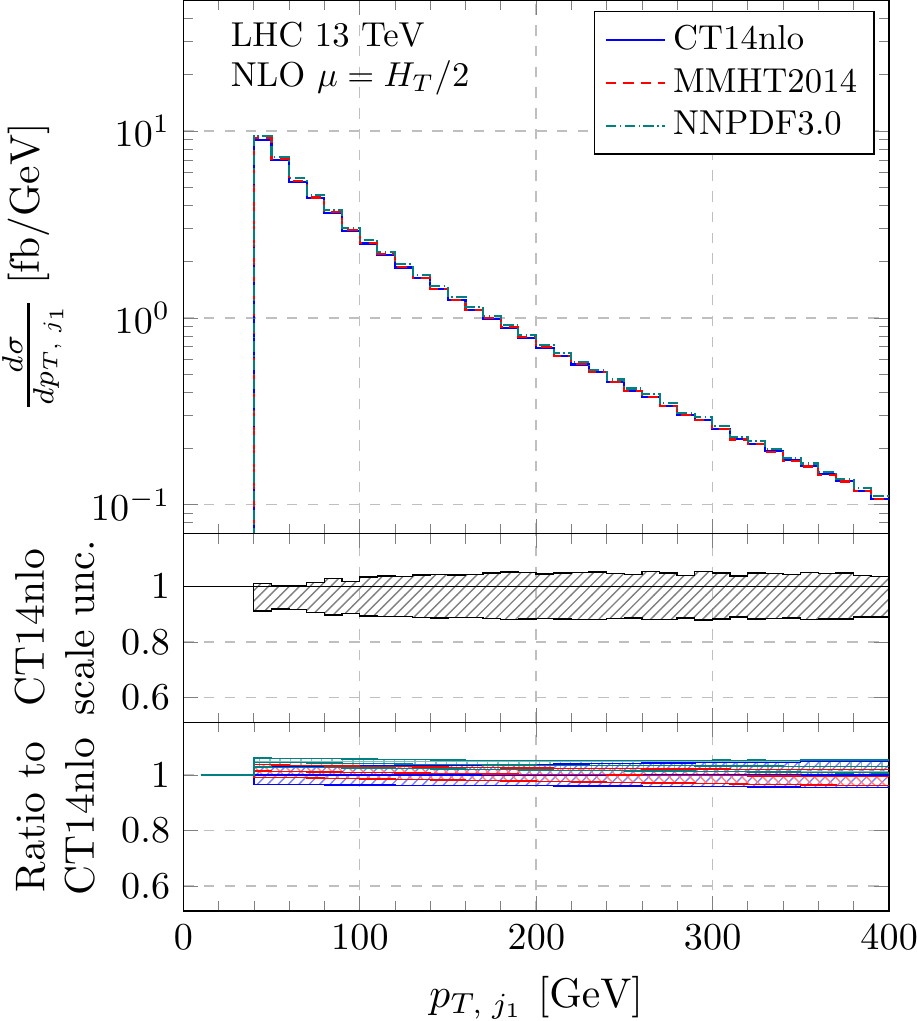}
\end{center}
\label{fig:pdf:1} 
\end{figure}
\begin{figure}[t!]
\caption{\it NLO differential cross section distributions as a
function of the invariant mass of the $t\bar{t}$ system, charged
leptons and bottom-jet and positron.  Also given is the averaged NLO
differential cross section distribution as a function of the invariant
mass of the top quark.  Results are shown for the $pp\to e^+ \nu_e
\mu^- \bar{\nu}_\mu b\bar{b}j +X$ process at the LHC run II with
$\sqrt{s} = 13$.}
\begin{center}
\includegraphics[width=0.45\textwidth]{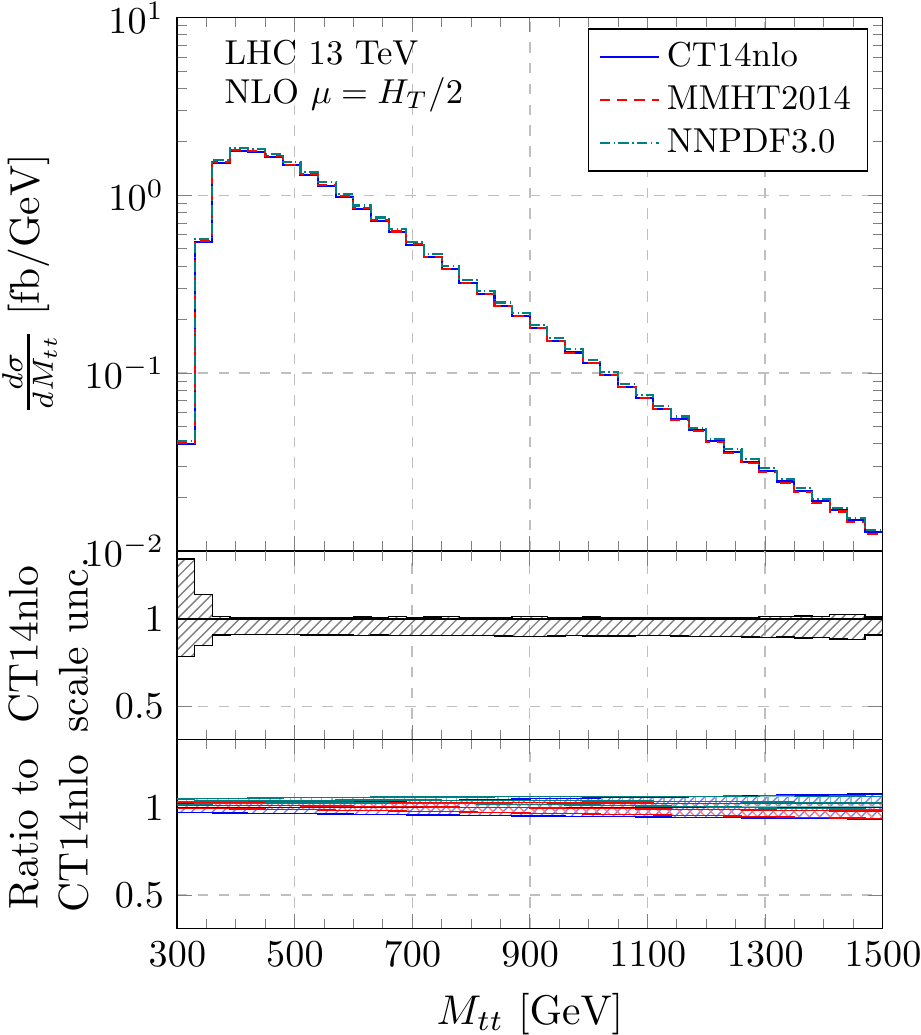}
\includegraphics[width=0.45\textwidth]{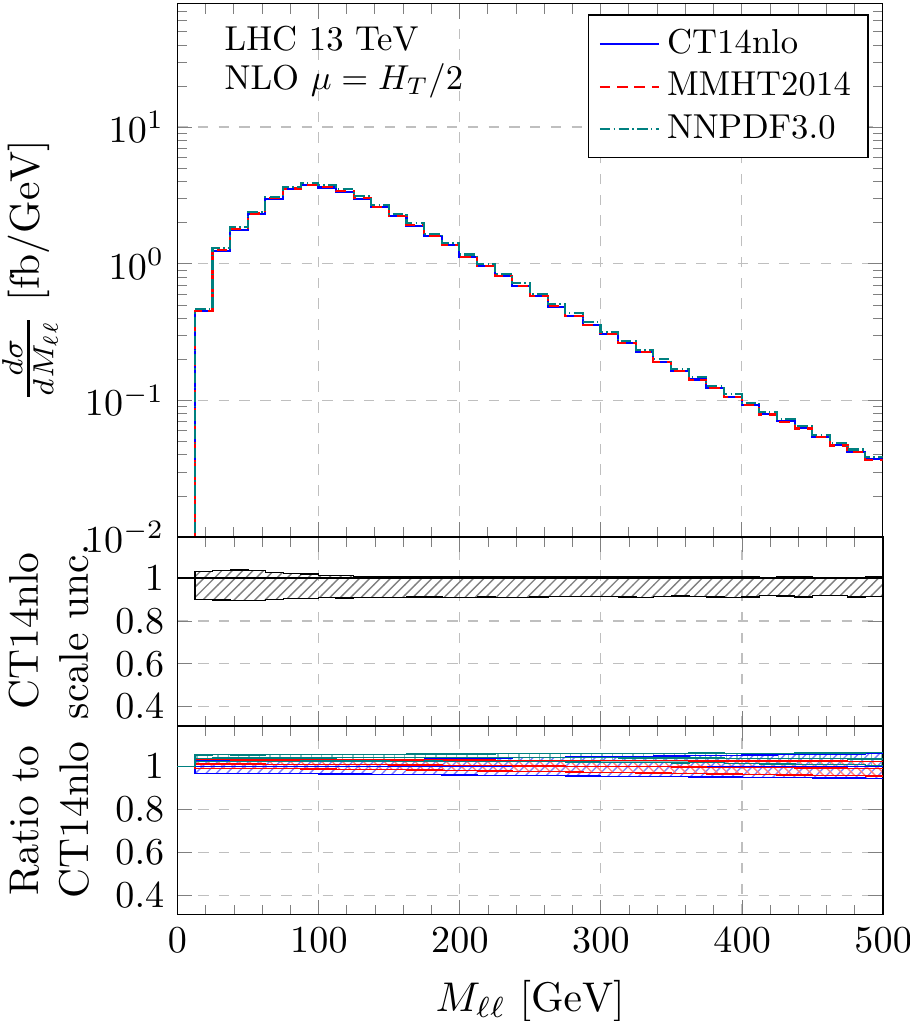}\\
\vspace{0.2cm}
\includegraphics[width=0.45\textwidth]{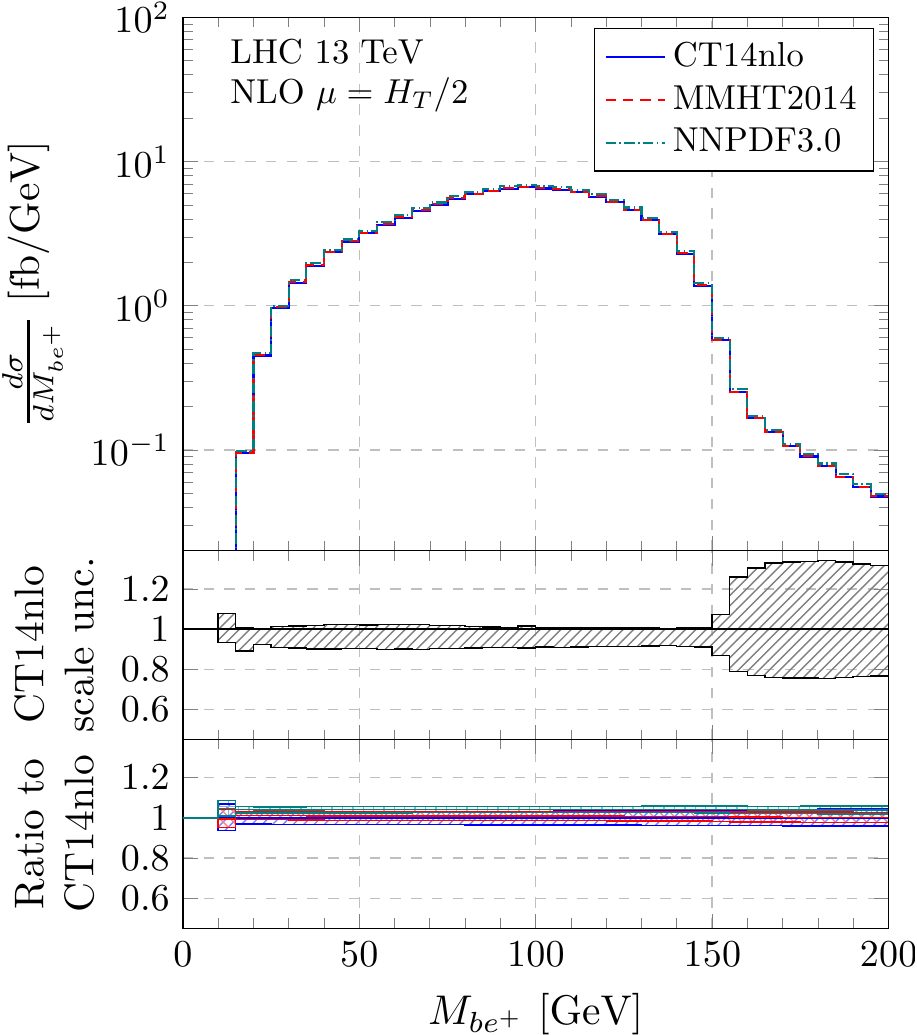}
\includegraphics[width=0.45\textwidth]{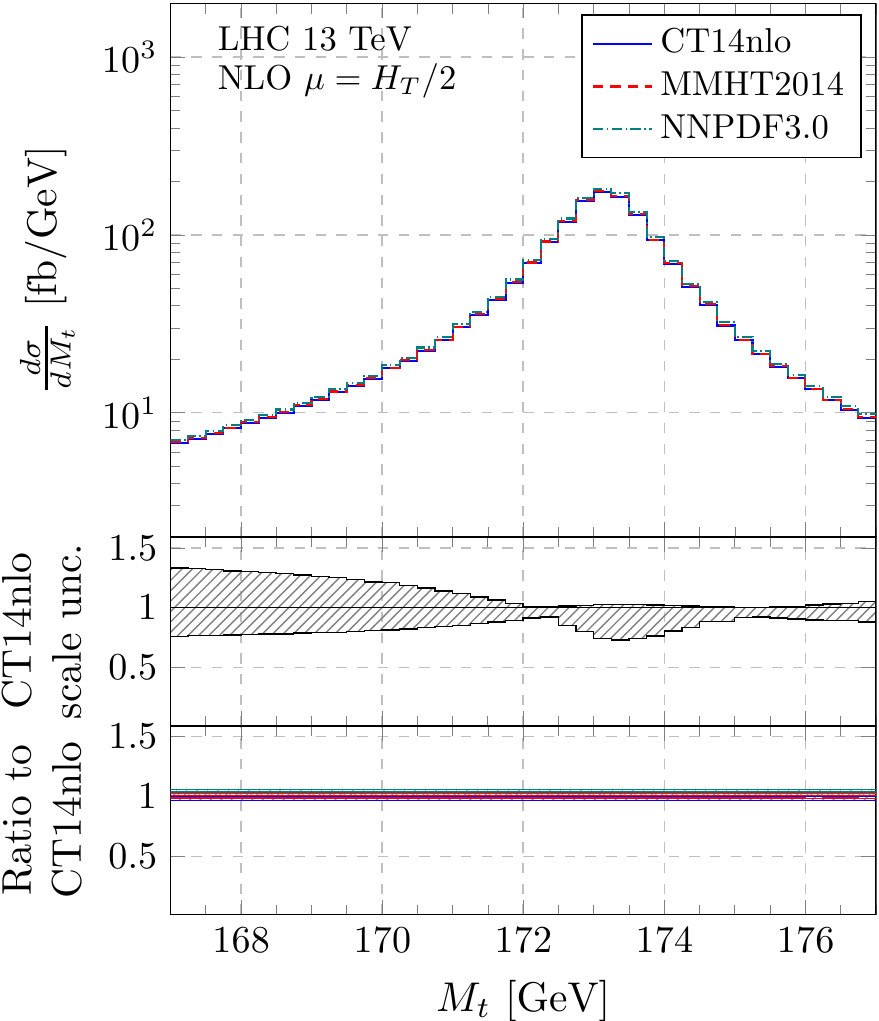}
\end{center}
\label{fig:pdf:2} 
\end{figure}
\begin{figure}[t!]
\caption{\it Averaged NLO differential cross section distributions as
a function of the rapidity of the top quark, bottom-jet and
charged lepton. Also given is the NLO differential cross section as a
function of the rapidity of the hardest light jet. Results are
shown for the $pp\to e^+ \nu_e \mu^- \bar{\nu}_\mu b\bar{b}j +X$
process at the LHC run II with $\sqrt{s} = 13$ TeV.}
\begin{center}
\includegraphics[width=0.45\textwidth]{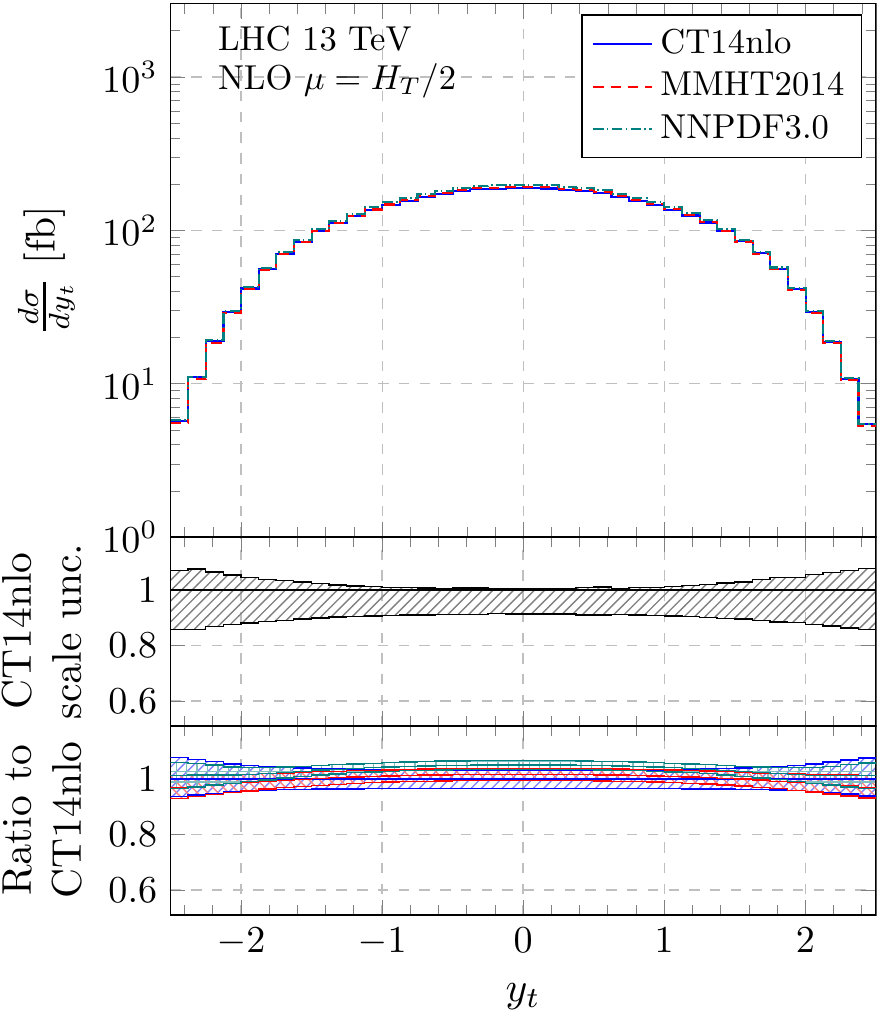}
\includegraphics[width=0.45\textwidth]{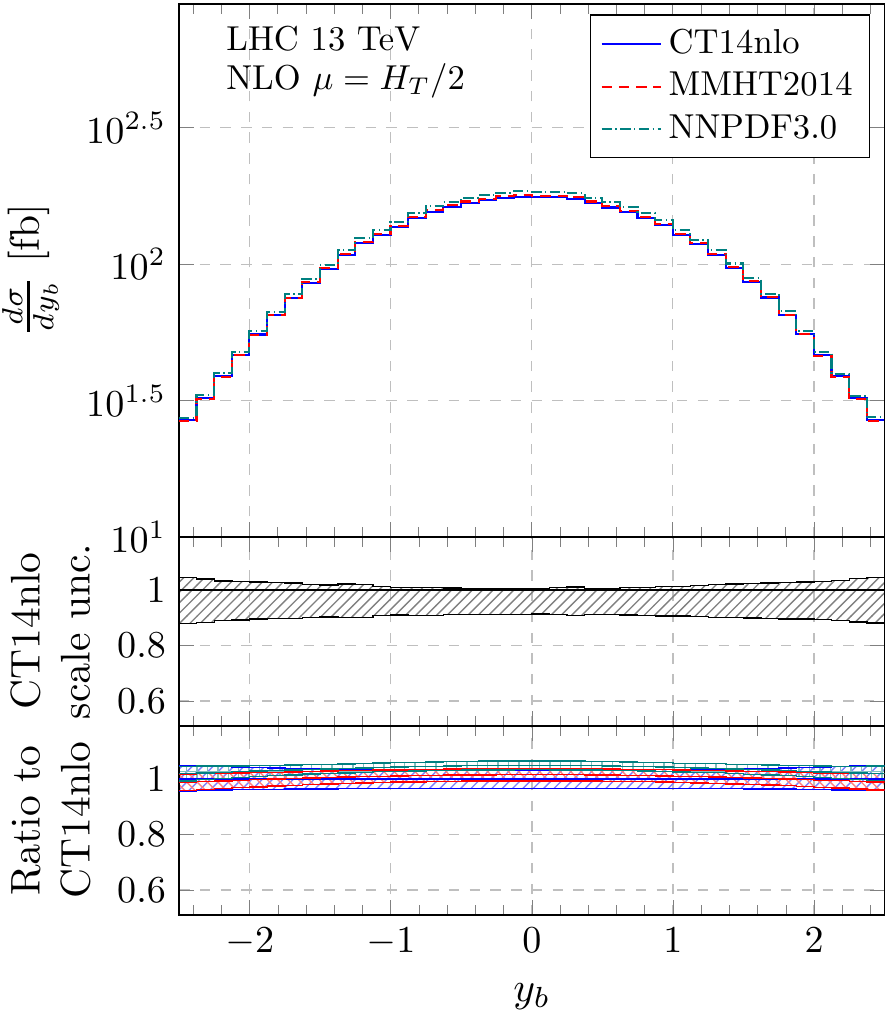}\\
\vspace{0.2cm}
\includegraphics[width=0.45\textwidth]{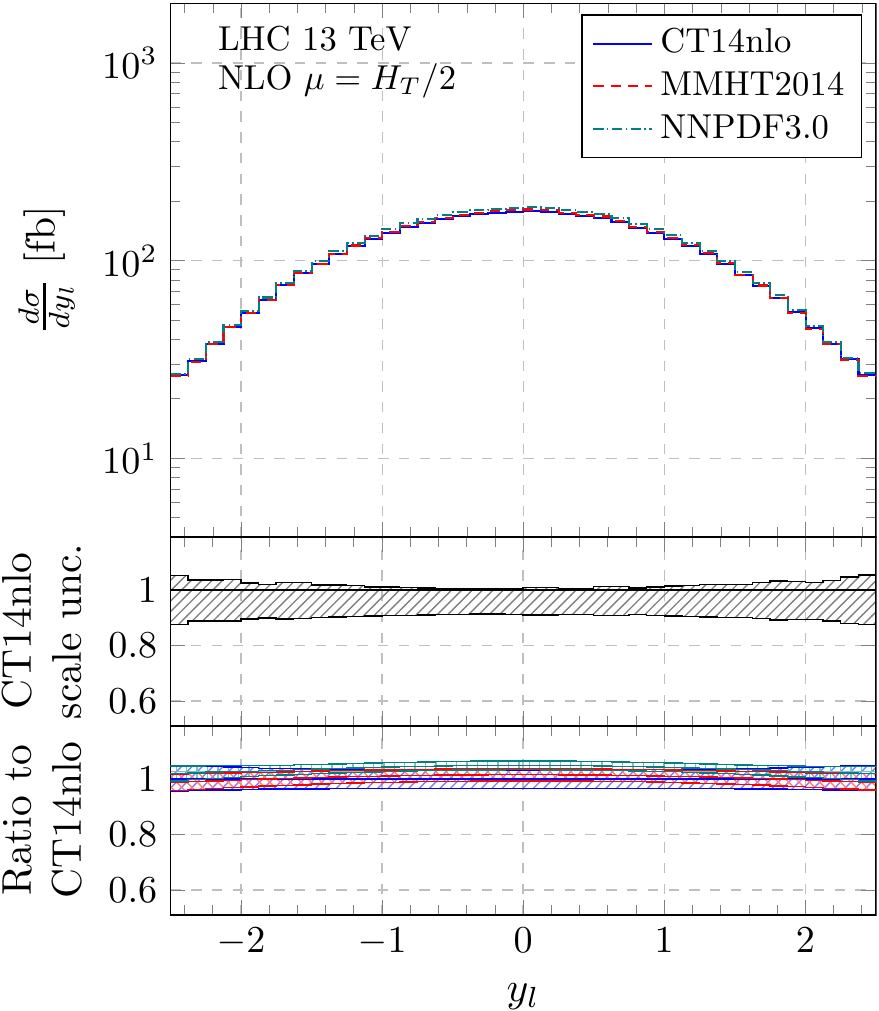}
\includegraphics[width=0.45\textwidth]{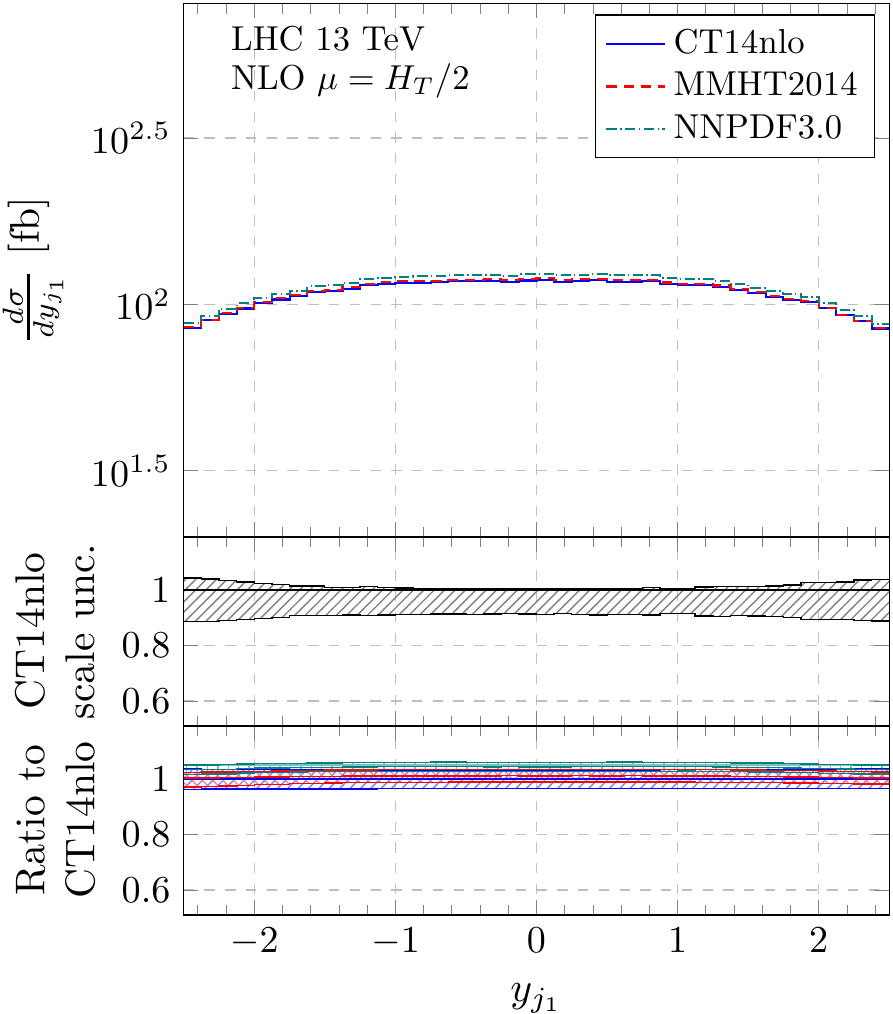}
\end{center}
\label{fig:pdf:3} 
\end{figure}
\begin{figure}[t!]
\caption{\it NLO differential cross section distributions as a
function of the total transverse momentum of the system, missing
transverse momentum, $\Delta R_{\ell\ell}$ and $\Delta \phi_{\ell
\ell}$.  Results are shown for the $pp\to e^+ \nu_e \mu^-
\bar{\nu}_\mu b\bar{b}j +X$ process at the LHC run II with $\sqrt{s} =
13$ TeV .}
\begin{center}
\includegraphics[width=0.45\textwidth]{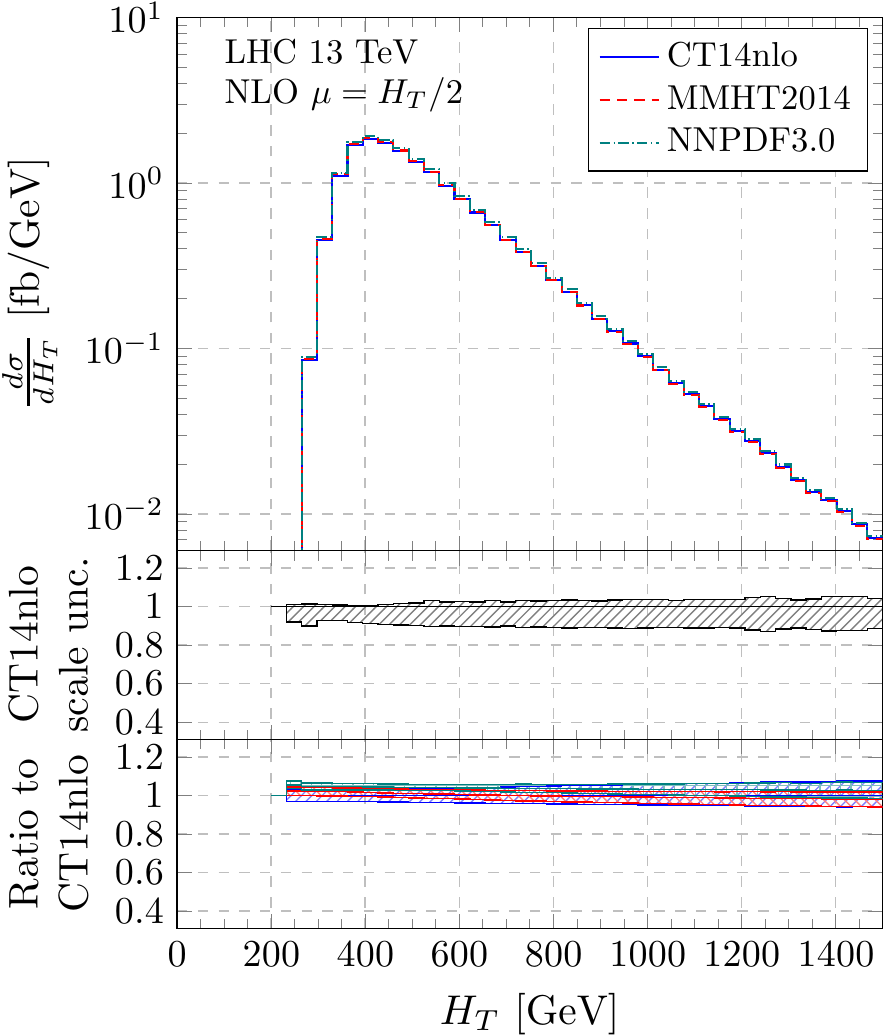}
\includegraphics[width=0.45\textwidth]{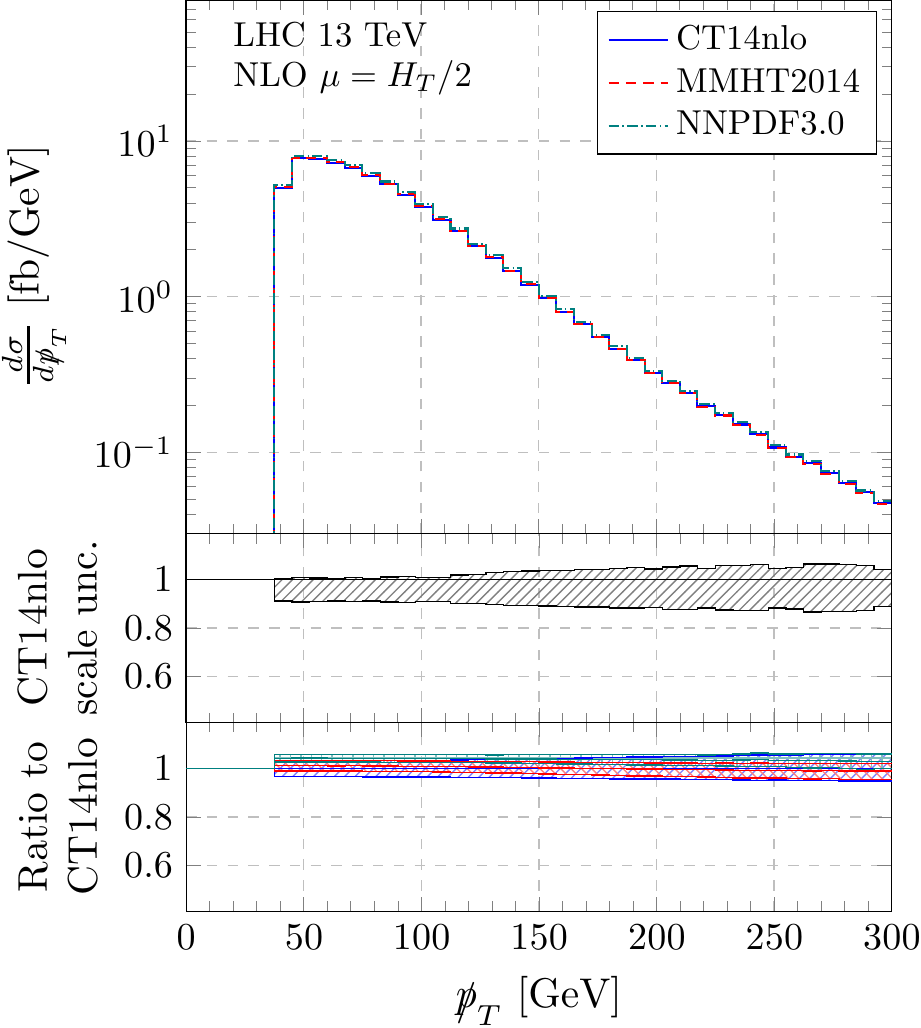}\\
\vspace{0.2cm}
\includegraphics[width=0.45\textwidth]{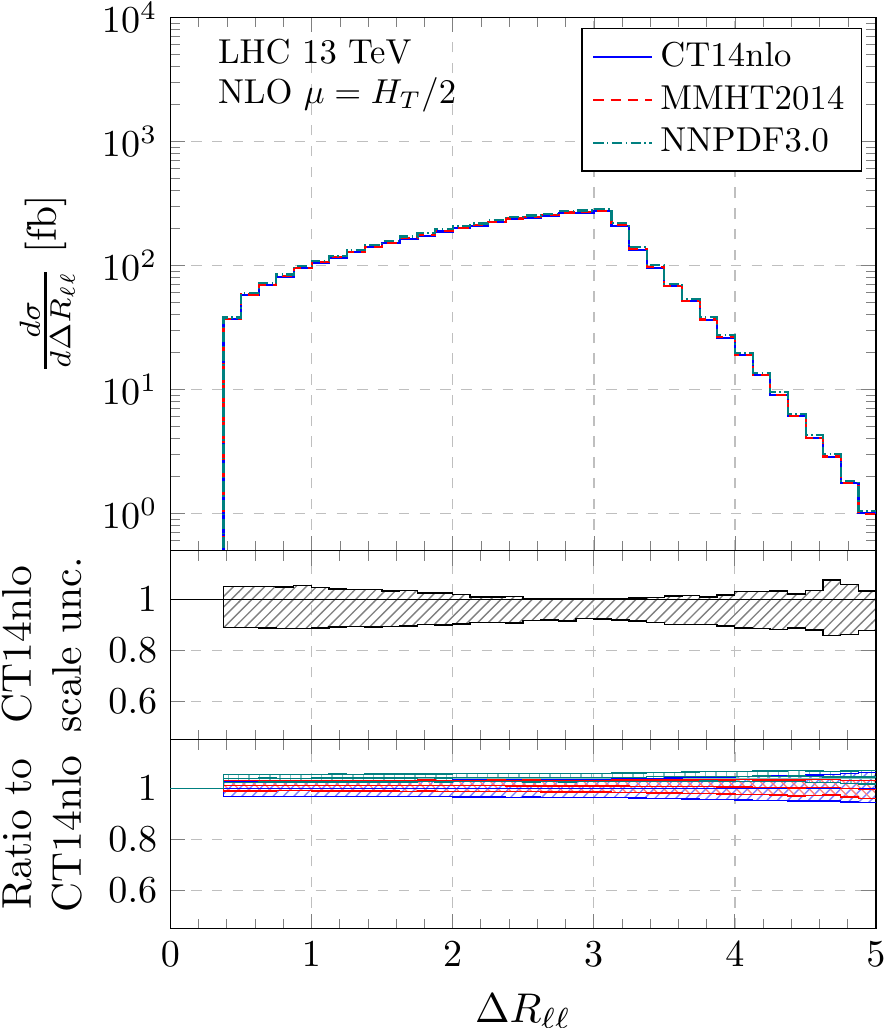}
\includegraphics[width=0.45\textwidth]{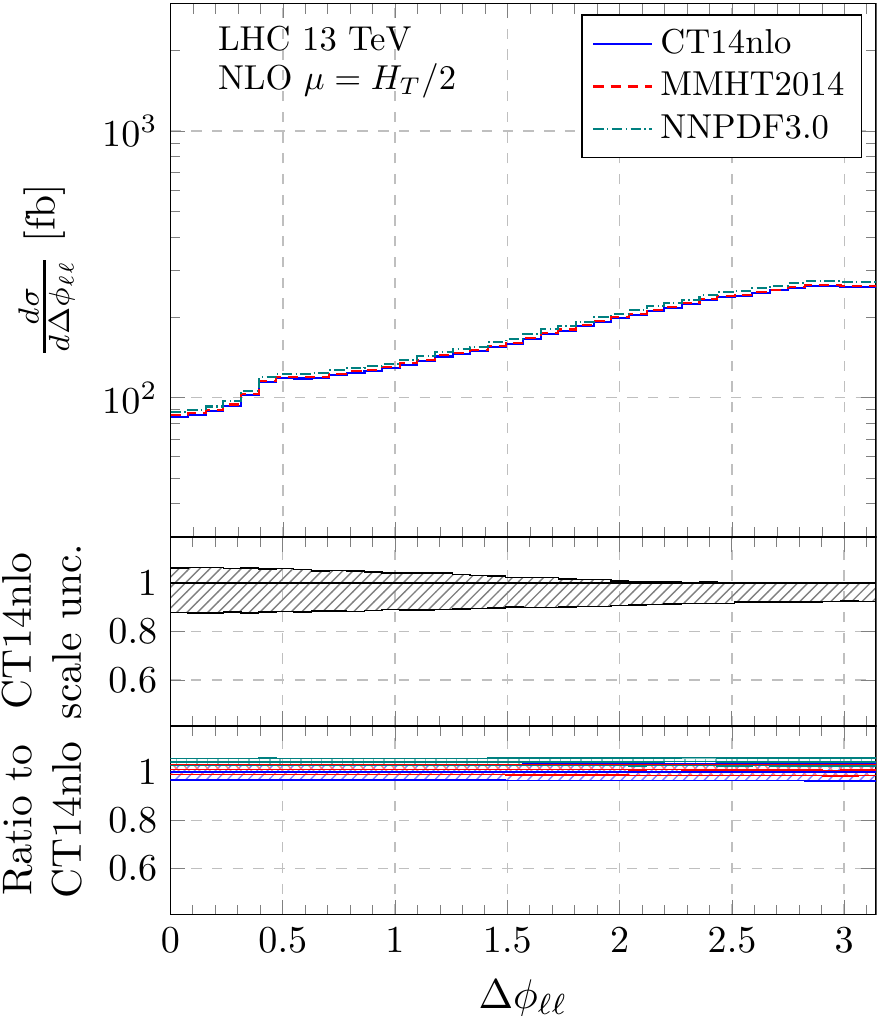}
\end{center}
\label{fig:pdf:4} 
\end{figure}

At this point we would like to fully assess the uncertainties inherent
in our NLO differential predictions. An extensive discussion of the
scale uncertainties has already been presented in the previous
section, based on a fixed PDF choice (CT14). In this section, we
complete our analysis by studying comparatively the impact of PDF and
scale variations on the overall theoretical uncertainty.  Judging by
the dependence of the  total cross section, the PDF uncertainties should be
below or of the same order as the theoretical uncertainties predicted
by the scale variation.  Nevertheless, we would like to examine this
carefully for all differential cross sections that we have presented in the
previous section.  To this end we plot afresh NLO differential cross
sections for our best (dynamical) scale choice, $\mu_0=H_T/2$, for
three different PDF sets, CT14, MMHT14 and NNPDF3.0. We shall start
with the averaged distribution of the transverse momentum of the top
quark, bottom-jet and charged lepton that are shown in
Figure~\ref{fig:pdf:1}. Also given there is the transverse momentum
distribution of the hardest light-jet. Each figure comprises three
parts; the upper panel shows the NLO prediction for three different
PDF sets at the central scale value, $\mu_R=\mu_F=\mu_0=H_T/2$, the
middle panel displays the NLO scale-dependence band normalised to the
central CT14 NLO prediction, whereas the lower panel gives the PDF
uncertainties obtained for each PDF set separately, again normalised
to the central NLO prediction as obtained with the CT14 PDF set.  For
each observable plotted in Figure~\ref{fig:pdf:1} we obtain
symmetrised scale uncertainties below $10\%$ (with respect to the
central value). To be more specific we have estimated $8\%$ scale
uncertainties for $p_{T,\,t}$ at the beginning of the spectrum, which
decreased down to $4\%$ in the tails. In these high $p_T$ regions,
however, PDF uncertainties are of a comparable size, i.e. they are of
the order of $6\%$ for CT14 and $3\%$ for MMHT14 and NNPDF3.0 (again
with respect to the corresponding central values). For the
transverse momentum distribution of the bottom-jet we have a different
behaviour, namely scale uncertainties have increased in the tails and
reached almost $10\%$ while PDF uncertainties stayed below $6\%$
($3\%$) for CT14 (for MMHT14 and NNPDF3.0).  For the transverse
momentum distribution of the charged lepton we find that the scale
variations are of the order of the error of the CT14 PDFs, i.e. below
$6\%$. The other two PDF sets show a smaller uncertainty, of the order
of $3\%$. For the hardest light jet in the whole plotted range CT14
PDF uncertainties are below scale uncertainties. The former are
estimated to be below $5\%$ the latter below $9\%$. For MMHT14 and
NNPDF3.0 PDF sets we have respectively $3\%$ and $2\%$ PDF
uncertainties.

In Figure~\ref{fig:pdf:2} we present the invariant mass of the
$t\bar{t}$ system, $M_{t\bar{t}}$, of two charged leptons,
$M_{\ell\ell}$, and of bottom-jet and positron, $M_{be^+}$, together
with the mass of the reconstructed top quark, $M_t$. We start with the
invariant mass of the $t\bar{t}$ pair. In the vicinity of the
$t\bar{t}$ threshold sizeable, of the order of $30\%$, scale
uncertainties are attained. However, starting from about $400$ GeV
almost constant $5\%$ uncertainties are noticed until the end of the
plotted spectrum, i.e. up to $1.5$ TeV. As for the PDF uncertainties
we observe $7\%$ effects in the tail of this distribution in case of
CT14 and $4\%$ for MMHT14 and NNPDF3.0. For the invariant mass of the
positron-muon system, $M_{\ell\ell}$, we observe $7\%$ scale
uncertainties for small values, decreasing down to $5\%$ after $100$
GeV. Thus, around $500$ GeV they are comparable to CT14 PDF
uncertainties that are of the order of $6\%$. Also here this effect is
smaller for MMHT14 and NNPDF3.0 ($4\%$ and $3\%$). The invariant mass
of the bottom-jet and positron, $M_{be^+}$, has clearly two distinct
ranges when it comes to scale uncertainties. Up to the kinematical
endpoint they are of the order of $6\%$, on the other hand, above this
point they reach $30\%$. PDF uncertainties, as expected, do not affect
$M_{be^+}^{\rm max}$ and are of the order of $2\%-4\%$ in the whole
range independently of the PDF set. The mass of the reconstructed top
quark, that is presented close to the resonance, has a more complex
pattern. Even for such a small range, i.e. $167-177$ GeV, we can
distinguish three different regions. Up to $170$ GeV scale
uncertainties are within the $20\%-30\%$ range, they are decreased
down to $10\%-15\%$ for $M_t\in (170-174)$ GeV and are further reduced
below $10\%$ for $M_t> 174$ GeV. One more time,  PDF uncertainties
remain the same in the whole plotted range and are of the order of
$2\%-3\%$, well below scale uncertainties, independently of the PDF
set used.
 
In Figure~\ref{fig:pdf:3} we show dimensionless  observables, namely
rapidity distributions for the top quark, bottom jet, charged lepton
and the hardest light-jet. In the central rapidity regions of $y_{t}$
scale uncertainties are of the order of $5\%$, whereas they reach
$10\%$ at the peripheral parts of the distribution. The CT14 (MMHT14,
NNPDF3.0) PDF uncertainties are at the level of $3\%$ ($2\%$) and
$7\%$ ($4\%$ and $5\%$) in these two distinct regions. In case of the rapidity
distribution of the bottom-jet scale uncertainties are below $8\%$ and
the PDF uncertainties are in the range $3\%-5\%$. A similar pattern
could be recognised for $y_{\ell}$ and $y_{j_1}$.

Finally, in Figure~\ref{fig:pdf:4} we plot the total transverse
momentum of the $t\bar{t}j$ system, the missing transverse momentum,
$\Delta R_{\ell\ell}$ and $\Delta \phi_{\ell\ell}$. For  the $H_T$
distribution, scale
uncertainties are below $10\%$. In the high $p_T$ region they are
comparable to the CT14 PDF uncertainties that are of the order of
$7\%$. For the other PDF sets we obtained PDF uncertainties below
$4\%$. Qualitatively comparable conclusions have been reached for the
$\slashed{p}_T$ distribution. Also for $\Delta R_{\ell\ell}$ and $\Delta
\phi_{\ell\ell} $ distributions, we have estimated scale uncertainties below
$10\%$. The PDF uncertainties for $\Delta R_{\ell\ell}$ have been
found to be below $6\%$, $3.5\%$ and $3\%$ for the CT14, MMHT14 and
NNPDF3.0 PDF sets respectively. In the case of $\Delta \phi_{\ell\ell} $
they are slightly smaller, i.e. $4\%$ for CT14 and $2\%$ for the
MMHT14 and NNPDF3.0 PDF sets.

%
\section{Conclusions}
\label{Conclusions}
%

In this paper,  we have presented a comprehensive NLO study of the
off-shell production of $t\bar{t}$ + jet with leptonic decays of the
top quarks. All results have been obtained by use of the package
\textsc{Helac-NLO}. We have shown predictions for total cross sections
and distributions for a variety of observables of phenomenological
interest for the LHC Run II energy of $13$ TeV. Also, we have
carefully assessed the theoretical uncertainties of our predictions
stemming from scale dependence and from different PDF
parametrizations. For our best scale choice, $\mu_R =\mu_F = H_T /2$,
the QCD corrections to the total cross section are positive and vary
from rather small to moderate. To be more specific, we have
obtained corrections of the order $15\%$ for the CT14 PDF set, $6\%$
for MMHT14 and $24\%$ for NNPDF3.0. As to the theoretical
uncertainties, taking them conservatively from the upper and lower
results, we have observed a reduction from $50\%$ at LO down to $10\%$
at the NLO. Using symmetrization, the scale uncertainties become
$40\%$ at LO and $6\%$ at NLO. The PDF uncertainties have been
assessed to be rather small at the inclusive level, within the range
of $2\% - 3\%$. Moreover, results have been found to be quite stable
for cuts on the $p_T$ of the hard jet ranging from $40$ GeV to $120$
GeV.

We have considered several differential distributions which are
relevant for the ongoing analyses at the LHC. Two different dynamical
scales have been considered for our analysis, $\mu_R = \mu_F = E_T /2$
and $\mu_R = \mu_F = H_T /2$, which proved both effective in
stabilizing the perturbative convergence in phase space regions far
away from the $2 m_t$ threshold. Of the two scales, $H_T /2$ is the
one which provides the smallest theoretical uncertainties as estimated
by the scale variation. The size of the QCD corrections varies
considerably from observable to observable. For the majority of cases
we have found that corrections are below $10\%-20\%$, yet they can
exceed $100\%$ for specific observables independently of the scale
choice. At the differential level,  PDF uncertainties are found to be of
comparable size, i.e. below $10\%$, thus they cannot simply be
 neglected. This fact is particularly evident using the CT14 PDF set,
while the uncertainties related to the MMHT14 and NNPDF3.0 sets are
within the scale dependence ones.

In the next step, we plan to use our predictions to study broad
phenomenological aspects of top quark physics at the LHC. Our priority
is to assist precise measurements of the top quark mass at the LHC,
where the impact of the off-shell effects has to be carefully examined
in order to assess realistically the  theoretical uncertainties. To this
end, a systematic comparison with predictions based on the
narrow-width approximation is required. We also plan to quantify the
impact of b-quark mass effects at NLO by means of comparisons between
the so-called Five-Flavour and Four-Flavour schemes.

\acknowledgments
The work of H. B. Hartanto and M. Worek was supported by the German
Research Foundation (DFG). M. Kraus acknowledges support by the German
Federal Ministry of Education and Research (BMBF). Authors would like
to thank Adam Kardos for help with the \textsc{Qgraf} program.

\end{document}